\documentclass[twocolumn,aps,prl,superscriptaddress]{revtex4-1}
\usepackage{amssymb} 
\usepackage{amsmath,bm} 
\usepackage{graphicx} 
\usepackage[normalem]{ulem} 
\usepackage{multirow} 
\usepackage[colorlinks,linkcolor=blue,urlcolor=blue,citecolor=blue]{hyperref} 
\usepackage{lipsum} 
\usepackage[usenames, dvipsnames]{xcolor} 
\usepackage{tensor} 
\usepackage{isotope} 
\usepackage{amsmath} 
\usepackage{comment} 
\usepackage{physics}

\usepackage{xspace}
\usepackage{epsfig} 
\usepackage{fancyhdr}
\usepackage{pslatex}   
\usepackage{xcolor}
\usepackage{color} 

\usepackage{mathtools}

\allowdisplaybreaks
\newcommand {\Li} {\mbox{$^4$Li}\xspace}

\renewcommand{\sout}{\bgroup \color{red} \ULdepth=-.5ex \ULset}

\begin{document}

\title{Global Spin Alignment of (Anti-)\Li~in Non-Central Heavy-Ion Collisions}
\author{Yun-Peng Zheng}
\affiliation{Key Laboratory of Nuclear Physics and Ion-beam Application~(MOE), Institute of Modern Physics, Fudan University, Shanghai $200433$, China} 
\affiliation{Shanghai Research Center for Theoretical Nuclear Physics, NSFC and Fudan University, Shanghai 200438, China}
\author{Dai-Neng Liu} 
\affiliation{Key Laboratory of Nuclear Physics and Ion-beam Application~(MOE), Institute of Modern Physics, Fudan University, Shanghai $200433$, China} 
\affiliation{Shanghai Research Center for Theoretical Nuclear Physics, NSFC and Fudan University, Shanghai 200438, China}

\author{Lie-Wen Chen}
\email{lwchen$@$sjtu.edu.cn}
\affiliation{State Key Laboratory of Dark Matter Physics, Key Laboratory for Particle Astrophysics and Cosmology (MOE), and Shanghai Key Laboratory for Particle Physics and Cosmology, School of Physics and Astronomy, Shanghai Jiao Tong University, Shanghai 200240, China}

\author{Jin-Hui Chen} 
\email{chenjinhui@fudan.edu.cn} 
\affiliation{Key Laboratory of Nuclear Physics and Ion-beam Application~(MOE), Institute of Modern Physics, Fudan University, Shanghai $200433$, China} 
\affiliation{Shanghai Research Center for Theoretical Nuclear Physics, NSFC and Fudan University, Shanghai 200438, China} 

\author{Che Ming Ko} 
\email{ko@comp.tamu.edu} 
\affiliation{Cyclotron Institute and Department of Physics and Astronomy, Texas A\&M University, College Station, Texas 77843, USA} 

\author{Yu-Gang Ma} 
\email{Correspond to: mayugang@fudan.edu.cn} 
\affiliation{Key Laboratory of Nuclear Physics and Ion-beam Application~(MOE), Institute of Modern Physics, Fudan University, Shanghai $200433$, China} 
\affiliation{Shanghai Research Center for Theoretical Nuclear Physics, NSFC and Fudan University, Shanghai 200438, China}

\author{Kai-Jia Sun} 
\email{Correspond to: kjsun@fudan.edu.cn} 
\affiliation{Key Laboratory of Nuclear Physics and Ion-beam Application~(MOE), Institute of Modern Physics, Fudan University, Shanghai $200433$, China} 
\affiliation{Shanghai Research Center for Theoretical Nuclear Physics, NSFC and Fudan University, Shanghai 200438, China}

\author{Jun Xu} 
\email{junxu@tongji.edu.cn} 
\affiliation{School of Physics Science and Engineering, Tongji University, Shanghai 200092, China}  
\date{\today}

\author{Bo Zhou} 
\email{zhou\_bo@fudan.edu.cn} 
\affiliation{Key Laboratory of Nuclear Physics and Ion-beam Application~(MOE), Institute of Modern Physics, Fudan University, Shanghai $200433$, China} 
\affiliation{Shanghai Research Center for Theoretical Nuclear Physics, NSFC and Fudan University, Shanghai 200438, China}
\date{\today}

\begin{abstract}    
Non-central heavy-ion collisions produce hot and dense nuclear matter with significant fluid vorticity, which can induce global polarizations or alignments of particles with non-zero spins along 
the direction of the total orbital angular momentum. This phenomenon has been observed for  hyperons and vector mesons in experiments.  In the present study, we demonstrate that polarized 
nucleons lead to global spin alignment of the unstable nucleus $^4$Li, which can be measured through its strong decays via $^4\text{Li} \rightarrow {^3\text{He}} + p$. Assuming that $^4$Li is 
formed through the coalescence of polarized nucleons at kinetic freeze-out, we obtain the angular distribution of the daughter particle $^3$He in the rest frame of the polarized $^4$Li. 
 Taking kinetically freeze-out nucleons from an isotropic and thermalized fireball of constant vorticity and including quantum corrections up to $\hbar^2$ in the coalescence 
calculation  through the Moyal star product, we find that the angular distribution of $^3$He has a $\cos(2\theta^*)$ dependence with $\theta^*$ being 
 its angle  with respect to the quantization axis of $^4$Li.  We also find that the $^3$He angular distribution 
depends on both the vorticity and the polarization of kinetically freeze-out nucleons. Future measurements on the spin alignment of $^4$Li in heavy-ion collisions thus offer a promising method to probe the spin dynamics, vortical structure, and spin-dependent equation-of-state of the nuclear matter produced in these collisions.
\end{abstract}
\pacs{12.38.Mh, 5.75.Ld, 25.75.-q, 24.10.Lx} 
\maketitle
\emph{Introduction.}{\bf ---} 
In non-central relativistic heavy-ion collisions with a very large orbital angular momentum of the colliding nuclei partially transferred to the produced quark-gluon plasma (QGP) in the form of fluid vorticity, the produced particles of non-zero spins  are expected to be partially polarized along the direction of the initial orbital angular momentum through their spin-orbit couplings~\cite{Liang:2004ph,Liang:2004xn,Voloshin:2004ha,Betz:2007kg,Becattini:2007sr,Gao:2007bc}. Experiments at the Relativistic Heavy-Ion Collider (RHIC)~\cite{STAR:2017ckg,STAR:2018gyt,STAR:2019erd} and the Large Hadron Collider (LHC)~\cite{ALICE:2019onw} have measured a global $\Lambda$ ($\bar{\Lambda}$) polarization of magnitude expected from a QGP that is the most vortical fluid with a vorticity of $\omega\approx (9\pm 1)\times 10^{21}s^{-1}$~\cite{STAR:2017ckg}. The collision energy dependence of measured  $\Lambda$ polarization can be well reproduced by many model calculations~\cite{Karpenko:2016jyx, Li:2017slc, Sun:2017xhx, Wei:2018zfb, Deng:2021miw,Sun:2025oib,Deng:2025xfo}. Intriguing polarization and alignment patterns have been observed for $\Xi$ and $\Omega$ hyperons~\cite{STAR:2020xbm} as well as  the $\phi$~\cite{STAR:2008lcm,ALICE:2019aid,STAR:2022fan,Sheng:2022wsy}, $J/\psi$~\cite{ALICE:2020iev,ALICE:2022dyy}, and $D^*$ mesons~\cite{ALICE:2025cdf}.  Comprehensive reviews of these very interesting phenomena on hadron polarizations can be found in Refs.~\cite{Becattini:2007nd,Huang:2020xyr,Becattini:2020ngo,Huang:2020dtn,Gao:2020vbh,Becattini:2021wqt,Hidaka:2022dmn,Chen:2023hnb,Becattini:2024uha,Niida:2024ntm,Chen:2024aom,Chen:2024afy,Huang:2024ffg}.

Beyond hyperons and vector mesons, loosely-bound states such as the deuteron, helium-3, and hypertriton may acquire non-zero polarization when formed via the coalescence of polarized nucleons at kinetic freeze-out~\cite{Liu:2023nkm,Sun:2025oib}. For stable proton, deuteron, and helium-3, direct measurement of their spin orientations is experimentally challenging.  It requires external polarimeters based on secondary interactions~\cite{Liang:2025owx}, which is not available in current heavy-ion experiments. 

In contrast, the polarization of unstable hypertriton and $^4$Li can be inferred from the angular distributions of their decay products. Recent studies have investigated the spin polarizations of hypertriton and anti-hypertriton~\cite{Sun:2025oib,Liu:2025kpp}. The spin alignment or tensor polarization of $^4$Li is of particular interest at low collision energies, where hyperons and hypernuclei are rarely produced.  The $^4$Li ground state has spin $J=2$ and decays via $^4\text{Li} \rightarrow {^3\text{He}} + p$ with 100\% branching ratio~\cite{Tilley:1992zz,Vovchenko:2020dmv}. Besides, it has three excited states all decaying into $p-^3$He pairs.  For its anti-particle  $^4\overline{\text{Li}}$, it could be reconstructed via momentum correlation function of $\bar{p}-^3\overline{\text{He}}$~\cite{Xi:2019vev}.

In the present study, we investigate the spin alignment of $^4$Li by analyzing the angular distribution of the decay product $^3$He in the rest frame of $^4$Li mainly through a spin-dependent coalescence model. We will also discuss the quantum corrections and thermal limit of $^4$Li spin alignment.

\emph{Spin alignment of the ground state of~\Li~with spin-parity $J^\pi=2^-$.}{\bf ---}
For the decay of ground state \Li$(2^-)$, the angular distribution  of decay products ($^3$He or $p$) in $^4\text{Li}(2^-) \rightarrow {^3\text{He}} + p$ at the \Li~rest frame is given by~\cite{pilkuhn1967interaction}
\begin{eqnarray}
\label{eq:Angular distribution} 
    \frac{dN}{d\Omega^*}=\frac{\text{Tr}[\hat{T}^\dagger \hat{\rho} \hat{T}]}{\int  d\Omega^* ~ \text{Tr}[\hat{T}^\dagger \hat{\rho} \hat{T}] },
\end{eqnarray}
where $\hat{T}$, $\hat{\rho} $, and $\Omega^*$ denote the two-body decay matrix, the density matrix of \Li, and the solid angle of $^3$He in the rest frame of \Li, respectively. Since the \Li nucleus can be viewed as a loosely bound state of a spin half Helium-3 and a spin half proton with a relative orbital angular momentum $L=1$, the angular dependence of the transition matrix $\hat{T}$ can be written as~\cite{Lee:1957he} 
\begin{widetext}
\begin{equation} 
\label{eq:angle}
    T(^4\text{Li}(2^-) \rightarrow {^3\text{He}} + p)=\sqrt{\frac{3}{8\pi}}\left.\left[\begin{matrix}-\sin\theta^* e^{i\phi^*} & 0&0&0\\\cos\theta^*&-\frac{1}{2}\sin\theta^* e^{i\phi^*}&-\frac{1}{2}\sin\theta^* e^{i\phi^*}&0\\\sqrt{\frac{1}{6}}\sin\theta^* e^{-i\phi^*}&\sqrt{\frac{2}{3}}\cos\theta^*&\sqrt{\frac{2}{3}}\cos\theta^*&-\sqrt{\frac{1}{6}}\sin\theta^* e^{i\phi^*}&\\0&\frac{1}{2}\sin\theta^* e^{-i\phi^*}&\frac{1}{2}\sin\theta^* e^{-i\phi^*}&\cos\theta^*&\\0&0&0&\sin\theta^* e^{-i\phi^*}&\end{matrix}
\right.\right], 
\end{equation}
\end{widetext}
where $\theta^*$ and $\phi^*$ are the polar and azimuthal angles between the momentum of the daughter particles and the quantization axis of \Li~nucleus, respectively.  The differential solid angle in Eq.~(\ref{eq:angle}) is given by $d\Omega^* = \sin\theta^* d\theta^*d\phi^*$.

The spin state of \Li($2^-$) can be described by a 5$\times$5 spin density matrix $\hat{\rho}$ of  unit trace. The diagonal elements of this matrix,  given by $\hat{\rho}_{2,2}$, $\hat{\rho}_{1,1}$, $\hat{\rho}_{0,0}$, $\hat{\rho}_{-1,-1}$, and $\hat{\rho}_{-2,-2}$, are probabilities for the spin component along a quantization axis to take the values of $2$, $1$, $0$, $-1$, and $-2$ respectively. For unpolarized \Li, all its diagonal elements take 1/5. For polarized \Li, we adopted the spin-dependent nucleon coalescence model to calculate its density matrix element. Assuming that the produced \Li propagates as plane waves, the invariant momentum distribution of \Li within the spin-dependent coalescence model~\cite{Sun:2025oib,Yang:2017sdk,Sheng:2020ghv} with $J_z = m$ ($m=\pm2,\pm 1, 0$) is given by
\begin{eqnarray}
\label{eq:Coal} 
E\frac{\text{d}^3N_{^4{\rm Li},J_z=m}}{\text{d}{\bf K}^3} &=& E\int \left[ \prod_{i=1,2,3,4}k_i^\mu\text{d}^3\sigma_{\mu}\frac{\text{d}^3k_i}{E_i}\bar{f}_i({\bf x}_i,{\bf k}_i) \right ]  \notag \\
&\times& \text{Tr}_s[\hat{W}(\mathbf{r_1^\prime},\mathbf{k_1^\prime},\mathbf{r_2^\prime},\mathbf{k_2^\prime},\mathbf{r_3^\prime},\mathbf{k_3^\prime};J_z=m)\notag\\&\times& \hat{\sigma}_{p_1p_2p_3n}] \times \delta({\bf K}-\sum_i{\bf k}_i).
\end{eqnarray}
In the above equation, $\text{d}^3\sigma_\mu$ denotes the differential volume of the freeze-out hypersurface $\Sigma^\mu$ of the emission source; $k_i^\mu$ denotes  the four momentum of  nucleon $i$;  and $\bar{f}_i=\frac{g_i}{(2\pi)^3}\big{[} \exp(p_i^\mu u_\mu/T)/\xi_i+1 \big{]}^{-1}$ is the unpolarized nucleon phase-space distribution function with  $g_i=2J_i+1=2$ and $\xi_i$ being  its spin degeneracy and fugacity factor, respectively, and $u^\mu$ being the local four flow velocity in the emission source. The $\hat{W}(\mathbf{r_1^\prime},\mathbf{k_1^\prime},\mathbf{r_2^\prime},\mathbf{k_2^\prime},\mathbf{r_3^\prime},\mathbf{k_3^\prime};J_z=m)$ in Eq.~(\ref{eq:Coal}) is the internal Wigner function matrix of $|2,m\rangle_\text{rel}$ state, and its element is defined as
\begin{align}
W^{m_1m_2m_3m_4}_{\tilde{m}_1\tilde{m}_2\tilde{m}_3\tilde{m}_4}&=\int d\mathbf{\eta_1^\prime}d\mathbf{\eta_2^\prime}d\mathbf{\eta_3^\prime}e^{-i(\mathbf{\eta_1^\prime}\mathbf{k_1^\prime}+\mathbf{\eta_2^\prime}\mathbf{k_2^\prime}+\mathbf{\eta_3^\prime}\mathbf{k_3^\prime})}\notag 
    \\&\times \Big\langle \mathbf{r_1^\prime}+\frac{\mathbf{\eta_1^\prime}}{2},\mathbf{r_2^\prime}+\frac{\mathbf{\eta_2^\prime}}{2},\mathbf{r_3^\prime}+\frac{\mathbf{\eta_3^\prime}}{2};\tilde{m}_1...\tilde{m}_4|2,m\Big\rangle_{\text{rel}} \notag \\
&\times\Big\langle 2,m| \ \mathbf{r_1^\prime}-\frac{\mathbf{\eta_1^\prime}}{2},\mathbf{r_2^\prime}-\frac{\mathbf{\eta_2^\prime}}{2},\mathbf{r_3^\prime}-\frac{\mathbf{\eta_3^\prime}}{2};m_1...m_4\Big\rangle_{\text{rel}}.
    \label{eq:Wigner function}
\end{align}
The uncorrelated four-particle spin density matrix is defined as $\hat{\sigma}_{p_1p_2p_3 n} = \hat{\sigma}_{p_1}\otimes \hat{\sigma}_{p_2} \otimes \hat{\sigma}_{p_3} \otimes \hat{\sigma}_{n}$ and the single-particle spin density matrix is given by $\hat{\sigma}_{j}=\text{diag}[\frac{1+\mathcal{P}_j}{2},\frac{1-\mathcal{P}_j}{2}]$ with the nucleon polarization defined as $\mathcal{P}_j=\frac{N_{j,\frac{1}{2}}-N_{j,-\frac{1}{2}}}{N_{j,\frac{1}{2}}+N_{j,-\frac{1}{2}}}$, where  $j=p_1, p_2, p_3, n$ denoting the three protons and one neutron.

For \Li~in the spin state $|\hat{J},\hat{J_z}\rangle=|2,2\rangle$, its wavefunction is approximately given by 
\begin{widetext}
\begin{align}
    \langle \mathbf{r_1},...,\mathbf{r_4}\left|2,2\right>&=\sqrt{\frac{1}{3!}}\left.\left[\begin{matrix}\phi_{M=1}^{L=1}(\mathbf{r_1})\left|\frac{1}{2},\frac{1}{2}\right>_{p_1}&\phi_{M=1}^{L=1}(\mathbf{r_2})\left|\frac{1}{2},\frac{1}{2}\right>_{p_2}&\phi_{M=1}^{L=1}(\mathbf{r_3})\left|\frac{1}{2},\frac{1}{2}\right>_{p_3}\\\phi_{M=0}^{L=0}(\mathbf{r_1})\left|\frac{1}{2},\frac{1}{2}\right>_{p_1}&\phi_{M=0}^{L=0}(\mathbf{r_2})\left|\frac{1}{2},\frac{1}{2}\right>_{p_2}&\phi_{M=0}^{L=0}(\mathbf{r_3})\left|\frac{1}{2},\frac{1}{2}\right>_{p_3}\\\phi_{M=0}^{L=0}(\mathbf{r_1})\left|\frac{1}{2},-\frac{1}{2}\right>_{p_1}&\phi_{M=0}^{L=0}(\mathbf{r_2})\left|\frac{1}{2},-\frac{1}{2}\right>_{p_2}&\phi_{M=0}^{L=0}(\mathbf{r_3})\left|\frac{1}{2},-\frac{1}{2}\right>_{p_3}\end{matrix}\right.\right]\phi_{M=0}^{L=0}(\mathbf{r_4})\left|\frac{1}{2},\frac{1}{2}\right>_{n}\notag
    \\&= \sqrt{\frac{1}{6}}\Big(\phi_{M=1}^{L=1}(\mathbf{r_1})\phi_{M=0}^{L=0}(\mathbf{r_2})\phi_{M=0}^{L=0}(\mathbf{r_3})\left|\frac{1}{2},\frac{1}{2}\right>_{p_1}\left|\frac{1}{2},\frac{1}{2}\right>_{p_2}\left|\frac{1}{2},-\frac{1}{2}\right>_{p_3}\notag
    \\&\quad-\phi_{M=1}^{L=1}(\mathbf{r_1})\phi_{M=0}^{L=0}(\mathbf{r_2})\phi_{M=0}^{L=0}(\mathbf{r_3})\left|\frac{1}{2},\frac{1}{2}\right>_{p_1}\left|\frac{1}{2},-\frac{1}{2}\right>_{p_2}\left|\frac{1}{2},\frac{1}{2}\right>_{p_3}\notag
    \\&\quad+\phi_{M=1}^{L=1}(\mathbf{r_3})\phi_{M=0}^{L=0}(\mathbf{r_1})\phi_{M=0}^{L=0}(\mathbf{r_2})\left|\frac{1}{2},\frac{1}{2}\right>_{p_3}\left|\frac{1}{2},\frac{1}{2}\right>_{p_1}\left|\frac{1}{2},-\frac{1}{2}\right>_{p_2}\notag
    \\&\quad-\phi_{M=1}^{L=1}(\mathbf{r_2})\phi_{M=0}^{L=0}(\mathbf{r_1})\phi_{M=0}^{L=0}(\mathbf{r_3})\left|\frac{1}{2},\frac{1}{2}\right>_{p_2}\left|\frac{1}{2},\frac{1}{2}\right>_{p_1}\left|\frac{1}{2},-\frac{1}{2}\right>_{p_3}\notag
    \\&\quad-\phi_{M=1}^{L=1}(\mathbf{r_3})\phi_{M=0}^{L=0}(\mathbf{r_2})\phi_{M=0}^{L=0}(\mathbf{r_1})\left|\frac{1}{2},\frac{1}{2}\right>_{p_3}\left|\frac{1}{2},\frac{1}{2}\right>_{p_2}\left|\frac{1}{2},-\frac{1}{2}\right>_{p_1}\notag
    \\&\quad+\phi_{M=1}^{L=1}(\mathbf{r_2})\phi_{M=0}^{L=0}(\mathbf{r_3})\phi_{M=0}^{L=0}(\mathbf{r_1})\left|\frac{1}{2},\frac{1}{2}\right>_{p_2}\left|\frac{1}{2},\frac{1}{2}\right>_{p_3}\left|\frac{1}{2},-\frac{1}{2}\right>_{p_1}\Big)\phi_{M=0}^{L=0}(\mathbf{r_4})\left|\frac{1}{2},\frac{1}{2}\right>_{n}\notag,
\end{align}
where $\phi_{M=0}^{L=0}(\mathbf{r_i})=\phi_0(x_i)\phi_0(y_i)\phi_0(z_i)$ and $\phi_{M=1}^{L=1}(\mathbf{r_i})=\frac{1}{\sqrt{2}}[\phi_{1}(x_i)\phi_{0}(y_i)+i\phi_{0}(x_i)\phi_{1}(y_i)]\phi_{0}(z_i)$~\cite{Baltz:1995tv} are wavefunctions of the single-particle ground state and the first excited state in a harmonic potential, respectively. Note that $\phi_0(x_i)=\frac{1}{\sqrt{b\sqrt{\pi}}}\exp\left[-\frac{x_i^2}{2b^2}\right]$ and $\phi_1(x_i)=\sqrt{\frac2{b\sqrt{\pi}}}\frac{x_i} b\exp\left[-\frac{x^2}{2b^2}\right]$ are the ground and first excited harmonic oscillator states in one dimension, where $b^2=\frac{\hbar}{m \omega_0}$ with $m$ and $\omega_0$ denoting the nucleon mass and the oscillator frequency in the harmonic potential, respectively.

With the transformations
\begin{align}
    &\mathbf{R}=\frac{\mathbf{r_1}+\mathbf{r_2}+ \mathbf{r_3}+ \mathbf{r_4}}{4}, \quad {\mathbf{r_1^\prime}}=\frac{\mathbf{r_1}-\mathbf{r_2}}{\sqrt{2}},
    \quad \mathbf{r_2^\prime}=\sqrt{\frac{2}{3}}\frac{\mathbf{r_1}+\mathbf{r_2}-2 \mathbf{r_3}}{2},
    \quad \mathbf{r_3^\prime}=\sqrt{\frac{3}{4}}\frac{\mathbf{r_1}+\mathbf{r_2}+\mathbf{r_3}-3\mathbf{r_4}}{3},
    \\&\mathbf{K}=\mathbf{k_1}+\mathbf{k_2}+ \mathbf{k_3}+ \mathbf{k_4}, \quad {\mathbf{k_1^\prime}}=\frac{\mathbf{k_1}-\mathbf{k_2}}{\sqrt{2}},
    \quad \mathbf{k_2^\prime}=\sqrt{\frac{2}{3}}\frac{\mathbf{k_1}+\mathbf{k_2}-2 \mathbf{k_3}}{2},
    \quad \mathbf{k_3^\prime}=\sqrt{\frac{3}{4}}\frac{\mathbf{k_1}+\mathbf{k_2}+\mathbf{k_3}-3\mathbf{k_4}}{3},
    \\&B^2=\frac{\hbar}{M \omega_0}, \quad {b_1^\prime}^2=\frac{\hbar}{\mu_1 \omega_0}, \quad {b_2^\prime}^2=\frac{\hbar}{\mu_2 \omega_0}, \quad {b_3^\prime}^2=\frac{\hbar}{\mu_3 \omega_0},
    \\&M=m_1+m_2+m_3+m_4, \quad \mu_1=2{\big{(}\frac{1}{m_1}+\frac{1}{m_2}\big{)}}^{-1}, \quad \mu_2=\frac{3}{2}{\big{(}\frac{1}{m_1+m_2}+\frac{1}{m_3}\big{)}}^{-1}, \quad \mu_3=\frac{4}{3}{\big{(}\frac{1}{m_1+m_2+m_3}+\frac{1}{m_4}\big{)}}^{-1},
\end{align}
the wave function $\langle \mathbf{r_1},...,\mathbf{r_4}\left|2,2\right>$ can be represented as
\begin{align}
    \langle \mathbf{r_1},...,\mathbf{r_4}\left|2,2\right>&=\langle \mathbf{R},\mathbf{r^\prime_1},\mathbf{r^\prime_2},\mathbf{r^\prime_3}\left|2,2\right>
    \\&=4^{\frac{3}{4}}(B^2 \pi  )^{-\frac{3}{4}}\exp\biggl[{-\frac{R^2}{2B^2}}\biggr] \notag
    \\&\quad\times \sqrt{\frac{1}{6}}({b_1^\prime}^2 {b_2^\prime}^2 {b_3^\prime}^2 \pi^3  )^{-\frac{3}{4}}\exp\biggl[-\frac{{r_1^\prime}^2}{2{b_1^\prime}^2}-\frac{{r_2^\prime}^2}{2{b_2^\prime}^2}-\frac{{r_3^\prime}^2}{2{b_3^\prime}^2}\biggr]\notag
     \times \Bigg(\sqrt{2}    \frac{x_1^\prime+i y_1^\prime}{b_1^\prime}\left|\frac{1}{2},\frac{1}{2}\right>_{p_1}\left|\frac{1}{2},\frac{1}{2}\right>_{p_2}\left|\frac{1}{2},-\frac{1}{2}\right>_{p_3}\left|\frac{1}{2},\frac{1}{2}\right>_{n}\notag
    \\&\quad +(-\frac{\sqrt{2}}{2}\frac{x_1^\prime+i y_1^\prime}{b_1^\prime}-\frac{\sqrt{6}}{2}\frac{x_2^\prime+i y_2^\prime}{b_2^\prime})\left|\frac{1}{2},\frac{1}{2}\right>_{p_1}\left|\frac{1}{2},-\frac{1}{2}\right>_{p_2}\left|\frac{1}{2},\frac{1}{2}\right>_{p_3}\left|\frac{1}{2},\frac{1}{2}\right>_{n}\notag
    \\&\quad +(-\frac{\sqrt{2}}{2}\frac{x_1^\prime+i y_1^\prime}{b_1^\prime}+\frac{\sqrt{6}}{2}\frac{x_2^\prime+i y_2^\prime}{b_2^\prime})\left|\frac{1}{2},\frac{1}{2}\right>_{p_1}\left|\frac{1}{2},\frac{1}{2}\right>_{p_2}\left|\frac{1}{2},-\frac{1}{2}\right>_{p_3}\left|\frac{1}{2},\frac{1}{2}\right>_{n}\Bigg).
\end{align}
The internal part of the above $^4$Li wave function is thus given by  
\begin{align}
    \langle \mathbf{r_1^\prime},\mathbf{r_2^\prime},\mathbf{r_3^\prime}\left|2,2\right>_{\text{rel}}&=\sqrt{\frac{1}{6}}({b_1^\prime}^2 {b_2^\prime}^2 {b_3^\prime}^2 \pi^3  )^{-\frac{3}{4}}\exp\biggl[-\frac{{r_1^\prime}^2}{2{b_1^\prime}^2}-\frac{{r_2^\prime}^2}{2{b_2^\prime}^2}-\frac{{r_3^\prime}^2}{2{b_3^\prime}^2}\biggr]\notag
    \\&\quad \times \Bigg(\sqrt{2}    \frac{x_1^\prime+i y_1^\prime}{b_1^\prime}\left|\frac{1}{2},\frac{1}{2}\right>_{p_1}\left|\frac{1}{2},\frac{1}{2}\right>_{p_2}\left|\frac{1}{2},-\frac{1}{2}\right>_{p_3}\left|\frac{1}{2},\frac{1}{2}\right>_{n}\notag
    \\&\quad +(-\frac{\sqrt{2}}{2}\frac{x_1^\prime+i y_1^\prime}{b_1^\prime}-\frac{\sqrt{6}}{2}\frac{x_2^\prime+i y_2^\prime}{b_2^\prime})\left|\frac{1}{2},\frac{1}{2}\right>_{p_1}\left|\frac{1}{2},-\frac{1}{2}\right>_{p_2}\left|\frac{1}{2},\frac{1}{2}\right>_{p_3}\left|\frac{1}{2},\frac{1}{2}\right>_{n}\notag
    \\&\quad +(-\frac{\sqrt{2}}{2}\frac{x_1^\prime+i y_1^\prime}{b_1^\prime}+\frac{\sqrt{6}}{2}\frac{x_2^\prime+i y_2^\prime}{b_2^\prime})\left|\frac{1}{2},\frac{1}{2}\right>_{p_1}\left|\frac{1}{2},\frac{1}{2}\right>_{p_2}\left|\frac{1}{2},-\frac{1}{2}\right>_{p_3}\left|\frac{1}{2},\frac{1}{2}\right>_{n}\Bigg).
\end{align}

Further assuming that protons and neutrons have the same polarization  $\mathcal{P}_p=\mathcal{P}_n=\mathcal{P}_N$~\cite{Becattini:2013fla,Becattini:2016gvu}, the trace in Eq.~(\ref{eq:Coal}) is simplified as
\begin{eqnarray} 
    \label{eq:rho22}
     &&\text{Tr}_s[\hat{W}(\mathbf{r_1^\prime},\mathbf{k_1^\prime},\mathbf{r_2^\prime},\mathbf{k_2^\prime},\mathbf{r_3^\prime},\mathbf{k_3^\prime};J_z=2) \times  \hat{\sigma}_{p_1p_2p_3n}] \notag \\
     &&=8^3 \frac{(1+\mathcal{P}_N)^3(1-\mathcal{P}_N)}{16}\exp\left[-\frac{{r_1^\prime}^2}{{b_1^\prime}^2}-{b_1^\prime}^2 {k_1^\prime}^2-\frac{{r_2^\prime}^2}{{b_2^\prime}^2}-{b_2^\prime}^2 {k_2^\prime}^2-\frac{{r_3^\prime}^2}{{b_3^\prime}^2}-{b_3^\prime}^2 {k_3^\prime}^2\right]\notag
    \\&&\times\frac{1}{2}\left[\Big(\frac{{x_1^\prime}^2+{y_1^\prime}^2}{{b_1^\prime}^2}-1+{b_1^\prime}^2({k_{x_1}^\prime}^2+{k_{y_1}^\prime}^2)+2({k_{y_1}^\prime} {x_1^\prime}-{k_{x_1}^\prime} {y_1^\prime})\Big) 
    +\Big(\frac{{x_2^\prime}^2+{y_2^\prime}^2}{{b_2^\prime}^2}-1+{b_2^\prime}^2({k_{x_2}^\prime}^2+{k_{y_2}^\prime}^2)+2({k_{y_2}^\prime} {x_2^\prime}-{k_{x_2}^\prime} {y_2^\prime})\Big)\right].
\end{eqnarray} 
Detailed expressions for  $\text{Tr}_s[\hat{W}(\mathbf{r_1^\prime},\mathbf{k_1^\prime},\mathbf{r_2^\prime},\mathbf{k_2^\prime},\mathbf{r_3^\prime},\mathbf{k_3^\prime};J_z=m) \times  \hat{\sigma}_{p_1p_2p_3n}]$ are given in Appendix A. 

For the density matrix of $^4\text{Li}(2^-)$, it is obtained as
\begin{equation}
\begin{aligned}
\hat{\rho}\big{(}^4\text{Li}(2^-)\big{)}=\text{diag} [&\hat{\rho}_{2,2},\hat{\rho}_{1,1},\hat{\rho}_{0,0},\hat{\rho}_{-1,-1},\hat{\rho}_{-2,-2}]\\
\\=\text{diag}  \Bigg[&\frac{3 (\mathcal{P}_N+1)^2 (2 \mathcal{P}_L (\mathcal{P}_L+1)+1)}{4 \left(2 \mathcal{P}_N^2+5\right) \mathcal{P}_L^2+20 \mathcal{P}_N \mathcal{P}_L+5 \left(\mathcal{P}_N^2+3\right)},\frac{3 \left(-\mathcal{P}_L (\mathcal{P}_L+1) \mathcal{P}_N^2+\mathcal{P}_N+\mathcal{P}_L^2+\mathcal{P}_L+1\right)}{4 \left(2 \mathcal{P}_N^2+5\right) \mathcal{P}_L^2+20 \mathcal{P}_N \mathcal{P}_L+5 \left(\mathcal{P}_N^2+3\right)},\\&\frac{-\mathcal{P}_N^2-4 \mathcal{P}_L \mathcal{P}_N+2 \left(\mathcal{P}_N^2+1\right) \mathcal{P}_L^2+3}{4 \left(2 \mathcal{P}_N^2+5\right) \mathcal{P}_L^2+20 \mathcal{P}_N \mathcal{P}_L+5 \left(\mathcal{P}_N^2+3\right)} ,-\frac{3 (\mathcal{P}_N-1) ((\mathcal{P}_N+1) (\mathcal{P}_L-1) \mathcal{P}_L+1)}{4 \left(2 \mathcal{P}_N^2+5\right) \mathcal{P}_L^2+20 \mathcal{P}_N \mathcal{P}_L+5 \left(\mathcal{P}_N^2+3\right)},\\&\frac{3 (\mathcal{P}_N-1)^2 (2 (\mathcal{P}_L-1) \mathcal{P}_L+1)}{4 \left(2 \mathcal{P}_N^2+5\right) \mathcal{P}_L^2+20 \mathcal{P}_N \mathcal{P}_L+5 \left(\mathcal{P}_N^2+3\right)}\Bigg],
\end{aligned}
\end{equation}
where $\mathcal{P}_L \approx \frac{\hbar\omega}{2T}$ characterizes the polarization of orbital motion, derived from the coalescence model in a vortical fluid (see Appendix B where we evaluate the orbital-motion polarization for a particle pair with angular momentum $L=1$ using this framework).  

Since the spin alignment of $^{4}\mathrm{Li}$ is proportional to $\hbar^{2}$ in the absence of nucleon spin correlations, quantum corrections up to order $\hbar^{2}$ must be included. To achieve this, the ordinary product in the coalescence integral [Eq.~(\ref{eq:Coal})] is replaced by the Moyal star ($\star$) product~\cite{Groenewold:1946kp,Moyal:1949sk}, as required by quantum mechanics in phase space~\cite{Curtright:2011vw} (see Appendices~B and C). This follows from the $\hbar$-expansion of the Moyal bracket between two observables $A$ and $B$ in terms of the ordinary Poisson bracket, $\{A,B\}_{\star} = \{A,B\} + \mathcal{O}(\hbar^{2})$, where the leading correction is of order $\hbar^{2}$.  The angular distribution is then obtained as
\begin{eqnarray}
    \frac{dN}{\sin \theta^* d\theta^*}&=&\frac{1}{2}+\left(\frac{3}{8}\hat{\rho}_{1,1}+\frac{3}{8}\hat{\rho}_{-1,-1}+\frac{1}{2}\hat{\rho}_{0,0}-\frac{1}{4}\right)(3\cos^2 \theta^*-1)\notag 
    \\&\approx&\frac{1}{2} \left[1-\frac{7}{30} \left(\mathcal{P}_N^2+4 \mathcal{P}_N \mathcal{P}_L+\mathcal{P}_L^2\right)(3\cos ^2\theta^* -1) \right].
\end{eqnarray}
In the above equation, the density matrix elements $\hat{\rho}_{1,1}$, $\hat{\rho}_{0,0}$ and $\hat{\rho}_{-1,-1}$ are  truncated to the second order of $\mathcal{P}_N$ and $\mathcal{P}_L$. The angular dependence is found to be described by the  Legendre  polynomial $P_2(\cos\theta^*)$.
\begin{table}[!t]
    \centering
    \begin{tabular}{|c|c|c|c|c|c|c|}
     \hline
     State&$E$ (MeV)&Structure&$L$&Decay mode&$\Gamma$(MeV)&$\frac{dN}{\sin \theta^* d\theta^*}$\\\hline$^4\text{Li}(^3P_2)$&g.s.&$^3\text{He}(\frac{1}{2}^+)-p(\frac{1}{2}^+)$&1&$^4\text{Li}\rightarrow {^3\text{He}} + p$&$6.03$&$\frac{1}{2} \left(1-\frac{7}{30} \left(\mathcal{P}_N^2+4 \mathcal{P}_N \mathcal{P}_L+\mathcal{P}_L^2\right)(3\cos ^2\theta^* -1) \right)$\\\hline$^4\text{Li}(^3P_1)$&0.32&$^3\text{He}(\frac{1}{2}^+)-p(\frac{1}{2}^+)$&1&$^4\text{Li}\rightarrow {^3\text{He}} + p$&$7.35$&$\frac{1}{2} \left(1-\frac{1}{6} \left(\mathcal{P}_N^2-4 \mathcal{P}_N \mathcal{P}_L+\mathcal{P}_L^2\right)(3 \cos ^2\theta^* -1)\right)$\\\hline$^4\text{Li}(^1P_0)$&2.08&$^3\text{He}(\frac{1}{2}^+)-p(\frac{1}{2}^+)$&1&$^4\text{Li}\rightarrow {^3\text{He}} + p$&$9.35$&$\frac{1}{2}$\\\hline$^4\text{Li}(^1P_1)$&2.85&$^3\text{He}(\frac{1}{2}^+)-p(\frac{1}{2}^+)$&1&$^4\text{Li}\rightarrow {^3\text{He}} + p$&$13.51$&$\frac{1}{2} \left(1-\frac{2}{3} \mathcal{P}_L^2 (3\cos ^2\theta^* -1)\right)$\\
     \hline
     \end{tabular}
     \caption{ Angular distributions of decay particles from different states of \Li.  The ground state (g.s.) energy of \Li state lies 4.07 MeV above $^3\text{He}$+p threshold masses~\cite{Tilley:1992zz,Vovchenko:2020dmv}.}
     \label{tab:result}
\end{table}

\emph{Spin polarization of excited states  of \Li.}{\bf ---}
For the first excited state of $^4\text{Li}$ with $J^{\pi}=1^-$, the decay matrix $\hat{T}$ is given by~\cite{Lee:1957he}
\begin{eqnarray}
    &\hat{T}(^4\text{Li}(^3P_1) \rightarrow {^3\text{He}} + p)&=\sqrt{\frac{3}{8\pi}}\left.\left[\begin{matrix}-\cos\theta^*&-\frac{1}{2}\sin\theta^* e^{i\phi^*}&-\frac{1}{2}\sin\theta^* e^{i\phi^*}&0\\-\sqrt{\frac{1}{2}}\sin\theta^* e^{-i\phi^*}&0&0&-\sqrt{\frac{1}{2}}\sin\theta^* e^{i\phi^*}&\\0&-\frac{1}{2}\sin\theta^* e^{-i\phi^*}&-\frac{1}{2}\sin\theta^* e^{-i\phi^*}&\cos\theta^*&\end{matrix}
\right.\right],
\end{eqnarray}
and the density matrix is given by
\begin{eqnarray}
\hat{\rho}\big{(}^4\text{Li}(^3P_1)\big{)}=\left.\left[\begin{matrix}\hat{\rho}_{1,1}&0&0\\0&\hat{\rho}_{0,0}&0\\0&0&\hat{\rho}_{-1,-1}\end{matrix}
\right.\right]=\left[
\begin{array}{ccc}
 \frac{-\mathcal{P}_L (\mathcal{P}_L+1) \mathcal{P}_N^2+\mathcal{P}_N+\mathcal{P}_L^2+\mathcal{P}_L+1}{(\mathcal{P}_N-2 \mathcal{P}_L)^2+3} & 0 & 0 \\
 0 & \frac{\mathcal{P}_N^2-4 \mathcal{P}_L \mathcal{P}_N+2 \left(\mathcal{P}_N^2+1\right) \mathcal{P}_L^2+1}{(\mathcal{P}_N-2 \mathcal{P}_L)^2+3} & 0 \\
 0 & 0 & -\frac{(\mathcal{P}_N-1) ((\mathcal{P}_N+1) (\mathcal{P}_L-1) \mathcal{P}_L+1)}{(\mathcal{P}_N-2 \mathcal{P}_L)^2+3} \\
\end{array}
\right].
\end{eqnarray}
\end{widetext}
The diagonal elements $\hat{\rho}_{1,1},\hat{\rho}_{0,0}$ and $\hat{\rho}_{-1,-1}$ for the unit-trace matrix are the relative intensities of the spin component $m$ to take the values 1, 0, and -1 respectively, which  all have the value 1/3 in the unpolarized case, and their specific expressions can be obtained using the same method as in the $^4\text{Li}(2^-)$ case. The normalized angular distribution of the daughter particles in the decay $^4\text{Li}(1^-) \rightarrow {^3\text{He}} + p$ from Eq.~(\ref{eq:Angular distribution}) is given by
\begin{eqnarray}
    &&\frac{dN}{\sin \theta^* d\theta^*}=\frac{1}{2}-\frac{3}{8}(\hat{\rho}_{0,0}-\frac{1}{3})(3\cos^2 \theta^*-1)\notag
    \\&&\approx\frac{1}{2} \left[1-\frac{1}{6} \left(\mathcal{P}_N^2-4 \mathcal{P}_N \mathcal{P}_L+\mathcal{P}_L^2\right)(3 \cos ^2\theta^* -1)\right],
\end{eqnarray}
where we have used 
\begin{eqnarray}
    \hat{\rho}_{0,0}&=&\frac{\left(\mathcal{P}_L^2+1\right) \mathcal{P}_N^2-4 \mathcal{P}_L \mathcal{P}_N+\mathcal{P}_L^2+1}{-\left(\mathcal{P}_L^2-1\right) \mathcal{P}_N^2-4 \mathcal{P}_L \mathcal{P}_N+\mathcal{P}_L^2+3}\notag
    \\&\approx&\frac{1}{3} \left(1+\frac{2 \mathcal{P}_N^2}{3}-\frac{8 \mathcal{P}_N \mathcal{P}_L}{3}+\frac{2 \mathcal{P}_L^2}{3}\right).
\end{eqnarray}

For the second excited state of \Li with spin-parity $J^\pi=0^-$, its decay particles have an isotropic angular distribution with respect to the reaction plane. The third excited state $^1P_1$~\cite{Tilley:1992zz,Vovchenko:2020dmv} has the same spin-parity as the first excited state $^3P_1$ but different spin coupling. Its decay matrix is given by~\cite{Lee:1957he}
\begin{eqnarray}
    &&\hat{T}(^4\text{Li}(^1P_1) \rightarrow {^3\text{He}} + p)\notag\\
    &=&\sqrt{\frac{3}{8\pi}} \left[\begin{matrix}0&-\sqrt{\frac{1}{2}}\sin \theta^* e^{i\phi^*}&\sqrt{\frac{1}{2}}\sin \theta^* e^{i\phi^*} &0\\0&\cos \theta^*&-\cos \theta^*&0\\0&\sqrt{\frac{1}{2}}\sin \theta^* e^{-i\phi^*}&-\sqrt{\frac{1}{2}}\sin \theta^* e^{-i\phi^*}&0
    \end{matrix}
 \right],
\end{eqnarray}
\\
and its density matrix is given by
\\
\begin{eqnarray}
\hat{\rho}\big{(}^4\text{Li}(^1P_1)\big{)}=\left[
\begin{array}{ccc}
 \frac{2 \mathcal{P}_L (\mathcal{P}_L+1)+1}{4 \mathcal{P}_L^2+3} & 0 & 0 \\
 0 & \frac{1}{4 \mathcal{P}_L^2+3} & 0 \\
 0 & 0 & \frac{2 (\mathcal{P}_L-1) \mathcal{P}_L+1}{4 \mathcal{P}_L^2+3} \\
\end{array}
\right],
\\  \notag
\end{eqnarray}

which is obtained by the same method as  that used for $^4\text{Li}(^3P_1)$. The angular distribution of its decay particle $^3\text{He}$  is then given by
\begin{eqnarray}
    \frac{dN}{\sin \theta^* d\theta^*}&=&\frac{1}{2}+\frac{3}{4}(\hat{\rho}_{0,0}-\frac{1}{3})(3\cos^2 \theta^*-1)\notag
    \\&\approx&\frac{1}{2} \left[1-\frac{2}{3} \mathcal{P}_L^2 (3\cos ^2\theta^* -1)\right],
\end{eqnarray}
after using 
\begin{eqnarray}
    \hat{\rho}_{0,0}= \frac{1}{4 \mathcal{P}_L^2+3} \approx\frac{1}{3}\left(1-\frac{4}{3}\mathcal{P}_L^2\right).
\end{eqnarray}

The angular distributions of the decay particles of \Li with different spin-parity  derived in the above are summarized in Table~\ref{tab:result}.  In experiments, both the ground and excited states of \Li~contribute to the measured angular distribution of the daughter particles from the decay $^4\text{Li}\rightarrow {^3\text{He}}+p$. Since the excitation energies of \Li are a few MeV, which are much smaller than the temperature of the hadronic matter produced in heavy-ion collisions, the yield of these states is essentially proportional to their spin degeneracies, i.e., 5:3:1:3. The averaged angular distribution of the decay particle is thus given by 

\begin{widetext}
\begin{eqnarray}
\label{coalescence average angular distribution}
    \frac{dN}{\sin \theta^* d\theta^*}&\approx&\frac{1}{12}\Big\{5\times\frac{1}{2} \left[1-\frac{7}{30} \left(\mathcal{P}_N^2+4 \mathcal{P}_N \mathcal{P}_L+\mathcal{P}_L^2\right)(3\cos ^2\theta^* -1) \right] \notag \\
    &+&3\times\frac{1}{2} \left[1-\frac{1}{6} \left(\mathcal{P}_N^2-4 \mathcal{P}_N \mathcal{P}_L+\mathcal{P}_L^2\right)(3 \cos ^2\theta^* -1)\right]+\frac{1}{2}+3\times\frac{1}{2} \left[1-\frac{2}{3} \mathcal{P}_L^2 (3\cos ^2\theta^* -1)\right]\Big\} \notag \\
    &\approx&\frac{1}{2} \left[1-\frac{1}{36}\left(5 \mathcal{P}_N^2+8 \mathcal{P}_N \mathcal{P}_L+11 \mathcal{P}_L^2\right)(3\cos ^2\theta^* -1) \right].
\end{eqnarray}
\end{widetext}
Further assuming  $\mathcal{P}_L=\mathcal{P}_N$ simplifies the average angular distribution  to
\begin{equation}
    \frac{dN}{\sin \theta^* d\theta^*}=\frac{1}{2} \left[1- \frac{2}{3}\mathcal{P}_N^2 (3\cos ^2\theta^* -1)\right].
    \label{eq:Angle1}
\end{equation}

\emph{Discussion.}{\bf ---}
Besides the coalescence model, the statistical model or thermal model has also been widely adopted to describe the production of light nuclei in heavy-ion collisions. In the thermal model, which assumes local thermal equilibrium for nucleons and \Li, the particle polarization in the non-relativistic limit is given by $\mathcal{P}=\frac{J+1}{3}\frac{\omega}{T}$ with $J$, $\omega$, and $T$ being the particle spin, vorticity, and local temperature, respectively~\cite{Becattini:2016gvu}. For nucleons, their polarization is given by $\mathcal{P}_N=\frac{\omega}{2T}$. The density matrices of \Li~ in this model are obtained as
\begin{widetext}
    \begin{eqnarray}
&&\hat{\rho}(^4\text{Li}({2}^-))=\frac{1}{1+2\cosh (\frac{\omega}{T})+2\cosh(\frac{2\omega}{T})}\left.\left[\begin{matrix}e^{\frac{2\omega}{T}}&0&0&0&0\\0&e^{\frac{\omega}{T}}&0&0&0\\0&0&1&0&0\\0&0&0&e^{-\frac{\omega}{T}}&0\\0&0&0&0&e^{-\frac{2\omega}{T}}\end{matrix}
\right.\right], \notag\\
&&\hat{\rho}(^4\text{Li}({1}^-))=\frac{1}{1+2\cosh (\frac{\omega}{T})}\left.\left[\begin{matrix}e^{\frac{\omega}{T}}&0&0\\0&1&0\\0&0&e^{-\frac{\omega}{T}}\end{matrix}
\right.\right], \notag\\
&&\hat{\rho}(^4\text{Li}({0}^-))=1. 
\end{eqnarray}
With $\mathcal{P}=\frac{\omega}{2T}$, one obtains the averaged angular distribution  in the thermal model as
\begin{eqnarray}
\label{thermal model average angular distribution}
    \frac{dN}{\sin \theta^* d\theta^*}&\approx&\frac{2 \mathcal{P}^6+29 \mathcal{P}^4+90 \mathcal{P}^2-3 \left(\mathcal{P}^4+10 \mathcal{P}^2+15\right) \mathcal{P}^2 \cos (2 \theta^* )+45}{6 \left(\mathcal{P}^2+3\right) \left(\mathcal{P}^4+10 \mathcal{P}^2+5\right)}
    \approx\frac{1}{2} \left[1- \frac{2}{3}\mathcal{P}^2 (3\cos ^2\theta^* -1)\right],
\end{eqnarray}
\end{widetext}
which is identical to  Eq.~(\ref{eq:Angle1}).

The equivalence between results from the coalescence model and the thermal model in our study is due to our assumption that kinetically freeze-out nucleons in our coalescence model calculations are taken from a thermalized and isotropic emission source. The coalescence model is, however, more general as it can also be used to study the $^4$Li polarization in heavy ion collisions by using the phase-space distributions of polarized nucleons from non-equlibrium spin-dependent transport models~\cite{Xu:2025uwd}.

\emph{Summary}{\bf ---}
In the present study, we investigate the global spin alignment of the unstable nucleus $^4\text{Li}$ in heavy-ion collisions. Assuming that polarized nucleons coalesce at kinetic freeze-out and neglecting spin correlations among nucleons, we derive the angular distribution of the daughter ${^3\text{He}}$. To include leading quantum corrections (up to $\hbar^2$) in the coalescence model for $^4$Li production, we incorporate the Moyal star product to calculate the overlap integral. We show that these spin states, $^4\text{Li}(^3P_2)$, $^4\text{Li}(^3P_1)$, and $^4\text{Li}(^1P_1)$, decay via $^4\text{Li}\rightarrow {^3\text{He}} + p$ with angular distributions proportional to $\mathcal{P}_N^2$ and $\cos 2\theta^*$, differing only in magnitude. We also show that the results from coalescence model calculation are consistent with thermal model for a thermalized medium. Our results indicate that measuring the spin alignment of short-lived light nuclei offers a viable probe of the proton spin polarization~\cite{Liu:2025vho,Sun:2025oib} and vortical structure of nuclear matter formed in heavy-ion collisions. It is also important for understanding the spin-dependent transport phenomenon~\cite{Liu:2023nkm,Xu:2025uwd}, such as the hadronic spin Hall effect~\cite{Liu:2020dxg}, as well as the spin-dependent equation of state of nuclear matter~\cite{Tachibana:2025wey}. 

In our study, $^{4}\mathrm{Li}$ is assumed to form through the coalescence of three protons and one neutron. Alternatively, one could consider $^{4}\mathrm{Li}$ formation via the coalescence of a proton and a $^{3}\mathrm{He}$ nucleus. In this alternative scenario, our results for the polarization of $^{4}\mathrm{Li}$ remain unchanged, and both approaches reproduce the same result from the thermal  model.   However, a broader question arises regarding whether the coalescence model is sufficiently accurate for describing the production of unstable particles. An alternative treatment would be to model $^{4}\mathrm{Li}$ formation and its decay as low-energy scattering, $^{3}\mathrm{He}(p,p)^{3}\mathrm{He}$, whose properties are constrained by phase-shift analyses~\cite{Tombrello:1965zz,Alley:1993zz}. In addition, the in-medium Mott effect~\cite{Wang:2025wim}  and re-scattering effect~\cite{Sun:2022xjr} could significantly influence $^{4}\mathrm{Li}$ production in high--baryon-density regions of heavy-ion collisions. Furthermore, spin correlations~\cite{Lv:2024uev,Chen:2024hki,Sheng:2025puj,Tang:2025oav,STAR:2025njp} among nucleons may strongly modify the spin-alignment pattern of $^{4}\mathrm{Li}$, similar to the spin alignment observed for the $\phi$ meson. These aspects highlight the need for further theoretical and experimental developments in the near future.  \\

\begin{acknowledgments} 
The authors thank helpful discussions with Mei Huang, Yu-Tie Liang, Shi Pu, Yi-Feng Sun, Rui Wang, Jun-Lin Wu, Guan-Nan Xie, Zhen Zhang, and Ya-Peng Zhang. This work was supported in part by the National Key Research and Development Project of China under Grant No. 2024YFA1612500;  the National Natural Science Foundation of China under contract No. 12375121, No. 12025501, No. 12147101, No. 11891070, No. 12375125, and No. 12235010;  the Natural Science Foundation of Shanghai under Grant No. 23JC1400200; the National
SKA Program of China No. 2020SKA0120300; the Guangdong Major Project of Basic and Applied Basic Research No. 2020B0301030008; the STCSM under Grant No. 23590780100; the Fundamental Research Funds for the Central Universities;  the Science and Technology Commission of Shanghai Municipality under Grant No. 23JC1402700; the U.S. Department of Energy under Award No. DE-SC0015266. The computations in this research were performed using the CFFF platform of Fudan University.
\end{acknowledgments}

\appendix

\section{Appendix A: Wigner function and density matrix of \Li }
\renewcommand{\theequation}{A.\arabic{equation}}
\setcounter{equation}{0}

We first construct a three-dimensional single-particle harmonic oscillator wave function  from the one-dimensional harmonic oscillator wave function. Then we construct the harmonic oscillator wave function of $^4\text{Li}$, thereby we can obtain the internal Wigner function of $^4\text{Li}$~\cite{Baltz:1995tv}. Finally, the density matrix of \Li is obtained from Eq.~(\ref{eq:Coal}).

In Cartesian coordinates, the one-dimensional wave functions of the ground and first excited states in a harmonic oscillator potential can be represented, respectively, as 
\begin{eqnarray}
    \phi_0(x)=\frac{1}{\sqrt{b\sqrt{\pi}}}\exp\biggl[-\frac{x^2}{2b^2}\biggr]
    \label{eq.phi0},
\end{eqnarray}
\begin{eqnarray}
    \phi_1(x)=\sqrt{\frac2{b\sqrt{\pi}}}\frac xb\exp\biggl[-\frac{x^2}{2b^2}\biggr]
    \label{eq.phi1},
\end{eqnarray}
where $b^2=\frac{\hbar}{m\omega_0}$ and $\omega_0$ denotes the oscillator frequency in the harmonic potential.  With the Wigner functions in one dimension defined as
\begin{eqnarray}
     W_i(x,k_x)=&\int d\eta e^{-i\eta k_x}\left<x+\frac{\eta}{2}|\phi_i(x)\right>\left<\phi_i(x)|x-\frac{\eta}{2}\right>,
\end{eqnarray}
we obtain
\begin{eqnarray}
    W_0(x,k_x)=2e^{-\frac{x^2}{b^2}-b^2k_x^2},
\end{eqnarray}
\begin{eqnarray}
    W_1(x,k_x)=2(2\frac{x^2}{b^2}-1+2b^2k_x^2)e^{-\frac{x^2}{b^2}-b^2k_x^2},
\end{eqnarray}
by using Eq.~(\ref{eq.phi0}) and Eq.~(\ref{eq.phi1}), respectively.   We can  construct the three-dimensional harmonic oscillator wave functions from one-dimensional wave functions $\phi_0$ and $\phi_1$~\cite{Baltz:1995tv},
\begin{eqnarray}
    \phi_{M=0}^{L=0}(\mathbf{r})=\phi_0(x)\phi_0(y)\phi_0(z),
\end{eqnarray}
\begin{eqnarray}
    \phi_{M=1}^{L=1}(\mathbf{r})=\frac{1}{\sqrt{2}}[\phi_{1}(x)\phi_{0}(y)+i\phi_{0}(x)\phi_{1}(y)]\phi_{0}(z),
\end{eqnarray}
\begin{eqnarray}
    \phi_{M=-1}^{L=1}(\mathbf{r})=\phi_{M=1}^{L=1}(\mathbf{r})^{*},
\end{eqnarray}
\begin{eqnarray}
    \phi_{M=0}^{L=1}(\mathbf{r})=\phi_0(x)\phi_0(y)\phi_1(z),
\end{eqnarray}
where $r^2=x^2+y^2+z^2$  and $ k^2=k_x^2+k_y^2+k_z^2$. The harmonic oscillator Wigner functions with different principal quantum numbers $n$, angular momentum $l$ and its $z-$component $m$ then are given by
\begin{align}
    \label{wigner function 000}
    W_{NLM=100}&=8e^{-\frac{r^2}{b^2}-b^2k^2},
\end{align}
\begin{align}
    W_{NLM=110}&=8\left(2\frac{z^2}{b^2}-1+2b^2k_z^2\right)e^{-\frac{r^2}{b^2}-b^2k^2},
\end{align}

\begin{align}
    &W_{NLM=111}\notag
    \\&=8\left[\frac{x^2+y^2}{b^2}-1+b^2(k_x^2+k_y^2)+2(k_y x-k_x y)\right]e^{-\frac{r^2}{b^2}-b^2k^2},
\end{align}
\begin{align}
    \label{wigner function 11-1}
    &W_{NLM=11-1}\notag
    \\&=8\left[\frac{x^2+y^2}{b^2}-1+b^2(k_x^2+k_y^2)-2(k_y x-k_x y)\right]e^{-\frac{r^2}{b^2}-b^2k^2}.
\end{align}

\begin{widetext}
With the single-particle harmonic oscillator wave function and its corresponding Wigner function, we can use them to construct the wave functions and Wigner functions of $^4\text{Li}$. $^4\text{Li}$ is composed of three identical protons and one neutron, which we assume that two protons and one neutron are in the $L=0 $ state and the third proton is in the $L = 1$ state. Its wave function, which satisfies the antisymmetry with respect to the exchange of proton coordinates, can be represented as 
\begin{align}
    \langle \mathbf{r_1},...,\mathbf{r_4}\left|2,2\right>&=\sqrt{\frac{1}{3!}}\left.\left[\begin{matrix}\phi_{M=1}^{L=1}(\mathbf{r_1})\left|\frac{1}{2},\frac{1}{2}\right>_{p_1}&\phi_{M=1}^{L=1}(\mathbf{r_2})\left|\frac{1}{2},\frac{1}{2}\right>_{p_2}&\phi_{M=1}^{L=1}(\mathbf{r_3})\left|\frac{1}{2},\frac{1}{2}\right>_{p_3}\\\phi_{M=0}^{L=0}(\mathbf{r_1})\left|\frac{1}{2},\frac{1}{2}\right>_{p_1}&\phi_{M=0}^{L=0}(\mathbf{r_2})\left|\frac{1}{2},\frac{1}{2}\right>_{p_2}&\phi_{M=0}^{L=0}(\mathbf{r_3})\left|\frac{1}{2},\frac{1}{2}\right>_{p_3}\\\phi_{M=0}^{L=0}(\mathbf{r_1})\left|\frac{1}{2},-\frac{1}{2}\right>_{p_1}&\phi_{M=0}^{L=0}(\mathbf{r_2})\left|\frac{1}{2},-\frac{1}{2}\right>_{p_2}&\phi_{M=0}^{L=0}(\mathbf{r_3})\left|\frac{1}{2},-\frac{1}{2}\right>_{p_3}\end{matrix}\right.\right]\phi_{M=0}^{L=0}(\mathbf{r_4})\left|\frac{1}{2},\frac{1}{2}\right>_{n}\notag
    \\&= \sqrt{\frac{1}{6}}\Big(\phi_{M=1}^{L=1}(\mathbf{r_1})\phi_{M=0}^{L=0}(\mathbf{r_2})\phi_{M=0}^{L=0}(\mathbf{r_3})\left|\frac{1}{2},\frac{1}{2}\right>_{p_1}\left|\frac{1}{2},\frac{1}{2}\right>_{p_2}\left|\frac{1}{2},-\frac{1}{2}\right>_{p_3}\notag
    \\&\quad-\phi_{M=1}^{L=1}(\mathbf{r_1})\phi_{M=0}^{L=0}(\mathbf{r_2})\phi_{M=0}^{L=0}(\mathbf{r_3})\left|\frac{1}{2},\frac{1}{2}\right>_{p_1}\left|\frac{1}{2},-\frac{1}{2}\right>_{p_2}\left|\frac{1}{2},\frac{1}{2}\right>_{p_3}\notag
    \\&\quad+\phi_{M=1}^{L=1}(\mathbf{r_3})\phi_{M=0}^{L=0}(\mathbf{r_1})\phi_{M=0}^{L=0}(\mathbf{r_2})\left|\frac{1}{2},\frac{1}{2}\right>_{p_3}\left|\frac{1}{2},\frac{1}{2}\right>_{p_1}\left|\frac{1}{2},-\frac{1}{2}\right>_{p_2}\notag
    \\&\quad-\phi_{M=1}^{L=1}(\mathbf{r_2})\phi_{M=0}^{L=0}(\mathbf{r_1})\phi_{M=0}^{L=0}(\mathbf{r_3})\left|\frac{1}{2},\frac{1}{2}\right>_{p_2}\left|\frac{1}{2},\frac{1}{2}\right>_{p_1}\left|\frac{1}{2},-\frac{1}{2}\right>_{p_3}\notag
    \\&\quad-\phi_{M=1}^{L=1}(\mathbf{r_3})\phi_{M=0}^{L=0}(\mathbf{r_2})\phi_{M=0}^{L=0}(\mathbf{r_1})\left|\frac{1}{2},\frac{1}{2}\right>_{p_3}\left|\frac{1}{2},\frac{1}{2}\right>_{p_2}\left|\frac{1}{2},-\frac{1}{2}\right>_{p_1}\notag
    \\&\quad+\phi_{M=1}^{L=1}(\mathbf{r_2})\phi_{M=0}^{L=0}(\mathbf{r_3})\phi_{M=0}^{L=0}(\mathbf{r_1})\left|\frac{1}{2},\frac{1}{2}\right>_{p_2}\left|\frac{1}{2},\frac{1}{2}\right>_{p_3}\left|\frac{1}{2},-\frac{1}{2}\right>_{p_1}\Big)\phi_{M=0}^{L=0}(\mathbf{r_4})\left|\frac{1}{2},\frac{1}{2}\right>_{n}
    \\&=4^{\frac{3}{4}}(B^2 \pi  )^{-\frac{3}{4}}\exp\biggl[{-\frac{R^2}{2B^2}}\biggr] \notag
    \\&\quad\times \sqrt{\frac{1}{6}}({b_1^\prime}^2 {b_2^\prime}^2 {b_3^\prime}^2 \pi^3  )^{-\frac{3}{4}}\exp\biggl[-\frac{{r_1^\prime}^2}{2{b_1^\prime}^2}-\frac{{r_2^\prime}^2}{2{b_2^\prime}^2}-\frac{{r_3^\prime}^2}{2{b_3^\prime}^2}\biggr]\notag
     \times \Bigg(\sqrt{2}    \frac{x_1^\prime+i y_1^\prime}{b_1^\prime}\left|\frac{1}{2},\frac{1}{2}\right>_{p_1}\left|\frac{1}{2},\frac{1}{2}\right>_{p_2}\left|\frac{1}{2},-\frac{1}{2}\right>_{p_3}\left|\frac{1}{2},\frac{1}{2}\right>_{n}\notag
    \\&\quad +(-\frac{\sqrt{2}}{2}\frac{x_1^\prime+i y_1^\prime}{b_1^\prime}-\frac{\sqrt{6}}{2}\frac{x_2^\prime+i y_2^\prime}{b_2^\prime})\left|\frac{1}{2},\frac{1}{2}\right>_{p_1}\left|\frac{1}{2},-\frac{1}{2}\right>_{p_2}\left|\frac{1}{2},\frac{1}{2}\right>_{p_3}\left|\frac{1}{2},\frac{1}{2}\right>_{n}\notag
    \\&\quad +(-\frac{\sqrt{2}}{2}\frac{x_1^\prime+i y_1^\prime}{b_1^\prime}+\frac{\sqrt{6}}{2}\frac{x_2^\prime+i y_2^\prime}{b_2^\prime})\left|\frac{1}{2},\frac{1}{2}\right>_{p_1}\left|\frac{1}{2},\frac{1}{2}\right>_{p_2}\left|\frac{1}{2},-\frac{1}{2}\right>_{p_3}\left|\frac{1}{2},\frac{1}{2}\right>_{n}\Bigg).
\end{align}
With the transformation
\begin{align}
    &\mathbf{R}=\frac{\mathbf{r_1}+\mathbf{r_2}+ \mathbf{r_3}+ \mathbf{r_4}}{4}, \quad {\mathbf{r_1^\prime}}=\frac{\mathbf{r_1}-\mathbf{r_2}}{\sqrt{2}},
    \quad \mathbf{r_2^\prime}=\sqrt{\frac{2}{3}}\frac{\mathbf{r_1}+\mathbf{r_2}-2 \mathbf{r_3}}{2},
    \quad \mathbf{r_3^\prime}=\sqrt{\frac{3}{4}}\frac{\mathbf{r_1}+\mathbf{r_2}+\mathbf{r_3}-3\mathbf{r_4}}{3},
    \\&\mathbf{K}=\mathbf{k_1}+\mathbf{k_2}+ \mathbf{k_3}+ \mathbf{k_4}, \quad {\mathbf{k_1^\prime}}=\frac{\mathbf{k_1}-\mathbf{k_2}}{\sqrt{2}},
    \quad \mathbf{k_2^\prime}=\sqrt{\frac{2}{3}}\frac{\mathbf{k_1}+\mathbf{k_2}-2 \mathbf{k_3}}{2},
    \quad \mathbf{k_3^\prime}=\sqrt{\frac{3}{4}}\frac{\mathbf{k_1}+\mathbf{k_2}+\mathbf{k_3}-3\mathbf{k_4}}{3},
\end{align}
the internal wave function of $^4$Li  is given by 
\begin{align}
    \langle \mathbf{r_1^\prime},\mathbf{r_2^\prime},\mathbf{r_3^\prime}\left|2,2\right>_{\text{rel}}&=\sqrt{\frac{1}{6}}({b_1^\prime}^2 {b_2^\prime}^2 {b_3^\prime}^2 \pi^3  )^{-\frac{3}{4}}\exp\biggl[-\frac{{r_1^\prime}^2}{2{b_1^\prime}^2}-\frac{{r_2^\prime}^2}{2{b_2^\prime}^2}-\frac{{r_3^\prime}^2}{2{b_3^\prime}^2}\biggr]\notag
    \\&\quad \times \Bigg(\sqrt{2}    \frac{x_1^\prime+i y_1^\prime}{b_1^\prime}\left|\frac{1}{2},\frac{1}{2}\right>_{p_1}\left|\frac{1}{2},\frac{1}{2}\right>_{p_2}\left|\frac{1}{2},-\frac{1}{2}\right>_{p_3}\left|\frac{1}{2},\frac{1}{2}\right>_{n}\notag
    \\&\quad +(-\frac{\sqrt{2}}{2}\frac{x_1^\prime+i y_1^\prime}{b_1^\prime}-\frac{\sqrt{6}}{2}\frac{x_2^\prime+i y_2^\prime}{b_2^\prime})\left|\frac{1}{2},\frac{1}{2}\right>_{p_1}\left|\frac{1}{2},-\frac{1}{2}\right>_{p_2}\left|\frac{1}{2},\frac{1}{2}\right>_{p_3}\left|\frac{1}{2},\frac{1}{2}\right>_{n}\notag
    \\&\quad +(-\frac{\sqrt{2}}{2}\frac{x_1^\prime+i y_1^\prime}{b_1^\prime}+\frac{\sqrt{6}}{2}\frac{x_2^\prime+i y_2^\prime}{b_2^\prime})\left|\frac{1}{2},\frac{1}{2}\right>_{p_1}\left|\frac{1}{2},\frac{1}{2}\right>_{p_2}\left|\frac{1}{2},-\frac{1}{2}\right>_{p_3}\left|\frac{1}{2},\frac{1}{2}\right>_{n}\Bigg).
\end{align}

The Wigner matrix  $\hat{W}(\mathbf{r_1^\prime},\mathbf{k_1^\prime},\mathbf{r_2^\prime},\mathbf{k_2^\prime},\mathbf{r_3^\prime},\mathbf{k_3^\prime};J_z=2)$ defined as Eq.~(\ref{eq:Wigner function}), is then given by
\begin{align}
    &\hat{W}(\mathbf{r_1^\prime},\mathbf{k_1^\prime},\mathbf{r_2^\prime},\mathbf{k_2^\prime},\mathbf{r_3^\prime},\mathbf{k_3^\prime};J_z=2) \notag
    \\&=\text{diag}[0,0,\int d\mathbf{\eta_1^\prime}d\mathbf{\eta_2^\prime}d\mathbf{\eta_3^\prime}e^{-i(\mathbf{\eta_1^\prime}\mathbf{k_1^\prime}+\mathbf{\eta_2^\prime}\mathbf{k_2^\prime}+\mathbf{\eta_3^\prime}\mathbf{k_3^\prime})}\notag 
    \\&\times \langle \mathbf{r_1^\prime}+\frac{\mathbf{\eta_1^\prime}}{2},\mathbf{r_2^\prime}+\frac{\mathbf{\eta_2^\prime}}{2},\mathbf{r_3^\prime}+\frac{\mathbf{\eta_3^\prime}}{2}|\sqrt{\frac{1}{6}}({b_1^\prime}^2 {b_2^\prime}^2 {b_3^\prime}^2 \pi^3  )^{-\frac{3}{4}}\exp\biggl[-\frac{{r_1^\prime}^2}{2{b_1^\prime}^2}-\frac{{r_2^\prime}^2}{2{b_2^\prime}^2}-\frac{{r_3^\prime}^2}{2{b_3^\prime}^2}\biggr]\sqrt{2}    \frac{x_1^\prime-i y_1^\prime}{b_1^\prime}\rangle\notag
    \\&\times\langle\sqrt{\frac{1}{6}}({b_1^\prime}^2 {b_2^\prime}^2 {b_3^\prime}^2 \pi^3  )^{-\frac{3}{4}}\exp\biggl[-\frac{{r_1^\prime}^2}{2{b_1^\prime}^2}-\frac{{r_2^\prime}^2}{2{b_2^\prime}^2}-\frac{{r_3^\prime}^2}{2{b_3^\prime}^2}\biggr]\sqrt{2}    \frac{x_1^\prime+i y_1^\prime}{b_1^\prime}|\mathbf{r_1^\prime}+\frac{\mathbf{\eta_1^\prime}}{2},\mathbf{r_2^\prime}+\frac{\mathbf{\eta_2^\prime}}{2},\mathbf{r_3^\prime}+\frac{\mathbf{\eta_3^\prime}}{2}\rangle\notag,0,\notag
    \notag\\&\int d\mathbf{\eta_1^\prime}d\mathbf{\eta_2^\prime}d\mathbf{\eta_3^\prime}e^{-i(\mathbf{\eta_1^\prime}\mathbf{k_1^\prime}+\mathbf{\eta_2^\prime}\mathbf{k_2^\prime}+\mathbf{\eta_3^\prime}\mathbf{k_3^\prime})}\notag 
    \\&\times \langle \mathbf{r_1^\prime}+\frac{\mathbf{\eta_1^\prime}}{2},\mathbf{r_2^\prime}+\frac{\mathbf{\eta_2^\prime}}{2},\mathbf{r_3^\prime}+\frac{\mathbf{\eta_3^\prime}}{2}|\sqrt{\frac{1}{6}}({b_1^\prime}^2 {b_2^\prime}^2 {b_3^\prime}^2 \pi^3  )^{-\frac{3}{4}}\exp\biggl[-\frac{{r_1^\prime}^2}{2{b_1^\prime}^2}-\frac{{r_2^\prime}^2}{2{b_2^\prime}^2}-\frac{{r_3^\prime}^2}{2{b_3^\prime}^2}\biggr](-\frac{\sqrt{2}}{2}\frac{x_1^\prime-i y_1^\prime}{b_1^\prime}-\frac{\sqrt{6}}{2}\frac{x_2^\prime-i y_2^\prime}{b_2^\prime})\rangle\notag
    \\&\times\langle \sqrt{\frac{1}{6}}({b_1^\prime}^2 {b_2^\prime}^2 {b_3^\prime}^2 \pi^3  )^{-\frac{3}{4}}\exp\biggl[-\frac{{r_1^\prime}^2}{2{b_1^\prime}^2}-\frac{{r_2^\prime}^2}{2{b_2^\prime}^2}-\frac{{r_3^\prime}^2}{2{b_3^\prime}^2}\biggr](-\frac{\sqrt{2}}{2}\frac{x_1^\prime+i y_1^\prime}{b_1^\prime}-\frac{\sqrt{6}}{2}\frac{x_2^\prime+i y_2^\prime}{b_2^\prime})|\mathbf{r_1^\prime}+\frac{\mathbf{\eta_1^\prime}}{2},\mathbf{r_2^\prime}+\frac{\mathbf{\eta_2^\prime}}{2},\mathbf{r_3^\prime}+\frac{\mathbf{\eta_3^\prime}}{2}\rangle \notag,0,0,0,
    \notag\\&\int d\mathbf{\eta_1^\prime}d\mathbf{\eta_2^\prime}d\mathbf{\eta_3^\prime}e^{-i(\mathbf{\eta_1^\prime}\mathbf{k_1^\prime}+\mathbf{\eta_2^\prime}\mathbf{k_2^\prime}+\mathbf{\eta_3^\prime}\mathbf{k_3^\prime})}\notag 
    \\&\times \langle \mathbf{r_1^\prime}+\frac{\mathbf{\eta_1^\prime}}{2},\mathbf{r_2^\prime}+\frac{\mathbf{\eta_2^\prime}}{2},\mathbf{r_3^\prime}+\frac{\mathbf{\eta_3^\prime}}{2}|\sqrt{\frac{1}{6}}({b_1^\prime}^2 {b_2^\prime}^2 {b_3^\prime}^2 \pi^3  )^{-\frac{3}{4}}\exp\biggl[-\frac{{r_1^\prime}^2}{2{b_1^\prime}^2}-\frac{{r_2^\prime}^2}{2{b_2^\prime}^2}-\frac{{r_3^\prime}^2}{2{b_3^\prime}^2}\biggr](-\frac{\sqrt{2}}{2}\frac{x_1^\prime-i y_1^\prime}{b_1^\prime}+\frac{\sqrt{6}}{2}\frac{x_2^\prime-i y_2^\prime}{b_2^\prime})\rangle
    \notag\\&\times\langle \sqrt{\frac{1}{6}}({b_1^\prime}^2 {b_2^\prime}^2 {b_3^\prime}^2 \pi^3  )^{-\frac{3}{4}}\exp\biggl[-\frac{{r_1^\prime}^2}{2{b_1^\prime}^2}-\frac{{r_2^\prime}^2}{2{b_2^\prime}^2}-\frac{{r_3^\prime}^2}{2{b_3^\prime}^2}\biggr](-\frac{\sqrt{2}}{2}\frac{x_1^\prime+i y_1^\prime}{b_1^\prime}+\frac{\sqrt{6}}{2}\frac{x_2^\prime+i y_2^\prime}{b_2^\prime})|\mathbf{r_1^\prime}+\frac{\mathbf{\eta_1^\prime}}{2},\mathbf{r_2^\prime}+\frac{\mathbf{\eta_2^\prime}}{2},\mathbf{r_3^\prime}+\frac{\mathbf{\eta_3^\prime}}{2}\rangle \notag,0,0,0,0,0,0,0]
    \\&+\hat{W}[off-diag],
\end{align}
where $\hat{W}[off-diag]$ denotes the off-diagonal elements of $\hat{W}(\mathbf{r_1^\prime},\mathbf{k_1^\prime},\mathbf{r_2^\prime},\mathbf{k_2^\prime},\mathbf{r_3^\prime},\mathbf{k_3^\prime};J_z=2)$. 

The trace in Eq.~(\ref{eq:Coal}) can be obtained as
\begin{align}
\label{trace: Jz=2}
    &\text{Tr}_s[\hat{W}(\mathbf{r_1^\prime},\mathbf{k_1^\prime},\mathbf{r_2^\prime},\mathbf{k_2^\prime},\mathbf{r_3^\prime},\mathbf{k_3^\prime};J_z=2) \times  \hat{\sigma}_{p_1p_2p_3n}]
    \notag\\&=\frac{(1+\mathcal{P}_N)^3(1-\mathcal{P}_N)}{16}\Big(\int d\mathbf{\eta_1^\prime}d\mathbf{\eta_2^\prime}d\mathbf{\eta_3^\prime}e^{-i(\mathbf{\eta_1^\prime}\mathbf{k_1^\prime}+\mathbf{\eta_2^\prime}\mathbf{k_2^\prime}+\mathbf{\eta_3^\prime}\mathbf{k_3^\prime})}\notag 
    \\&\times \langle \mathbf{r_1^\prime}+\frac{\mathbf{\eta_1^\prime}}{2},\mathbf{r_2^\prime}+\frac{\mathbf{\eta_2^\prime}}{2},\mathbf{r_3^\prime}+\frac{\mathbf{\eta_3^\prime}}{2}|\sqrt{\frac{1}{6}}({b_1^\prime}^2 {b_2^\prime}^2 {b_3^\prime}^2 \pi^3  )^{-\frac{3}{4}}\exp\biggl[-\frac{{r_1^\prime}^2}{2{b_1^\prime}^2}-\frac{{r_2^\prime}^2}{2{b_2^\prime}^2}-\frac{{r_3^\prime}^2}{2{b_3^\prime}^2}\biggr]\sqrt{2}    \frac{x_1^\prime-i y_1^\prime}{b_1^\prime}\rangle\notag
    \\&\times\langle\sqrt{\frac{1}{6}}({b_1^\prime}^2 {b_2^\prime}^2 {b_3^\prime}^2 \pi^3  )^{-\frac{3}{4}}\exp\biggl[-\frac{{r_1^\prime}^2}{2{b_1^\prime}^2}-\frac{{r_2^\prime}^2}{2{b_2^\prime}^2}-\frac{{r_3^\prime}^2}{2{b_3^\prime}^2}\biggr]\sqrt{2}    \frac{x_1^\prime+i y_1^\prime}{b_1^\prime}|\mathbf{r_1^\prime}+\frac{\mathbf{\eta_1^\prime}}{2},\mathbf{r_2^\prime}+\frac{\mathbf{\eta_2^\prime}}{2},\mathbf{r_3^\prime}+\frac{\mathbf{\eta_3^\prime}}{2}\rangle\notag
    \\&+\int d\mathbf{\eta_1^\prime}d\mathbf{\eta_2^\prime}d\mathbf{\eta_3^\prime}e^{-i(\mathbf{\eta_1^\prime}\mathbf{k_1^\prime}+\mathbf{\eta_2^\prime}\mathbf{k_2^\prime}+\mathbf{\eta_3^\prime}\mathbf{k_3^\prime})}\notag 
    \\&\times \langle \mathbf{r_1^\prime}+\frac{\mathbf{\eta_1^\prime}}{2},\mathbf{r_2^\prime}+\frac{\mathbf{\eta_2^\prime}}{2},\mathbf{r_3^\prime}+\frac{\mathbf{\eta_3^\prime}}{2}|\sqrt{\frac{1}{6}}({b_1^\prime}^2 {b_2^\prime}^2 {b_3^\prime}^2 \pi^3  )^{-\frac{3}{4}}\exp\biggl[-\frac{{r_1^\prime}^2}{2{b_1^\prime}^2}-\frac{{r_2^\prime}^2}{2{b_2^\prime}^2}-\frac{{r_3^\prime}^2}{2{b_3^\prime}^2}\biggr](-\frac{\sqrt{2}}{2}\frac{x_1^\prime-i y_1^\prime}{b_1^\prime}-\frac{\sqrt{6}}{2}\frac{x_2^\prime-i y_2^\prime}{b_2^\prime})\rangle\notag
    \\&\times\langle \sqrt{\frac{1}{6}}({b_1^\prime}^2 {b_2^\prime}^2 {b_3^\prime}^2 \pi^3  )^{-\frac{3}{4}}\exp\biggl[-\frac{{r_1^\prime}^2}{2{b_1^\prime}^2}-\frac{{r_2^\prime}^2}{2{b_2^\prime}^2}-\frac{{r_3^\prime}^2}{2{b_3^\prime}^2}\biggr](-\frac{\sqrt{2}}{2}\frac{x_1^\prime+i y_1^\prime}{b_1^\prime}-\frac{\sqrt{6}}{2}\frac{x_2^\prime+i y_2^\prime}{b_2^\prime})|\mathbf{r_1^\prime}+\frac{\mathbf{\eta_1^\prime}}{2},\mathbf{r_2^\prime}+\frac{\mathbf{\eta_2^\prime}}{2},\mathbf{r_3^\prime}+\frac{\mathbf{\eta_3^\prime}}{2}\rangle \notag
    \\&+\int d\mathbf{\eta_1^\prime}d\mathbf{\eta_2^\prime}d\mathbf{\eta_3^\prime}e^{-i(\mathbf{\eta_1^\prime}\mathbf{k_1^\prime}+\mathbf{\eta_2^\prime}\mathbf{k_2^\prime}+\mathbf{\eta_3^\prime}\mathbf{k_3^\prime})}\notag 
    \\&\times \langle \mathbf{r_1^\prime}+\frac{\mathbf{\eta_1^\prime}}{2},\mathbf{r_2^\prime}+\frac{\mathbf{\eta_2^\prime}}{2},\mathbf{r_3^\prime}+\frac{\mathbf{\eta_3^\prime}}{2}|\sqrt{\frac{1}{6}}({b_1^\prime}^2 {b_2^\prime}^2 {b_3^\prime}^2 \pi^3  )^{-\frac{3}{4}}\exp\biggl[-\frac{{r_1^\prime}^2}{2{b_1^\prime}^2}-\frac{{r_2^\prime}^2}{2{b_2^\prime}^2}-\frac{{r_3^\prime}^2}{2{b_3^\prime}^2}\biggr](-\frac{\sqrt{2}}{2}\frac{x_1^\prime-i y_1^\prime}{b_1^\prime}+\frac{\sqrt{6}}{2}\frac{x_2^\prime-i y_2^\prime}{b_2^\prime})\rangle
    \notag\\&\times\langle \sqrt{\frac{1}{6}}({b_1^\prime}^2 {b_2^\prime}^2 {b_3^\prime}^2 \pi^3  )^{-\frac{3}{4}}\exp\biggl[-\frac{{r_1^\prime}^2}{2{b_1^\prime}^2}-\frac{{r_2^\prime}^2}{2{b_2^\prime}^2}-\frac{{r_3^\prime}^2}{2{b_3^\prime}^2}\biggr](-\frac{\sqrt{2}}{2}\frac{x_1^\prime+i y_1^\prime}{b_1^\prime}+\frac{\sqrt{6}}{2}\frac{x_2^\prime+i y_2^\prime}{b_2^\prime})|\mathbf{r_1^\prime}+\frac{\mathbf{\eta_1^\prime}}{2},\mathbf{r_2^\prime}+\frac{\mathbf{\eta_2^\prime}}{2},\mathbf{r_3^\prime}+\frac{\mathbf{\eta_3^\prime}}{2}\rangle \notag 
   \Big)    \notag
    \\&=8^3 \frac{(1+\mathcal{P}_N)^3(1-\mathcal{P}_N)}{16}\exp\left[-\frac{{r_1^\prime}^2}{{b_1^\prime}^2}-{b_1^\prime}^2 {k_1^\prime}^2-\frac{{r_2^\prime}^2}{{b_2^\prime}^2}-{b_2^\prime}^2 {k_2^\prime}^2-\frac{{r_3^\prime}^2}{{b_3^\prime}^2}-{b_3^\prime}^2 {k_3^\prime}^2\right]\notag
    \\&\times\frac{1}{2}\left[\Big(\frac{{x_1^\prime}^2+{y_1^\prime}^2}{{b_1^\prime}^2}-1+{b_1^\prime}^2({k_{x_1}^\prime}^2+{k_{y_1}^\prime}^2)+2({k_{y_1}^\prime} {x_1^\prime}-{k_{x_1}^\prime} {y_1^\prime})\Big) 
    +\Big(\frac{{x_2^\prime}^2+{y_2^\prime}^2}{{b_2^\prime}^2}-1+{b_2^\prime}^2({k_{x_2}^\prime}^2+{k_{y_2}^\prime}^2)+2({k_{y_2}^\prime} {x_2^\prime}-{k_{x_2}^\prime} {y_2^\prime})\Big)\right].
\end{align}

Substituting Eq.~(\ref{trace: Jz=2}) into Eq.~(\ref{eq:Coal}) and performing the coalescence calculation in a vortical fluid, as shown in  Appendix B, we obtain the number of $^4$Li in $J_z=2$ state as 
\begin{align}
    N_{^4{Li},J_z=2}\approx  8 \frac{N_p^3 N_n}{V^3}\left(\frac{2\pi}{mT}\right)^{\frac{9}{2}}  \frac{(1+\mathcal{P}_N)^3(1-\mathcal{P}_N)}{16}\left(1+2\mathcal{P}_L+2\mathcal{P}_L^2\right).
\end{align}

Similarly, we obtain the wave functions, traces of Wigner functions and numbers of $^4\text{Li}$ with $J_z$ being 1, 0, -1, and -2. For the wavefunctions, they are
\begin{align}
    \langle \mathbf{r_1},...,\mathbf{r_4}\left|2,1\right>&=\frac{1}{2}\sqrt{\frac{1}{3!}}\left.\left[\begin{matrix}\phi_{M=1}^{L=1}(\mathbf{r_1})\left|\frac{1}{2},\frac{1}{2}\right>_{p_1}&\phi_{M=1}^{L=1}(\mathbf{r_2})\left|\frac{1}{2},\frac{1}{2}\right>_{p_2}&\phi_{M=1}^{L=1}(\mathbf{r_3})\left|\frac{1}{2},\frac{1}{2}\right>_{p_3}\\\phi_{M=0}^{L=0}(\mathbf{r_1})\left|\frac{1}{2},\frac{1}{2}\right>_{p_1}&\phi_{M=0}^{L=0}(\mathbf{r_2})\left|\frac{1}{2},\frac{1}{2}\right>_{p_2}&\phi_{M=0}^{L=0}(\mathbf{r_3})\left|\frac{1}{2},\frac{1}{2}\right>_{p_3}\\\phi_{M=0}^{L=0}(\mathbf{r_1})\left|\frac{1}{2},-\frac{1}{2}\right>_{p_1}&\phi_{M=0}^{L=0}(\mathbf{r_2})\left|\frac{1}{2},-\frac{1}{2}\right>_{p_2}&\phi_{M=0}^{L=0}(\mathbf{r_3})\left|\frac{1}{2},-\frac{1}{2}\right>_{p_3}\end{matrix}\right.\right]\phi_{M=0}^{L=0}(\mathbf{r_4})\left|\frac{1}{2},-\frac{1}{2}\right>_{n}\notag\\&+\frac{1}{2}\sqrt{\frac{1}{3!}}\left.\left[\begin{matrix}\phi_{M=1}^{L=1}(\mathbf{r_1})\left|\frac{1}{2},-\frac{1}{2}\right>_{p_1}&\phi_{M=1}^{L=1}(\mathbf{r_2})\left|\frac{1}{2},-\frac{1}{2}\right>_{p_2}&\phi_{M=1}^{L=1}(\mathbf{r_3})\left|\frac{1}{2},-\frac{1}{2}\right>_{p_3}\\\phi_{M=0}^{L=0}(\mathbf{r_1})\left|\frac{1}{2},\frac{1}{2}\right>_{p_1}&\phi_{M=0}^{L=0}(\mathbf{r_2})\left|\frac{1}{2},\frac{1}{2}\right>_{p_2}&\phi_{M=0}^{L=0}(\mathbf{r_3})\left|\frac{1}{2},\frac{1}{2}\right>_{p_3}\\\phi_{M=0}^{L=0}(\mathbf{r_1})\left|\frac{1}{2},-\frac{1}{2}\right>_{p_1}&\phi_{M=0}^{L=0}(\mathbf{r_2})\left|\frac{1}{2},-\frac{1}{2}\right>_{p_2}&\phi_{M=0}^{L=0}(\mathbf{r_3})\left|\frac{1}{2},-\frac{1}{2}\right>_{p_3}\end{matrix}\right.\right]\phi_{M=0}^{L=0}(\mathbf{r_4})\left|\frac{1}{2},\frac{1}{2}\right>_{n}\notag\\&+\sqrt{\frac{1}{2}}\sqrt{\frac{1}{3!}}\left.\left[\begin{matrix}\phi_{M=0}^{L=1}(\mathbf{r_1})\left|\frac{1}{2},\frac{1}{2}\right>_{p_1}&\phi_{M=0}^{L=1}(\mathbf{r_2})\left|\frac{1}{2},\frac{1}{2}\right>_{p_2}&\phi_{M=0}^{L=1}(\mathbf{r_3})\left|\frac{1}{2},\frac{1}{2}\right>_{p_3}\\\phi_{M=0}^{L=0}(\mathbf{r_1})\left|\frac{1}{2},\frac{1}{2}\right>_{p_1}&\phi_{M=0}^{L=0}(\mathbf{r_2})\left|\frac{1}{2},\frac{1}{2}\right>_{p_2}&\phi_{M=0}^{L=0}(\mathbf{r_3})\left|\frac{1}{2},\frac{1}{2}\right>_{p_3}\\\phi_{M=0}^{L=0}(\mathbf{r_1})\left|\frac{1}{2},-\frac{1}{2}\right>_{p_1}&\phi_{M=0}^{L=0}(\mathbf{r_2})\left|\frac{1}{2},-\frac{1}{2}\right>_{p_2}&\phi_{M=0}^{L=0}(\mathbf{r_3})\left|\frac{1}{2},-\frac{1}{2}\right>_{p_3}\end{matrix}\right.\right]\phi_{M=0}^{L=0}(\mathbf{r_4})\left|\frac{1}{2},\frac{1}{2}\right>_{n}\notag\\&=\sqrt{\frac{1}{24}}(\phi_{M=1}^{L=1}(\mathbf{r_1})\phi_{M=0}^{L=0}(\mathbf{r_2})\phi_{M=0}^{L=0}(\mathbf{r_3})\left|\frac{1}{2},\frac{1}{2}\right>_{p_1}\left|\frac{1}{2},\frac{1}{2}\right>_{p_2}\left|\frac{1}{2},-\frac{1}{2}\right>_{p_3}\notag\\&-\phi_{M=1}^{L=1}(\mathbf{r_1})\phi_{M=0}^{L=0}(\mathbf{r_2})\phi_{M=0}^{L=0}(\mathbf{r_3})\left|\frac{1}{2},\frac{1}{2}\right>_{p_1}\left|\frac{1}{2},-\frac{1}{2}\right>_{p_2}\left|\frac{1}{2},\frac{1}{2}\right>_{p_3}\notag\\&+\phi_{M=1}^{L=1}(\mathbf{r_3})\phi_{M=0}^{L=0}(\mathbf{r_1})\phi_{M=0}^{L=0}(\mathbf{r_2})\left|\frac{1}{2},\frac{1}{2}\right>_{p_3}\left|\frac{1}{2},\frac{1}{2}\right>_{p_1}\left|\frac{1}{2},-\frac{1}{2}\right>_{p_2}\notag\\&-\phi_{M=1}^{L=1}(\mathbf{r_2})\phi_{M=0}^{L=0}(\mathbf{r_1})\phi_{M=0}^{L=0}(\mathbf{r_3})\left|\frac{1}{2},\frac{1}{2}\right>_{p_2}\left|\frac{1}{2},\frac{1}{2}\right>_{p_1}\left|\frac{1}{2},-\frac{1}{2}\right>_{p_3}\notag\\&-\phi_{M=1}^{L=1}(\mathbf{r_3})\phi_{M=0}^{L=0}(\mathbf{r_2})\phi_{M=0}^{L=0}(\mathbf{r_1})\left|\frac{1}{2},\frac{1}{2}\right>_{p_3}\left|\frac{1}{2},\frac{1}{2}\right>_{p_2}\left|\frac{1}{2},-\frac{1}{2}\right>_{p_1}\notag\\&+\phi_{M=1}^{L=1}(\mathbf{r_2})\phi_{M=0}^{L=0}(\mathbf{r_3})\phi_{M=0}^{L=0}(\mathbf{r_1})\left|\frac{1}{2},\frac{1}{2}\right>_{p_2}\left|\frac{1}{2},\frac{1}{2}\right>_{p_3}\left|\frac{1}{2},-\frac{1}{2}\right>_{p_1})\phi_{M=0}^{L=0}(\mathbf{r_4})\left|\frac{1}{2},-\frac{1}{2}\right>_{n}\notag\\&+\sqrt{\frac{1}{24}}(\phi_{M=1}^{L=1}(\mathbf{r_1})\phi_{M=0}^{L=0}(\mathbf{r_2})\phi_{M=0}^{L=0}(\mathbf{r_3})\left|\frac{1}{2},-\frac{1}{2}\right>_{p_1}\left|\frac{1}{2},\frac{1}{2}\right>_{p_2}\left|\frac{1}{2},-\frac{1}{2}\right>_{p_3}\notag\\&-\phi_{M=1}^{L=1}(\mathbf{r_1})\phi_{M=0}^{L=0}(\mathbf{r_2})\phi_{M=0}^{L=0}(\mathbf{r_3})\left|\frac{1}{2},-\frac{1}{2}\right>_{p_1}\left|\frac{1}{2},-\frac{1}{2}\right>_{p_2}\left|\frac{1}{2},\frac{1}{2}\right>_{p_3}\notag\\&+\phi_{M=1}^{L=1}(\mathbf{r_3})\phi_{M=0}^{L=0}(\mathbf{r_1})\phi_{M=0}^{L=0}(\mathbf{r_2})\left|\frac{1}{2},-\frac{1}{2}\right>_{p_3}\left|\frac{1}{2},\frac{1}{2}\right>_{p_1}\left|\frac{1}{2},-\frac{1}{2}\right>_{p_2}\notag\\&-\phi_{M=1}^{L=1}(\mathbf{r_2})\phi_{M=0}^{L=0}(\mathbf{r_1})\phi_{M=0}^{L=0}(\mathbf{r_3})\left|\frac{1}{2},-\frac{1}{2}\right>_{p_2}\left|\frac{1}{2},\frac{1}{2}\right>_{p_1}\left|\frac{1}{2},-\frac{1}{2}\right>_{p_3}\notag\\&-\phi_{M=1}^{L=1}(\mathbf{r_3})\phi_{M=0}^{L=0}(\mathbf{r_2})\phi_{M=0}^{L=0}(\mathbf{r_1})\left|\frac{1}{2},-\frac{1}{2}\right>_{p_3}\left|\frac{1}{2},\frac{1}{2}\right>_{p_2}\left|\frac{1}{2},-\frac{1}{2}\right>_{p_1}\notag\\&+\phi_{M=1}^{L=1}(\mathbf{r_2})\phi_{M=0}^{L=0}(\mathbf{r_3})\phi_{M=0}^{L=0}(\mathbf{r_1})\left|\frac{1}{2},-\frac{1}{2}\right>_{p_2}\left|\frac{1}{2},\frac{1}{2}\right>_{p_3}\left|\frac{1}{2},-\frac{1}{2}\right>_{p_1})\phi_{M=0}^{L=0}(\mathbf{r_4})\left|\frac{1}{2},\frac{1}{2}\right>_{n}\allowdisplaybreaks\notag\\&+\sqrt{\frac{1}{12}}(\phi_{M=0}^{L=1}(\mathbf{r_1})\phi_{M=0}^{L=0}(\mathbf{r_2})\phi_{M=0}^{L=0}(\mathbf{r_3})\left|\frac{1}{2},\frac{1}{2}\right>_{p_1}\left|\frac{1}{2},\frac{1}{2}\right>_{p_2}\left|\frac{1}{2},-\frac{1}{2}\right>_{p_3}\notag\\&-\phi_{M=0}^{L=1}(\mathbf{r_1})\phi_{M=0}^{L=0}(\mathbf{r_2})\phi_{M=0}^{L=0}(\mathbf{r_3})\left|\frac{1}{2},\frac{1}{2}\right>_{p_1}\left|\frac{1}{2},-\frac{1}{2}\right>_{p_2}\left|\frac{1}{2},\frac{1}{2}\right>_{p_3}\notag\\&+\phi_{M=0}^{L=1}(\mathbf{r_3})\phi_{M=0}^{L=0}(\mathbf{r_1})\phi_{M=0}^{L=0}(\mathbf{r_2})\left|\frac{1}{2},\frac{1}{2}\right>_{p_3}\left|\frac{1}{2},\frac{1}{2}\right>_{p_1}\left|\frac{1}{2},-\frac{1}{2}\right>_{p_2}\notag\\&-\phi_{M=0}^{L=1}(\mathbf{r_2})\phi_{M=0}^{L=0}(\mathbf{r_1})\phi_{M=0}^{L=0}(\mathbf{r_3})\left|\frac{1}{2},\frac{1}{2}\right>_{p_2}\left|\frac{1}{2},\frac{1}{2}\right>_{p_1}\left|\frac{1}{2},-\frac{1}{2}\right>_{p_3}\notag\\&-\phi_{M=0}^{L=1}(\mathbf{r_3})\phi_{M=0}^{L=0}(\mathbf{r_2})\phi_{M=0}^{L=0}(\mathbf{r_1})\left|\frac{1}{2},\frac{1}{2}\right>_{p_3}\left|\frac{1}{2},\frac{1}{2}\right>_{p_2}\left|\frac{1}{2},-\frac{1}{2}\right>_{p_1}\notag\\&+\phi_{M=0}^{L=1}(\mathbf{r_2})\phi_{M=0}^{L=0}(\mathbf{r_3})\phi_{M=0}^{L=0}(\mathbf{r_1})\left|\frac{1}{2},\frac{1}{2}\right>_{p_2}\left|\frac{1}{2},\frac{1}{2}\right>_{p_3}\left|\frac{1}{2},-\frac{1}{2}\right>_{p_1})\phi_{M=0}^{L=0}(\mathbf{r_4})\left|\frac{1}{2},\frac{1}{2}\right>_{n},
\end{align}
\begin{align}
    \langle \mathbf{r_1},...,\mathbf{r_4}\left|2,0\right>&=\sqrt{\frac{1}{12}}\sqrt{\frac{1}{3!}}\left.\left[\begin{matrix}\phi_{M=1}^{L=1}(\mathbf{r_1})\left|\frac{1}{2},-\frac{1}{2}\right>_{p_1}&\phi_{M=1}^{L=1}(\mathbf{r_2})\left|\frac{1}{2},-\frac{1}{2}\right>_{p_2}&\phi_{M=1}^{L=1}(\mathbf{r_3})\left|\frac{1}{2},-\frac{1}{2}\right>_{p_3}\\\phi_{M=0}^{L=0}(\mathbf{r_1})\left|\frac{1}{2},\frac{1}{2}\right>_{p_1}&\phi_{M=0}^{L=0}(\mathbf{r_2})\left|\frac{1}{2},\frac{1}{2}\right>_{p_2}&\phi_{M=0}^{L=0}(\mathbf{r_3})\left|\frac{1}{2},\frac{1}{2}\right>_{p_3}\\\phi_{M=0}^{L=0}(\mathbf{r_1})\left|\frac{1}{2},-\frac{1}{2}\right>_{p_1}&\phi_{M=0}^{L=0}(\mathbf{r_2})\left|\frac{1}{2},-\frac{1}{2}\right>_{p_2}&\phi_{M=0}^{L=0}(\mathbf{r_3})\left|\frac{1}{2},-\frac{1}{2}\right>_{p_3}\end{matrix}\right.\right]\phi_{M=0}^{L=0}(\mathbf{r_4})\left|\frac{1}{2},-\frac{1}{2}\right>_{n}\notag\\&+\sqrt{\frac{1}{6}}\sqrt{\frac{1}{3!}}\left.\left[\begin{matrix}\phi_{M=0}^{L=1}(\mathbf{r_1})\left|\frac{1}{2},\frac{1}{2}\right>_{p_1}&\phi_{M=0}^{L=1}(\mathbf{r_2})\left|\frac{1}{2},\frac{1}{2}\right>_{p_2}&\phi_{M=0}^{L=1}(\mathbf{r_3})\left|\frac{1}{2},\frac{1}{2}\right>_{p_3}\\\phi_{M=0}^{L=0}(\mathbf{r_1})\left|\frac{1}{2},\frac{1}{2}\right>_{p_1}&\phi_{M=0}^{L=0}(\mathbf{r_2})\left|\frac{1}{2},\frac{1}{2}\right>_{p_2}&\phi_{M=0}^{L=0}(\mathbf{r_3})\left|\frac{1}{2},\frac{1}{2}\right>_{p_3}\\\phi_{M=0}^{L=0}(\mathbf{r_1})\left|\frac{1}{2},-\frac{1}{2}\right>_{p_1}&\phi_{M=0}^{L=0}(\mathbf{r_2})\left|\frac{1}{2},-\frac{1}{2}\right>_{p_2}&\phi_{M=0}^{L=0}(\mathbf{r_3})\left|\frac{1}{2},-\frac{1}{2}\right>_{p_3}\end{matrix}\right.\right]\phi_{M=0}^{L=0}(\mathbf{r_4})\left|\frac{1}{2},-\frac{1}{2}\right>_{n}\notag\\&+\sqrt{\frac{1}{6}}\sqrt{\frac{1}{3!}}\left.\left[\begin{matrix}\phi_{M=0}^{L=1}(\mathbf{r_1})\left|\frac{1}{2},-\frac{1}{2}\right>_{p_1}&\phi_{M=0}^{L=1}(\mathbf{r_2})\left|\frac{1}{2},-\frac{1}{2}\right>_{p_2}&\phi_{M=0}^{L=1}(\mathbf{r_3})\left|\frac{1}{2},-\frac{1}{2}\right>_{p_3}\\\phi_{M=0}^{L=0}(\mathbf{r_1})\left|\frac{1}{2},\frac{1}{2}\right>_{p_1}&\phi_{M=0}^{L=0}(\mathbf{r_2})\left|\frac{1}{2},\frac{1}{2}\right>_{p_2}&\phi_{M=0}^{L=0}(\mathbf{r_3})\left|\frac{1}{2},\frac{1}{2}\right>_{p_3}\\\phi_{M=0}^{L=0}(\mathbf{r_1})\left|\frac{1}{2},-\frac{1}{2}\right>_{p_1}&\phi_{M=0}^{L=0}(\mathbf{r_2})\left|\frac{1}{2},-\frac{1}{2}\right>_{p_2}&\phi_{M=0}^{L=0}(\mathbf{r_3})\left|\frac{1}{2},-\frac{1}{2}\right>_{p_3}\end{matrix}\right.\right]\phi_{M=0}^{L=0}(\mathbf{r_4})\left|\frac{1}{2},\frac{1}{2}\right>_{n}\notag\\&+\sqrt{\frac{1}{12}}\sqrt{\frac{1}{3!}}\left.\left[\begin{matrix}\phi_{M=-1}^{L=1}(\mathbf{r_1})\left|\frac{1}{2},\frac{1}{2}\right>_{p_1}&\phi_{M=-1}^{L=1}(\mathbf{r_2})\left|\frac{1}{2},\frac{1}{2}\right>_{p_2}&\phi_{M=-1}^{L=1}(\mathbf{r_3})\left|\frac{1}{2},\frac{1}{2}\right>_{p_3}\\\phi_{M=0}^{L=0}(\mathbf{r_1})\left|\frac{1}{2},\frac{1}{2}\right>_{p_1}&\phi_{M=0}^{L=0}(\mathbf{r_2})\left|\frac{1}{2},\frac{1}{2}\right>_{p_2}&\phi_{M=0}^{L=0}(\mathbf{r_3})\left|\frac{1}{2},\frac{1}{2}\right>_{p_3}\\\phi_{M=0}^{L=0}(\mathbf{r_1})\left|\frac{1}{2},-\frac{1}{2}\right>_{p_1}&\phi_{M=0}^{L=0}(\mathbf{r_2})\left|\frac{1}{2},-\frac{1}{2}\right>_{p_2}&\phi_{M=0}^{L=0}(\mathbf{r_3})\left|\frac{1}{2},-\frac{1}{2}\right>_{p_3}\end{matrix}\right.\right]\phi_{M=0}^{L=0}(\mathbf{r_4})\left|\frac{1}{2},\frac{1}{2}\right>_{n}\notag\\&=\sqrt{\frac{1}{36}}(\phi_{M=1}^{L=1}(\mathbf{r_1})\phi_{M=0}^{L=0}(\mathbf{r_2})\phi_{M=0}^{L=0}(\mathbf{r_3})\left|\frac{1}{2},-\frac{1}{2}\right>_{p_1}\left|\frac{1}{2},\frac{1}{2}\right>_{p_2}\left|\frac{1}{2},-\frac{1}{2}\right>_{p_3}\notag\\&-\phi_{M=1}^{L=1}(\mathbf{r_1})\phi_{M=0}^{L=0}(\mathbf{r_2})\phi_{M=0}^{L=0}(\mathbf{r_3})\left|\frac{1}{2},-\frac{1}{2}\right>_{p_1}\left|\frac{1}{2},-\frac{1}{2}\right>_{p_2}\left|\frac{1}{2},\frac{1}{2}\right>_{p_3}\notag\\&+\phi_{M=1}^{L=1}(\mathbf{r_3})\phi_{M=0}^{L=0}(\mathbf{r_1})\phi_{M=0}^{L=0}(\mathbf{r_2})\left|\frac{1}{2},-\frac{1}{2}\right>_{p_3}\left|\frac{1}{2},\frac{1}{2}\right>_{p_1}\left|\frac{1}{2},-\frac{1}{2}\right>_{p_2}\notag\\&-\phi_{M=1}^{L=1}(\mathbf{r_2})\phi_{M=0}^{L=0}(\mathbf{r_1})\phi_{M=0}^{L=0}(\mathbf{r_3})\left|\frac{1}{2},-\frac{1}{2}\right>_{p_2}\left|\frac{1}{2},\frac{1}{2}\right>_{p_1}\left|\frac{1}{2},-\frac{1}{2}\right>_{p_3}\notag\\&-\phi_{M=1}^{L=1}(\mathbf{r_3})\phi_{M=0}^{L=0}(\mathbf{r_2})\phi_{M=0}^{L=0}(\mathbf{r_1})\left|\frac{1}{2},-\frac{1}{2}\right>_{p_3}\left|\frac{1}{2},\frac{1}{2}\right>_{p_2}\left|\frac{1}{2},-\frac{1}{2}\right>_{p_1}\notag\\&+\phi_{M=1}^{L=1}(\mathbf{r_2})\phi_{M=0}^{L=0}(\mathbf{r_3})\phi_{M=0}^{L=0}(\mathbf{r_1})\left|\frac{1}{2},-\frac{1}{2}\right>_{p_2}\left|\frac{1}{2},\frac{1}{2}\right>_{p_3}\left|\frac{1}{2},-\frac{1}{2}\right>_{p_1})\phi_{M=0}^{L=0}(\mathbf{r_4})\left|\frac{1}{2},-\frac{1}{2}\right>_{n}\notag\\&+\sqrt{\frac{1}{18}}(\phi_{M=0}^{L=1}(\mathbf{r_1})\phi_{M=0}^{L=0}(\mathbf{r_2})\phi_{M=0}^{L=0}(\mathbf{r_3})\left|\frac{1}{2},\frac{1}{2}\right>_{p_1}\left|\frac{1}{2},\frac{1}{2}\right>_{p_2}\left|\frac{1}{2},-\frac{1}{2}\right>_{p_3}\notag\\&-\phi_{M=0}^{L=1}(\mathbf{r_1})\phi_{M=0}^{L=0}(\mathbf{r_2})\phi_{M=0}^{L=0}(\mathbf{r_3})\left|\frac{1}{2},\frac{1}{2}\right>_{p_1}\left|\frac{1}{2},-\frac{1}{2}\right>_{p_2}\left|\frac{1}{2},\frac{1}{2}\right>_{p_3}\notag\\&+\phi_{M=0}^{L=1}(\mathbf{r_3})\phi_{M=0}^{L=0}(\mathbf{r_1})\phi_{M=0}^{L=0}(\mathbf{r_2})\left|\frac{1}{2},\frac{1}{2}\right>_{p_3}\left|\frac{1}{2},\frac{1}{2}\right>_{p_1}\left|\frac{1}{2},-\frac{1}{2}\right>_{p_2}\notag\\&-\phi_{M=0}^{L=1}(\mathbf{r_2})\phi_{M=0}^{L=0}(\mathbf{r_1})\phi_{M=0}^{L=0}(\mathbf{r_3})\left|\frac{1}{2},\frac{1}{2}\right>_{p_2}\left|\frac{1}{2},\frac{1}{2}\right>_{p_1}\left|\frac{1}{2},-\frac{1}{2}\right>_{p_3}\notag\\&-\phi_{M=0}^{L=1}(\mathbf{r_3})\phi_{M=0}^{L=0}(\mathbf{r_2})\phi_{M=0}^{L=0}(\mathbf{r_1})\left|\frac{1}{2},\frac{1}{2}\right>_{p_3}\left|\frac{1}{2},\frac{1}{2}\right>_{p_2}\left|\frac{1}{2},-\frac{1}{2}\right>_{p_1}\notag\\&+\phi_{M=0}^{L=1}(\mathbf{r_2})\phi_{M=0}^{L=0}(\mathbf{r_3})\phi_{M=0}^{L=0}(\mathbf{r_1})\left|\frac{1}{2},\frac{1}{2}\right>_{p_2}\left|\frac{1}{2},\frac{1}{2}\right>_{p_3}\left|\frac{1}{2},-\frac{1}{2}\right>_{p_1})\phi_{M=0}^{L=0}(\mathbf{r_4})\left|\frac{1}{2},-\frac{1}{2}\right>_{n}\notag\\&+\sqrt{\frac{1}{18}}(\phi_{M=0}^{L=1}(\mathbf{r_1})\phi_{M=0}^{L=0}(\mathbf{r_2})\phi_{M=0}^{L=0}(\mathbf{r_3})\left|\frac{1}{2},-\frac{1}{2}\right>_{p_1}\left|\frac{1}{2},\frac{1}{2}\right>_{p_2}\left|\frac{1}{2},-\frac{1}{2}\right>_{p_3}\notag\\&-\phi_{M=0}^{L=1}(\mathbf{r_1})\phi_{M=0}^{L=0}(\mathbf{r_2})\phi_{M=0}^{L=0}(\mathbf{r_3})\left|\frac{1}{2},-\frac{1}{2}\right>_{p_1}\left|\frac{1}{2},-\frac{1}{2}\right>_{p_2}\left|\frac{1}{2},\frac{1}{2}\right>_{p_3}\notag\\&+\phi_{M=0}^{L=1}(\mathbf{r_3})\phi_{M=0}^{L=0}(\mathbf{r_1})\phi_{M=0}^{L=0}(\mathbf{r_2})\left|\frac{1}{2},-\frac{1}{2}\right>_{p_3}\left|\frac{1}{2},\frac{1}{2}\right>_{p_1}\left|\frac{1}{2},-\frac{1}{2}\right>_{p_2}\notag\\&-\phi_{M=0}^{L=1}(\mathbf{r_2})\phi_{M=0}^{L=0}(\mathbf{r_1})\phi_{M=0}^{L=0}(\mathbf{r_3})\left|\frac{1}{2},-\frac{1}{2}\right>_{p_2}\left|\frac{1}{2},\frac{1}{2}\right>_{p_1}\left|\frac{1}{2},-\frac{1}{2}\right>_{p_3}\notag\\&-\phi_{M=0}^{L=1}(\mathbf{r_3})\phi_{M=0}^{L=0}(\mathbf{r_2})\phi_{M=0}^{L=0}(\mathbf{r_1})\left|\frac{1}{2},-\frac{1}{2}\right>_{p_3}\left|\frac{1}{2},\frac{1}{2}\right>_{p_2}\left|\frac{1}{2},-\frac{1}{2}\right>_{p_1}\notag\\&+\phi_{M=0}^{L=1}(\mathbf{r_2})\phi_{M=0}^{L=0}(\mathbf{r_3})\phi_{M=0}^{L=0}(\mathbf{r_1})\left|\frac{1}{2},-\frac{1}{2}\right>_{p_2}\left|\frac{1}{2},\frac{1}{2}\right>_{p_3}\left|\frac{1}{2},-\frac{1}{2}\right>_{p_1})\phi_{M=0}^{L=0}(\mathbf{r_4})\left|\frac{1}{2},\frac{1}{2}\right>_{n}\notag\\&+\sqrt{\frac{1}{36}}(\phi_{M=-1}^{L=1}(\mathbf{r_1})\phi_{M=0}^{L=0}(\mathbf{r_2})\phi_{M=0}^{L=0}(\mathbf{r_3})\left|\frac{1}{2},\frac{1}{2}\right>_{p_1}\left|\frac{1}{2},\frac{1}{2}\right>_{p_2}\left|\frac{1}{2},-\frac{1}{2}\right>_{p_3}\notag\\&-\phi_{M=-1}^{L=1}(\mathbf{r_1})\phi_{M=0}^{L=0}(\mathbf{r_2})\phi_{M=0}^{L=0}(\mathbf{r_3})\left|\frac{1}{2},\frac{1}{2}\right>_{p_1}\left|\frac{1}{2},-\frac{1}{2}\right>_{p_2}\left|\frac{1}{2},\frac{1}{2}\right>_{p_3}\notag\\&+\phi_{M=-1}^{L=1}(\mathbf{r_3})\phi_{M=0}^{L=0}(\mathbf{r_1})\phi_{M=0}^{L=0}(\mathbf{r_2})\left|\frac{1}{2},\frac{1}{2}\right>_{p_3}\left|\frac{1}{2},\frac{1}{2}\right>_{p_1}\left|\frac{1}{2},-\frac{1}{2}\right>_{p_2}\notag\\&-\phi_{M=-1}^{L=1}(\mathbf{r_2})\phi_{M=0}^{L=0}(\mathbf{r_1})\phi_{M=0}^{L=0}(\mathbf{r_3})\left|\frac{1}{2},\frac{1}{2}\right>_{p_2}\left|\frac{1}{2},\frac{1}{2}\right>_{p_1}\left|\frac{1}{2},-\frac{1}{2}\right>_{p_3}\notag\\&-\phi_{M=-1}^{L=1}(\mathbf{r_3})\phi_{M=0}^{L=0}(\mathbf{r_2})\phi_{M=0}^{L=0}(\mathbf{r_1})\left|\frac{1}{2},\frac{1}{2}\right>_{p_3}\left|\frac{1}{2},\frac{1}{2}\right>_{p_2}\left|\frac{1}{2},-\frac{1}{2}\right>_{p_1}\notag\\&+\phi_{M=-1}^{L=1}(\mathbf{r_2})\phi_{M=0}^{L=0}(\mathbf{r_3})\phi_{M=0}^{L=0}(\mathbf{r_1})\left|\frac{1}{2},\frac{1}{2}\right>_{p_2}\left|\frac{1}{2},\frac{1}{2}\right>_{p_3}\left|\frac{1}{2},-\frac{1}{2}\right>_{p_1})\phi_{M=0}^{L=0}(\mathbf{r_4})\left|\frac{1}{2},\frac{1}{2}\right>_{n},
\end{align}
\begin{align}
    \langle \mathbf{r_1},...,\mathbf{r_4}\left|2,-1\right>&=\frac{1}{2}\sqrt{\frac{1}{3!}}\left.\left[\begin{matrix}\phi_{M=-1}^{L=1}(\mathbf{r_1})\left|\frac{1}{2},\frac{1}{2}\right>_{p_1}&\phi_{M=-1}^{L=1}(\mathbf{r_2})\left|\frac{1}{2},\frac{1}{2}\right>_{p_2}&\phi_{M=-1}^{L=1}(\mathbf{r_3})\left|\frac{1}{2},\frac{1}{2}\right>_{p_3}\\\phi_{M=0}^{L=0}(\mathbf{r_1})\left|\frac{1}{2},\frac{1}{2}\right>_{p_1}&\phi_{M=0}^{L=0}(\mathbf{r_2})\left|\frac{1}{2},\frac{1}{2}\right>_{p_2}&\phi_{M=0}^{L=0}(\mathbf{r_3})\left|\frac{1}{2},\frac{1}{2}\right>_{p_3}\\\phi_{M=0}^{L=0}(\mathbf{r_1})\left|\frac{1}{2},-\frac{1}{2}\right>_{p_1}&\phi_{M=0}^{L=0}(\mathbf{r_2})\left|\frac{1}{2},-\frac{1}{2}\right>_{p_2}&\phi_{M=0}^{L=0}(\mathbf{r_3})\left|\frac{1}{2},-\frac{1}{2}\right>_{p_3}\end{matrix}\right.\right]\phi_{M=0}^{L=0}(\mathbf{r_4})\left|\frac{1}{2},-\frac{1}{2}\right>_{n}\notag\\&+\frac{1}{2}\sqrt{\frac{1}{3!}}\left.\left[\begin{matrix}\phi_{M=-1}^{L=1}(\mathbf{r_1})\left|\frac{1}{2},-\frac{1}{2}\right>_{p_1}&\phi_{M=-1}^{L=1}(\mathbf{r_2})\left|\frac{1}{2},-\frac{1}{2}\right>_{p_2}&\phi_{M=-1}^{L=1}(\mathbf{r_3})\left|\frac{1}{2},-\frac{1}{2}\right>_{p_3}\\\phi_{M=0}^{L=0}(\mathbf{r_1})\left|\frac{1}{2},\frac{1}{2}\right>_{p_1}&\phi_{M=0}^{L=0}(\mathbf{r_2})\left|\frac{1}{2},\frac{1}{2}\right>_{p_2}&\phi_{M=0}^{L=0}(\mathbf{r_3})\left|\frac{1}{2},\frac{1}{2}\right>_{p_3}\\\phi_{M=0}^{L=0}(\mathbf{r_1})\left|\frac{1}{2},-\frac{1}{2}\right>_{p_1}&\phi_{M=0}^{L=0}(\mathbf{r_2})\left|\frac{1}{2},-\frac{1}{2}\right>_{p_2}&\phi_{M=0}^{L=0}(\mathbf{r_3})\left|\frac{1}{2},-\frac{1}{2}\right>_{p_3}\end{matrix}\right.\right]\phi_{M=0}^{L=0}(\mathbf{r_4})\left|\frac{1}{2},\frac{1}{2}\right>_{n}\notag\\&+\sqrt{\frac{1}{2}}\sqrt{\frac{1}{3!}}\left.\left[\begin{matrix}\phi_{M=0}^{L=1}(\mathbf{r_1})\left|\frac{1}{2},-\frac{1}{2}\right>_{p_1}&\phi_{M=0}^{L=1}(\mathbf{r_2})\left|\frac{1}{2},-\frac{1}{2}\right>_{p_2}&\phi_{M=0}^{L=1}(\mathbf{r_3})\left|\frac{1}{2},-\frac{1}{2}\right>_{p_3}\\\phi_{M=0}^{L=0}(\mathbf{r_1})\left|\frac{1}{2},\frac{1}{2}\right>_{p_1}&\phi_{M=0}^{L=0}(\mathbf{r_2})\left|\frac{1}{2},\frac{1}{2}\right>_{p_2}&\phi_{M=0}^{L=0}(\mathbf{r_3})\left|\frac{1}{2},\frac{1}{2}\right>_{p_3}\\\phi_{M=0}^{L=0}(\mathbf{r_1})\left|\frac{1}{2},-\frac{1}{2}\right>_{p_1}&\phi_{M=0}^{L=0}(\mathbf{r_2})\left|\frac{1}{2},-\frac{1}{2}\right>_{p_2}&\phi_{M=0}^{L=0}(\mathbf{r_3})\left|\frac{1}{2},-\frac{1}{2}\right>_{p_3}\end{matrix}\right.\right]\phi_{M=0}^{L=0}(\mathbf{r_4})\left|\frac{1}{2},-\frac{1}{2}\right>_{n}\notag\\&=\sqrt{\frac{1}{24}}(\phi_{M=-1}^{L=1}(\mathbf{r_1})\phi_{M=0}^{L=0}(\mathbf{r_2})\phi_{M=0}^{L=0}(\mathbf{r_3})\left|\frac{1}{2},\frac{1}{2}\right>_{p_1}\left|\frac{1}{2},\frac{1}{2}\right>_{p_2}\left|\frac{1}{2},-\frac{1}{2}\right>_{p_3}\notag\\&-\phi_{M=-1}^{L=1}(\mathbf{r_1})\phi_{M=0}^{L=0}(\mathbf{r_2})\phi_{M=0}^{L=0}(\mathbf{r_3})\left|\frac{1}{2},\frac{1}{2}\right>_{p_1}\left|\frac{1}{2},-\frac{1}{2}\right>_{p_2}\left|\frac{1}{2},\frac{1}{2}\right>_{p_3}\notag\\&+\phi_{M=-1}^{L=1}(\mathbf{r_3})\phi_{M=0}^{L=0}(\mathbf{r_1})\phi_{M=0}^{L=0}(\mathbf{r_2})\left|\frac{1}{2},\frac{1}{2}\right>_{p_3}\left|\frac{1}{2},\frac{1}{2}\right>_{p_1}\left|\frac{1}{2},-\frac{1}{2}\right>_{p_2}\notag\\&-\phi_{M=-1}^{L=1}(\mathbf{r_2})\phi_{M=0}^{L=0}(\mathbf{r_1})\phi_{M=0}^{L=0}(\mathbf{r_3})\left|\frac{1}{2},\frac{1}{2}\right>_{p_2}\left|\frac{1}{2},\frac{1}{2}\right>_{p_1}\left|\frac{1}{2},-\frac{1}{2}\right>_{p_3}\notag\\&-\phi_{M=-1}^{L=1}(\mathbf{r_3})\phi_{M=0}^{L=0}(\mathbf{r_2})\phi_{M=0}^{L=0}(\mathbf{r_1})\left|\frac{1}{2},\frac{1}{2}\right>_{p_3}\left|\frac{1}{2},\frac{1}{2}\right>_{p_2}\left|\frac{1}{2},-\frac{1}{2}\right>_{p_1}\notag\\&+\phi_{M=-1}^{L=1}(\mathbf{r_2})\phi_{M=0}^{L=0}(\mathbf{r_3})\phi_{M=0}^{L=0}(\mathbf{r_1})\left|\frac{1}{2},\frac{1}{2}\right>_{p_2}\left|\frac{1}{2},\frac{1}{2}\right>_{p_3}\left|\frac{1}{2},-\frac{1}{2}\right>_{p_1})\phi_{M=0}^{L=0}(\mathbf{r_4})\left|\frac{1}{2},-\frac{1}{2}\right>_{n}\notag\\&+\sqrt{\frac{1}{24}}(\phi_{M=-1}^{L=1}(\mathbf{r_1})\phi_{M=0}^{L=0}(\mathbf{r_2})\phi_{M=0}^{L=0}(\mathbf{r_3})\left|\frac{1}{2},-\frac{1}{2}\right>_{p_1}\left|\frac{1}{2},\frac{1}{2}\right>_{p_2}\left|\frac{1}{2},-\frac{1}{2}\right>_{p_3}\notag\\&-\phi_{M=-1}^{L=1}(\mathbf{r_1})\phi_{M=0}^{L=0}(\mathbf{r_2})\phi_{M=0}^{L=0}(\mathbf{r_3})\left|\frac{1}{2},-\frac{1}{2}\right>_{p_1}\left|\frac{1}{2},-\frac{1}{2}\right>_{p_2}\left|\frac{1}{2},\frac{1}{2}\right>_{p_3}\notag\\&+\phi_{M=-1}^{L=1}(\mathbf{r_3})\phi_{M=0}^{L=0}(\mathbf{r_1})\phi_{M=0}^{L=0}(\mathbf{r_2})\left|\frac{1}{2},-\frac{1}{2}\right>_{p_3}\left|\frac{1}{2},\frac{1}{2}\right>_{p_1}\left|\frac{1}{2},-\frac{1}{2}\right>_{p_2}\notag\\&-\phi_{M=-1}^{L=1}(\mathbf{r_2})\phi_{M=0}^{L=0}(\mathbf{r_1})\phi_{M=0}^{L=0}(\mathbf{r_3})\left|\frac{1}{2},-\frac{1}{2}\right>_{p_2}\left|\frac{1}{2},\frac{1}{2}\right>_{p_1}\left|\frac{1}{2},-\frac{1}{2}\right>_{p_3}\notag\\&-\phi_{M=-1}^{L=1}(\mathbf{r_3})\phi_{M=0}^{L=0}(\mathbf{r_2})\phi_{M=0}^{L=0}(\mathbf{r_1})\left|\frac{1}{2},-\frac{1}{2}\right>_{p_3}\left|\frac{1}{2},\frac{1}{2}\right>_{p_2}\left|\frac{1}{2},-\frac{1}{2}\right>_{p_1}\notag\\&+\phi_{M=-1}^{L=1}(\mathbf{r_2})\phi_{M=0}^{L=0}(\mathbf{r_3})\phi_{M=0}^{L=0}(\mathbf{r_1})\left|\frac{1}{2},-\frac{1}{2}\right>_{p_2}\left|\frac{1}{2},\frac{1}{2}\right>_{p_3}\left|\frac{1}{2},-\frac{1}{2}\right>_{p_1})\phi_{M=0}^{L=0}(\mathbf{r_4})\left|\frac{1}{2},\frac{1}{2}\right>_{n}\allowdisplaybreaks\notag\\&+\sqrt{\frac{1}{12}}(\phi_{M=0}^{L=1}(\mathbf{r_1})\phi_{M=0}^{L=0}(\mathbf{r_2})\phi_{M=0}^{L=0}(\mathbf{r_3})\left|\frac{1}{2},-\frac{1}{2}\right>_{p_1}\left|\frac{1}{2},\frac{1}{2}\right>_{p_2}\left|\frac{1}{2},-\frac{1}{2}\right>_{p_3}\notag\\&-\phi_{M=0}^{L=1}(\mathbf{r_1})\phi_{M=0}^{L=0}(\mathbf{r_2})\phi_{M=0}^{L=0}(\mathbf{r_3})\left|\frac{1}{2},-\frac{1}{2}\right>_{p_1}\left|\frac{1}{2},-\frac{1}{2}\right>_{p_2}\left|\frac{1}{2},\frac{1}{2}\right>_{p_3}\notag\\&+\phi_{M=0}^{L=1}(\mathbf{r_3})\phi_{M=0}^{L=0}(\mathbf{r_1})\phi_{M=0}^{L=0}(\mathbf{r_2})\left|\frac{1}{2},-\frac{1}{2}\right>_{p_3}\left|\frac{1}{2},\frac{1}{2}\right>_{p_1}\left|\frac{1}{2},-\frac{1}{2}\right>_{p_2}\notag\\&-\phi_{M=0}^{L=1}(\mathbf{r_2})\phi_{M=0}^{L=0}(\mathbf{r_1})\phi_{M=0}^{L=0}(\mathbf{r_3})\left|\frac{1}{2},-\frac{1}{2}\right>_{p_2}\left|\frac{1}{2},\frac{1}{2}\right>_{p_1}\left|\frac{1}{2},-\frac{1}{2}\right>_{p_3}\notag\\&-\phi_{M=0}^{L=1}(\mathbf{r_3})\phi_{M=0}^{L=0}(\mathbf{r_2})\phi_{M=0}^{L=0}(\mathbf{r_1})\left|\frac{1}{2},-\frac{1}{2}\right>_{p_3}\left|\frac{1}{2},\frac{1}{2}\right>_{p_2}\left|\frac{1}{2},-\frac{1}{2}\right>_{p_1}\notag\\&+\phi_{M=0}^{L=1}(\mathbf{r_2})\phi_{M=0}^{L=0}(\mathbf{r_3})\phi_{M=0}^{L=0}(\mathbf{r_1})\left|\frac{1}{2},-\frac{1}{2}\right>_{p_2}\left|\frac{1}{2},\frac{1}{2}\right>_{p_3}\left|\frac{1}{2},-\frac{1}{2}\right>_{p_1})\phi_{M=0}^{L=0}(\mathbf{r_4})\left|\frac{1}{2},-\frac{1}{2}\right>_{n},
\end{align}
\begin{align}
    \langle \mathbf{r_1},...,\mathbf{r_4}\left|2,-2\right>&=\sqrt{\frac{1}{3!}}\left.\left[\begin{matrix}\phi_{M=-1}^{L=1}(\mathbf{r_1})\left|\frac{1}{2},-\frac{1}{2}\right>_{p_1}&\phi_{M=-1}^{L=1}(\mathbf{r_2})\left|\frac{1}{2},-\frac{1}{2}\right>_{p_2}&\phi_{M=-1}^{L=1}(\mathbf{r_3})\left|\frac{1}{2},-\frac{1}{2}\right>_{p_3}\\\phi_{M=0}^{L=0}(\mathbf{r_1})\left|\frac{1}{2},\frac{1}{2}\right>_{p_1}&\phi_{M=0}^{L=0}(\mathbf{r_2})\left|\frac{1}{2},\frac{1}{2}\right>_{p_2}&\phi_{M=0}^{L=0}(\mathbf{r_3})\left|\frac{1}{2},\frac{1}{2}\right>_{p_3}\\\phi_{M=0}^{L=0}(\mathbf{r_1})\left|\frac{1}{2},-\frac{1}{2}\right>_{p_1}&\phi_{M=0}^{L=0}(\mathbf{r_2})\left|\frac{1}{2},-\frac{1}{2}\right>_{p_2}&\phi_{M=0}^{L=0}(\mathbf{r_3})\left|\frac{1}{2},-\frac{1}{2}\right>_{p_3}\end{matrix}\right.\right]\phi_{M=0}^{L=0}(\mathbf{r_4})\left|\frac{1}{2},-\frac{1}{2}\right>_{n}\notag\\&=\sqrt{\frac{1}{6}}(\phi_{M=-1}^{L=1}(\mathbf{r_1})\phi_{M=0}^{L=0}(\mathbf{r_2})\phi_{M=0}^{L=0}(\mathbf{r_3})\left|\frac{1}{2},-\frac{1}{2}\right>_{p_1}\left|\frac{1}{2},\frac{1}{2}\right>_{p_2}\left|\frac{1}{2},-\frac{1}{2}\right>_{p_3}\notag\\&-\phi_{M=-1}^{L=1}(\mathbf{r_1})\phi_{M=0}^{L=0}(\mathbf{r_2})\phi_{M=0}^{L=0}(\mathbf{r_3})\left|\frac{1}{2},-\frac{1}{2}\right>_{p_1}\left|\frac{1}{2},-\frac{1}{2}\right>_{p_2}\left|\frac{1}{2},\frac{1}{2}\right>_{p_3}\notag\\&+\phi_{M=-1}^{L=1}(\mathbf{r_3})\phi_{M=0}^{L=0}(\mathbf{r_1})\phi_{M=0}^{L=0}(\mathbf{r_2})\left|\frac{1}{2},-\frac{1}{2}\right>_{p_3}\left|\frac{1}{2},\frac{1}{2}\right>_{p_1}\left|\frac{1}{2},-\frac{1}{2}\right>_{p_2}\notag\\&-\phi_{M=-1}^{L=1}(\mathbf{r_2})\phi_{M=0}^{L=0}(\mathbf{r_1})\phi_{M=0}^{L=0}(\mathbf{r_3})\left|\frac{1}{2},-\frac{1}{2}\right>_{p_2}\left|\frac{1}{2},\frac{1}{2}\right>_{p_1}\left|\frac{1}{2},-\frac{1}{2}\right>_{p_3}\notag\\&-\phi_{M=-1}^{L=1}(\mathbf{r_3})\phi_{M=0}^{L=0}(\mathbf{r_2})\phi_{M=0}^{L=0}(\mathbf{r_1})\left|\frac{1}{2},-\frac{1}{2}\right>_{p_3}\left|\frac{1}{2},\frac{1}{2}\right>_{p_2}\left|\frac{1}{2},-\frac{1}{2}\right>_{p_1}\notag\\&+\phi_{M=-1}^{L=1}(\mathbf{r_2})\phi_{M=0}^{L=0}(\mathbf{r_3})\phi_{M=0}^{L=0}(\mathbf{r_1})\left|\frac{1}{2},-\frac{1}{2}\right>_{p_2}\left|\frac{1}{2},\frac{1}{2}\right>_{p_3}\left|\frac{1}{2},-\frac{1}{2}\right>_{p_1})\phi_{M=0}^{L=0}(\mathbf{r_4})\left|\frac{1}{2},-\frac{1}{2}\right>_{n}.
\end{align}

For traces of Wigner functions, they are
\begin{align}
   &\text{Tr}_s[\hat{W}(\mathbf{r_1^\prime},\mathbf{k_1^\prime},\mathbf{r_2^\prime},\mathbf{k_2^\prime},\mathbf{r_3^\prime},\mathbf{k_3^\prime};J_z=1) \times  \hat{\sigma}_{p_1p_2p_3n}]
    \notag\\&=\frac{1}{2}\frac{(1+\mathcal{P}_N)^2(1-\mathcal{P}_N)^2}{16} 8^3 \exp\left[-\frac{{r_1^\prime}^2}{{b_1^\prime}^2}-{b_1^\prime}^2 {k_1^\prime}^2-\frac{{r_2^\prime}^2}{{b_2^\prime}^2}-{b_2^\prime}^2 {k_2^\prime}^2-\frac{{r_3^\prime}^2}{{b_3^\prime}^2}-{b_3^\prime}^2 {k_3^\prime}^2\right]
    \\&\times\left[\frac{1}{2}\Big(\frac{{x_1^\prime}^2+{y_1^\prime}^2}{{b_1^\prime}^2}-1+{b_1^\prime}^2({k_{x_1}^\prime}^2+{k_{y_1}^\prime}^2)+2({k_{y_1}^\prime} {x_1^\prime}-{k_{x_1}^\prime} {y_1^\prime})\Big) \notag
    +\Big(\frac{{x_2^\prime}^2+{y_2^\prime}^2}{{b_2^\prime}^2}-1+{b_2^\prime}^2({k_{x_2}^\prime}^2+{k_{y_2}^\prime}^2)+2({k_{y_2}^\prime} {x_2^\prime}-{k_{x_2}^\prime} {y_2^\prime})\Big)\right]
    \notag\\&+\frac{1}{2}\frac{(1+\mathcal{P}_N)^3(1-\mathcal{P}_N)}{16}8^3 \exp\left[-\frac{{r_1^\prime}^2}{{b_1^\prime}^2}-{b_1^\prime}^2 {k_1^\prime}^2-\frac{{r_2^\prime}^2}{{b_2^\prime}^2}-{b_2^\prime}^2 {k_2^\prime}^2-\frac{{r_3^\prime}^2}{{b_3^\prime}^2}-{b_3^\prime}^2 {k_3^\prime}^2\right]\notag
    \\&\times\left[\frac{1}{2}\Big(\frac{2{z_1^\prime}^2}{{b_1^\prime}^2}-1+ 2{b_1^\prime}^2{k_{z_1}^\prime}^2\Big) \notag
    +\Big(\frac{2{z_2^\prime}^2}{{b_2^\prime}^2}-1+2{b_2^\prime}^2{k_{z_2}^\prime}^2\Big)\right],
\end{align}
\begin{align}
    &\text{Tr}_s[\hat{W}(\mathbf{r_1^\prime},\mathbf{k_1^\prime},\mathbf{r_2^\prime},\mathbf{k_2^\prime},\mathbf{r_3^\prime},\mathbf{k_3^\prime};J_z=0) \times  \hat{\sigma}_{p_1p_2p_3n}]
    \notag\\&=\frac{1}{6}\frac{(1+\mathcal{P}_N)(1-\mathcal{P}_N)^3}{16}8^3 \exp\left[-\frac{{r_1^\prime}^2}{{b_1^\prime}^2}-{b_1^\prime}^2 {k_1^\prime}^2-\frac{{r_2^\prime}^2}{{b_2^\prime}^2}-{b_2^\prime}^2 {k_2^\prime}^2-\frac{{r_3^\prime}^2}{{b_3^\prime}^2}-{b_3^\prime}^2 {k_3^\prime}^2\right]
    \\&\times\left[\frac{1}{2}\left(\frac{{x_1^\prime}^2+{y_1^\prime}^2}{{b_1^\prime}^2}-1+{b_1^\prime}^2({k_{x_1}^\prime}^2+{k_{y_1}^\prime}^2)+2({k_{y_1}^\prime} {x_1^\prime}-{k_{x_1}^\prime} {y_1^\prime})\right) \notag
    +\left(\frac{{x_2^\prime}^2+{y_2^\prime}^2}{{b_2^\prime}^2}-1+{b_2^\prime}^2({k_{x_2}^\prime}^2+{k_{y_2}^\prime}^2)+2({k_{y_2}^\prime} {x_2^\prime}-{k_{x_2}^\prime} {y_2^\prime})\right)\right]
    \notag\\&+\frac{2}{3}\frac{(1+\mathcal{P}_N)^2(1-\mathcal{P}_N)^2}{16}8^3 \exp\left[-\frac{{r_1^\prime}^2}{{b_1^\prime}^2}-{b_1^\prime}^2 {k_1^\prime}^2-\frac{{r_2^\prime}^2}{{b_2^\prime}^2}-{b_2^\prime}^2 {k_2^\prime}^2-\frac{{r_3^\prime}^2}{{b_3^\prime}^2}-{b_3^\prime}^2 {k_3^\prime}^2\right]\notag
    \\&\times\left[\frac{1}{2}\left(\frac{2{z_1^\prime}^2}{{b_1^\prime}^2}-1+ 2{b_1^\prime}^2{k_{z_1}^\prime}^2\right) \notag
    +\left(\frac{2{z_2^\prime}^2}{{b_2^\prime}^2}-1+2{b_2^\prime}^2{k_{z_2}^\prime}^2\right)\right]
    \notag\\&+\frac{1}{6}\frac{(1+\mathcal{P}_N)^3(1-\mathcal{P}_N)}{16} 8^3 \exp\left[-\frac{{r_1^\prime}^2}{{b_1^\prime}^2}-{b_1^\prime}^2 {k_1^\prime}^2-\frac{{r_2^\prime}^2}{{b_2^\prime}^2}-{b_2^\prime}^2 {k_2^\prime}^2-\frac{{r_3^\prime}^2}{{b_3^\prime}^2}-{b_3^\prime}^2 {k_3^\prime}^2\right]\notag
    \\&\times\left[\frac{1}{2}\left(\frac{{x_1^\prime}^2+{y_1^\prime}^2}{{b_1^\prime}^2}-1+{b_1^\prime}^2({k_{x_1}^\prime}^2+{k_{y_1}^\prime}^2)-2({k_{y_1}^\prime} {x_1^\prime}-{k_{x_1}^\prime} {y_1^\prime})\right) \notag
    +\left(\frac{{x_2^\prime}^2+{y_2^\prime}^2}{{b_2^\prime}^2}-1+{b_2^\prime}^2({k_{x_2}^\prime}^2+{k_{y_2}^\prime}^2)-2({k_{y_2}^\prime} {x_2^\prime}-{k_{x_2}^\prime} {y_2^\prime})\right)\right],
\end{align}
\begin{align}
    &\text{Tr}_s[\hat{W}(\mathbf{r_1^\prime},\mathbf{k_1^\prime},\mathbf{r_2^\prime},\mathbf{k_2^\prime},\mathbf{r_3^\prime},\mathbf{k_3^\prime};J_z=-1) \times  \hat{\sigma}_{p_1p_2p_3n}]
    \notag\\&=\frac{1}{2}\frac{(1+\mathcal{P}_N)^2(1-\mathcal{P}_N)^2}{16}8^3 \exp\left[-\frac{{r_1^\prime}^2}{{b_1^\prime}^2}-{b_1^\prime}^2 {k_1^\prime}^2-\frac{{r_2^\prime}^2}{{b_2^\prime}^2}-{b_2^\prime}^2 {k_2^\prime}^2-\frac{{r_3^\prime}^2}{{b_3^\prime}^2}-{b_3^\prime}^2 {k_3^\prime}^2\right]\notag
    \\&\times\left[\frac{1}{2}\left(\frac{{x_1^\prime}^2+{y_1^\prime}^2}{{b_1^\prime}^2}-1+{b_1^\prime}^2({k_{x_1}^\prime}^2+{k_{y_1}^\prime}^2)-2({k_{y_1}^\prime} {x_1^\prime}-{k_{x_1}^\prime} {y_1^\prime})\right) \notag
    +\left(\frac{{x_2^\prime}^2+{y_2^\prime}^2}{{b_2^\prime}^2}-1+{b_2^\prime}^2({k_{x_2}^\prime}^2+{k_{y_2}^\prime}^2)-2({k_{y_2}^\prime} {x_2^\prime}-{k_{x_2}^\prime} {y_2^\prime})\right)\right]
    \notag\\&+\frac{1}{2}\frac{(1+\mathcal{P}_N)(1-\mathcal{P}_N)^3}{16}8^3 \exp\left[-\frac{{r_1^\prime}^2}{{b_1^\prime}^2}-{b_1^\prime}^2 {k_1^\prime}^2-\frac{{r_2^\prime}^2}{{b_2^\prime}^2}-{b_2^\prime}^2 {k_2^\prime}^2-\frac{{r_3^\prime}^2}{{b_3^\prime}^2}-{b_3^\prime}^2 {k_3^\prime}^2\right]\notag
    \\&\times\left[\frac{1}{2}\left(\frac{2{z_1^\prime}^2}{{b_1^\prime}^2}-1+ 2{b_1^\prime}^2{k_{z_1}^\prime}^2\right) \notag
    +\left(\frac{2{z_2^\prime}^2}{{b_2^\prime}^2}-1+2{b_2^\prime}^2{k_{z_2}^\prime}^2\right)\right],
\end{align}
\begin{align}
    &\text{Tr}_s[\hat{W}(\mathbf{r_1^\prime},\mathbf{k_1^\prime},\mathbf{r_2^\prime},\mathbf{k_2^\prime},\mathbf{r_3^\prime},\mathbf{k_3^\prime};J_z=-2) \times  \hat{\sigma}_{p_1p_2p_3n}]
    \notag\\&=\frac{(1+\mathcal{P}_N)(1-\mathcal{P}_N)^3}{16}8^3 \exp\left[-\frac{{r_1^\prime}^2}{{b_1^\prime}^2}-{b_1^\prime}^2 {k_1^\prime}^2-\frac{{r_2^\prime}^2}{{b_2^\prime}^2}-{b_2^\prime}^2 {k_2^\prime}^2-\frac{{r_3^\prime}^2}{{b_3^\prime}^2}-{b_3^\prime}^2 {k_3^\prime}^2\right]\notag
    \\&\times\left[\frac{1}{2}\left(\frac{{x_1^\prime}^2+{y_1^\prime}^2}{{b_1^\prime}^2}-1+{b_1^\prime}^2({k_{x_1}^\prime}^2+{k_{y_1}^\prime}^2)-2({k_{y_1}^\prime} {x_1^\prime}-{k_{x_1}^\prime} {y_1^\prime})\right) \notag
    +\left(\frac{{x_2^\prime}^2+{y_2^\prime}^2}{{b_2^\prime}^2}-1+{b_2^\prime}^2({k_{x_2}^\prime}^2+{k_{y_2}^\prime}^2)-2({k_{y_2}^\prime} {x_2^\prime}-{k_{x_2}^\prime} {y_2^\prime})\right)\right].
\end{align}

The numbers of $^4$Li in different $J_z$ states are obtained as
\begin{align}
    N_{^4{Li},J_z=1}\approx 8 \frac{N_p^3 N_n}{V^3}\left(\frac{2\pi}{mT}\right)^{\frac{9}{2}}\left( \frac{1}{2}\frac{(1+\mathcal{P}_N)^2(1-\mathcal{P}_N)^2}{16}\left(1+2\mathcal{P}_L+2\mathcal{P}_L^2\right)+\frac{1}{2}\frac{(1+\mathcal{P}_N)^3(1-\mathcal{P}_N)}{16}\right),
\end{align}
\begin{align}
    N_{^4{Li},J_z=0}&\approx 8 \frac{N_p^3 N_n}{V^3}\left(\frac{2\pi}{mT}\right)^{\frac{9}{2}} \Big(\frac{1}{6}\frac{(1+\mathcal{P}_N)^3(1-\mathcal{P}_N)}{16}\left(1-2\mathcal{P}_L+2\mathcal{P}_L^2\right)\notag
    \\&\quad+\frac{2}{3}\frac{(1+\mathcal{P}_N)^2(1-\mathcal{P}_N)^2}{16}+\frac{1}{6}\frac{(1+\mathcal{P}_N)(1-\mathcal{P}_N)^3}{16}\left(1+2\mathcal{P}_L+2\mathcal{P}_L^2\right)\Big),
\end{align}
\begin{align}
    N_{^4{Li},J_z=-1}\approx 8 \frac{N_p^3 N_n}{V^3}\left(\frac{2\pi}{mT}\right)^{\frac{9}{2}} \Big(\frac{1}{2}\frac{(1+\mathcal{P}_N)^2(1-\mathcal{P}_N)^2}{16}\left(1-2\mathcal{P}_L+2\mathcal{P}_L^2\right)+\frac{1}{2}\frac{(1+\mathcal{P}_N)(1-\mathcal{P}_N)^3}{16}\Big),
\end{align}
\begin{align}
    N_{^4{Li},J_z=-2}\approx 8 \frac{N_p^3 N_n}{V^3}\left(\frac{2\pi}{mT}\right)^{\frac{9}{2}} \frac{(1+\mathcal{P}_N)(1-\mathcal{P}_N)^3}{16}\left(1-2\mathcal{P}_L+2\mathcal{P}_L^2\right).
\end{align}

The spin density matrix of $^4\text{Li}({2^-})$ is then given by
\begin{align}
    \hat{\rho}\big{(}^4\text{Li}(2^-)\big{)}=\text{diag}[&\hat{\rho}_{2,2},\hat{\rho}_{1,1},\hat{\rho}_{0,0},\hat{\rho}_{-1,-1},\hat{\rho}_{-2,-2}]
    \notag\\=\text{diag}\Bigg[&\frac{3 (\mathcal{P}_N+1)^2 (\mathcal{P}_L+1)^2}{\left(3 \mathcal{P}_L^2+5\right) \mathcal{P}_N^2+20 \mathcal{P}_L \mathcal{P}_N+5 \left(\mathcal{P}_L^2+3\right)},\frac{-3 (\mathcal{P}_N+1) (\mathcal{P}_L+1) (\mathcal{P}_N \mathcal{P}_L-1)}{\left(3 \mathcal{P}_L^2+5\right) \mathcal{P}_N^2+20 \mathcal{P}_L \mathcal{P}_N+5 \left(\mathcal{P}_L^2+3\right)},\notag\\
    &\frac{\left(3 \mathcal{P}_L^2-1\right) \mathcal{P}_N^2-4 \mathcal{P}_L \mathcal{P}_N-\mathcal{P}_L^2+3}{\left(3 \mathcal{P}_L^2+5\right) \mathcal{P}_N^2+20 \mathcal{P}_L \mathcal{P}_N+5 \left(\mathcal{P}_L^2+3\right)}, \frac{-3 (\mathcal{P}_N-1) (\mathcal{P}_L-1) (\mathcal{P}_N \mathcal{P}_L-1)}{\left(3 \mathcal{P}_L^2+5\right) \mathcal{P}_N^2+20 \mathcal{P}_L \mathcal{P}_N+5 \left(\mathcal{P}_L^2+3\right)}, \notag\\
  & \frac{3 (\mathcal{P}_N-1)^2 (\mathcal{P}_L-1)^2}{\left(3 \mathcal{P}_L^2+5\right) \mathcal{P}_N^2+20 \mathcal{P}_L \mathcal{P}_N+5 \left(\mathcal{P}_L^2+3\right)}\Bigg],
\end{align}
where the diagonal element $\hat{\rho}_{i,i}=\frac{N_{^4{Li},J_z=i}}{\sum_{m=-2}^{m=2} N_{^4{Li},J_z=m}}$ denotes the relative intensities of the spin component along the direction of the vortex field to take the value $i$.

\section{Appendix B: two-body coalescence in a vortical fluid up to $\hbar^2$}
\renewcommand{\theequation}{B.\arabic{equation}}
\setcounter{equation}{0}

For simplicity, we consider a cluster with an internal orbital angular momentum $L=1$ from an emission source of particles with spin zero, its number in the coalescence model is obtained from the overlap of the particle phase-space distribution functions $f(\mathbf{r,k})$ with the Wigner function $W(\mathbf{r,k})$ of the cluster internal wave function, i.e. \cite{Sun:2018jhg}
\begin{align}
    \label{eq:n-p coalescence2}
    N=&\int d^3\mathbf{r_1} d^3\mathbf{k_1}d^3\mathbf{r_2} d^3\mathbf{k_2} f(\mathbf{r_1},\mathbf{k_1})f(\mathbf{r_2},\mathbf{k_2})W\left(\mathbf{r_1}-\mathbf{r_2},\frac{\mathbf{k_1}-\mathbf{k_2}}{2}\right).
\end{align}
For particles emitted from an isotropic and thermalized fireball of an effective temperature $T$ (after taking into account the flow effect) and volume $V$ and uniformly distributed in space, their non-relativistic distribution functions in a vortical field $\omega$ are  given by
\begin{align}
    f(\mathbf{r}_i,\mathbf{k}_i)=\frac{\xi'_i}{(2\pi)^3} e^{-\frac{(\mathbf{k}_i-m \mathbf{\omega}\times \mathbf{r}_i)^2}{2mT}},
\end{align}
where $\xi'_i=\xi_i e^{-m/T}$ and $\omega=\frac{1}{2}\mathbf{\nabla} \times \mathbf{u}$ denote the effective fugacity and vorticity of the fluid, respectively. In obtaining the above distribution function, non-relativistic and high-temperature approximation are assumed. The distribution functions are normalized to particle numbers according to
\begin{align}
    N_i=&\int d^3\mathbf{r}_i d^3\mathbf{k}_i f(\mathbf{r}_i,\mathbf{k}_i)=\xi'_i V {\left(\frac{mT}{2\pi}\right)^\frac{3}{2}}.  
\end{align}

Using the harmonic oscillator wave functions for the internal wave function of a particle pair and with the coordinate and momentum transformations, 
\begin{align*}
    \mathbf{R}={\frac{\mathbf{r}_{1}+\mathbf{r}_{2}}{2}},\quad \mathbf{r}=\mathbf{r}_{1}-\mathbf{r}_{2},\quad \mathbf{K}=\mathbf{k}_{1}+\mathbf{k}_{2},\quad \mathbf{k}={\frac{\mathbf{k}_{1}-\mathbf{k}_{2}}{2}},
\end{align*}
the Wigner function for states of different $(L,M)$  are given by  Eqs. (\ref{wigner function 000}) - (\ref{wigner function 11-1}). The yield for state $(L,M)=(1,1)$  in Eq.~(\ref{eq:n-p coalescence2}) with the assumption $\mathbf{\omega}=\omega_z\hat{z}$ is given by
\begin{align} 
    \label{111}
    N_{|1,1\rangle}=&\quad  \frac{\xi'^2 }{(2\pi)^6} \int d^3\mathbf{R} d^3\mathbf{K} d^3\mathbf{r} d^3\mathbf{k}
   \notag\\&\times\exp_\star \left(-\frac{1}{2 m T}\left[\frac{1}{2}(4k^2+K^2)-2m \omega_z(K_yX-K_xY+k_y x-k_x y)+\frac{1}{2}m^2 \omega_z^2(4X^2+4Y^2+x^2+y^2)\right]\right)
    \notag\\&\times\exp\left(-\frac{r^2}{b^2}-b^2 k^2 \right)
    \notag\\&\times8\left[\frac{x^2+y^2}{b^2}-1+b^2(k_x^2+k_y^2)+2(k_y x-k_x y)\right],
\end{align}
where $\exp_\star(...)$ denotes the Moyal star product. If only counting the quantum effect caused by the Moyal star product for the relative coordinates and momentum, the Moyal star product term can be characterized as 
\begin{align}
    &\exp_\star \left(-\frac{1}{2 m T}\left[\frac{1}{2}(4k^2+K^2)-2m \omega_z(K_yX-K_xY+k_y x-k_x y)+\frac{1}{2}m^2 \omega_z^2(4X^2+4Y^2+x^2+y^2)\right]\right)\notag
    \\&\approx\exp \left(-\frac{1}{2 m T}\left[\frac{1}{2}(4k^2+K^2)-2m \omega_z(K_yX-K_xY)+\frac{1}{2}m^2 \omega_z^2(4X^2+4Y^2)\right]\right)\exp \left(\frac{m \omega_z^2( x^2+y^2)}{4 T}\right)\exp_\star \left(\frac{\omega_z(k_y x-k_x y)}{T}\right)\notag
    \\&\approx\exp \left(-\frac{1}{2 m T}\left[\frac{1}{2}(4k^2+K^2)-2m \omega_z(K_yX-K_xY)+\frac{1}{2}m^2 \omega_z^2(4X^2+4Y^2)\right]\right)\exp \left(\frac{m \omega_z^2( x^2+y^2)}{4 T}\right)\notag
    \\&\quad\times\left(1+\frac{\omega_z(k_y x-k_x y)}{T}+\frac{\omega_z^2(k_y x-k_x y)^2}{2T}-\frac{\omega_z^2}{4 T^2}\right)\notag
    \\&\approx\exp \left(-\frac{1}{2 m T}\left[\frac{1}{2}(4k^2+K^2)-2m \omega_z(K_yX-K_xY)+\frac{1}{2}m^2 \omega_z^2(4X^2+4Y^2)\right]\right) \notag
    \\&\quad\times\left(1+\frac{\omega_z(k_y x-k_x y)}{T}+\frac{\omega_z^2(k_y x-k_x y)^2}{2T}-\frac{\omega_z^2}{4 T^2}\right).
\end{align}
In  the above,  the term $\exp \left(\frac{m \omega_z^2( x^2+y^2)}{4 T}\right)$ corresponds to the kinetic energy of relative rotational motion. We can rewrite the formula as $e^ \frac{L^2}{2 I T}$, where $L$ denotes the angular momentum of the system and $I=\frac{1}{2}m (x^2+y^2)$ denotes the moment of inertia. This term is approximately equal to 1 for $\frac{\omega}{T} \ll 1$. Carrying out the integrations in  Eq.~(\ref{111}) then gives
\begin{align} 
    N_ {|1,1\rangle}\approx  \frac{4}{\sqrt{2}}\frac{N_1 N_2}{V}\left(\frac{2\pi}{mT}\right)^\frac{3}{2}\left(1+\frac{\omega_z}{T}+\frac{\omega_z^2}{2 T^2}\right).
\end{align}
Letting $\mathcal{P}_L=\frac{\omega}{2T}$ be the polarization due to the orbital motion leads to 
\begin{align} 
    N_ {|1,1\rangle}\approx  \frac{4}{\sqrt{2}}\frac{N_1 N_2}{V}\left(\frac{2\pi}{mT}\right)^\frac{3}{2}\left(1+2\mathcal{P}_L+2\mathcal{P}_L^2\right).
\end{align}

Similarly, we have the following numbers for the cases of $(L,M)=(1,0)$ and $(L,M)=(1,-1)$, 
\begin{align} 
    N_{|1,0\rangle}\approx  \frac{4}{\sqrt{2}}\frac{N_1 N_2}{V}\left(\frac{2\pi}{mT}\right)^\frac{3}{2},
\end{align}
\begin{align} 
   N_{|1,-1\rangle}\approx  \frac{4}{\sqrt{2}}\frac{N_1 N_2}{V}\left(\frac{2\pi}{mT}\right)^\frac{3}{2}\left(1-2\mathcal{P}_L+2\mathcal{P}_L^2\right).
\end{align}
The spin density matrix for a particle pair is then given by 
\begin{align}
\hat{\rho}&=\left.\left[\begin{matrix}\hat{\rho}_{1,1}&0&0\\0&\hat{\rho}_{0,0}&0\\0&0&\hat{\rho}_{-1,-1}\end{matrix}
\right.\right]=
    \left.\left[\begin{matrix}\frac{1+2\mathcal{P}_L+2\mathcal{P}_L^2}{3+4\mathcal{P}_L^2}&0&0\\0&\frac{1}{3+4\mathcal{P}_L^2}&0\\0&0&\frac{1-2\mathcal{P}_L+2\mathcal{P}_L^2}{3+4\mathcal{P}_L^2}\end{matrix}\right.\right], 
\end{align}
where the diagonal element $\hat{\rho}_{i,i}$ denotes the relative intensities of the orbital angular momentum component along the direction of the  vortical field to have the value $i$. 

\section{Appendix C: Properties of Moyal star product}
\renewcommand{\theequation}{C.\arabic{equation}}
\setcounter{equation}{0}

For a pure state $\psi(x)$, its Wigner density in phase space is defined as~\cite{Curtright:2011vw}
\begin{align}
    W(x, p)=\int d y \langle x+\frac{y}{2}|\psi\rangle \langle\psi |x-\frac{y}{2}  \rangle e^{-i p y/\hbar } =\int d y \psi^*\left(x-\frac{y}{2}\right) e^{-i p y/\hbar } \psi\left(x+\frac{y}{2}\right),
\end{align} 
which satisfies 
\begin{align}
    \int \frac{d p}{2 \pi \hbar} W(x, p)&=|\psi(x)|^2, \quad \int d x W(x, p) = |\phi(p)|^2,\\
    \int \frac{dx d p }{2 \pi \hbar} W(x, p)&=\int d x |\psi(x)|^2= \int \frac{d p}{2\pi \hbar} |\phi(p)|^2=1.
\end{align}

For clarity, we retain $\hbar$ in the equations in this section. The dynamical evolution of the Wigner function is specified by the Moyal equation
\begin{align}
    \frac{\partial W(x, p)}{\partial t}=\frac{H \star W(x, p)-W(x, p)\star H}{i \hbar},
\end{align}
where $H$ is the Hamiltonian and the Moyal $\star$-product is defined as 
\begin{align}
    \star\equiv e^{\frac{i \hbar}{2}(\overleftarrow{\partial_x}\overrightarrow{\partial_p}-\overleftarrow{\partial_p}\overrightarrow{\partial_x})}.
\end{align}
The $\star$-product is the Weyl correspondence of the Hilbert space operator product
\begin{align}
    A(x, p)\star B(x, p)&=A(x, p) \exp{\frac{i \hbar}{2}(\overleftarrow{\partial_x}\overrightarrow{\partial_p}-\overleftarrow{\partial_p}\overrightarrow{\partial_x})} B(x, p)  \notag
    \\&=\int \frac{d x_1 d p_1}{2 \pi \hbar} \frac{d x_2 d p_2}{2 \pi \hbar} A(x+\frac{x_1}{2}, p+p_1) B(x+\frac{x_2}{2}, p+p_2) e^{i(x_1 p_2 -x_2 p_1)/\hbar} \notag
    \\&=\int \frac{d x_1 d p_1}{2 \pi \hbar} \frac{d x_2 d p_2}{2 \pi \hbar}     \int d y_1 \langle x+\frac{x_1}{2}+\frac{y_1}{2}|\hat{A} |x+\frac{x_1}{2}-\frac{y_1}{2}  \rangle e^{-i (p+p_1) y_1 /\hbar} \notag
    \\& \quad \times \int d y_2 \langle x+\frac{x_2}{2}+\frac{y_2}{2}|\hat{B} |x+\frac{x_2}{2}-\frac{y_2}{2}  \rangle e^{-i (p+p_2) y_2 /\hbar}  e^{i(x_1 p_2 -x_2 p_1)/\hbar} \notag
    \\&=\int \frac{d x_1 d p_1}{2 \pi \hbar} \frac{d x_2 d p_2}{2 \pi \hbar}     \int d y_1 \langle x+\frac{x_1}{2}+\frac{y_1}{2}|\hat{A} |x+\frac{x_1}{2}-\frac{y_1}{2}  \rangle  \notag
    \\& \quad \times \int d y_2 \langle x+\frac{x_2}{2}+\frac{y_2}{2}|\hat{B} |x+\frac{x_2}{2}-\frac{y_2}{2}  \rangle e^{-i p_1(y_1+x_2)/\hbar} e^{-i p_2(y_2-x_1)/\hbar} e^{-ip(y_1+y_2)/\hbar} \notag
    \\&=\int d x_1 d x_2   \int d y_1 \langle x+\frac{x_1}{2}+\frac{y_1}{2}|\hat{A} |x+\frac{x_1}{2}-\frac{y_1}{2}  \rangle  \notag
    \\& \quad \times \int d y_2 \langle x+\frac{x_2}{2}+\frac{y_2}{2}|\hat{B} |x+\frac{x_2}{2}-\frac{y_2}{2}  \rangle \delta (y_1+x_2) \delta(y_2-x_1) e^{-ip(y_1+y_2)/\hbar} \notag
    \\&=\int d x_1 d x_2     \langle x+\frac{x_1-x_2}{2}|\hat{A} |x+\frac{x_1+x_2}{2}  \rangle  \notag
    \\& \quad \times  \langle x+\frac{x_1+x_2}{2}|\hat{B} |x+\frac{x_2-x_1}{2} \rangle  e^{-ip(x_1-x_2)/\hbar} \notag 
    \\&=\int d \rho d\sigma \langle x+\frac{\rho }{2}|\hat{A} |x+\sigma  \rangle  \notag
        \langle x+\sigma|\hat{B} |x-\frac{\rho }{2} \rangle  e^{-ip \rho/\hbar } \notag
    \\&=\int d \rho  \langle x+\frac{\rho }{2}|\hat{A} \hat{B} |x-\frac{\rho }{2} \rangle  e^{-ip \rho/\hbar } ,
\end{align}
where 
\begin{align}
    A(x, p)=\int d y_1 \langle x+\frac{y_1}{2}|\hat{A} |x-\frac{y_1}{2}  \rangle e^{-i p y_1 /\hbar} ,
\end{align}
\begin{align}
    B(x, p)=\int d y_2 \langle x+\frac{y_2}{2}|\hat{B} |x-\frac{y_2}{2}  \rangle e^{-i p y_2/\hbar } ,
\end{align}
and
\begin{align}
    \sigma=\frac{x_1+x_2}{2},\quad \rho=x_1-x_2.
\end{align}

With the association rule mapping phase-space functions $A(x, p)$ to operators $\hat{A}$  given by
\begin{align}
    \hat{A}(\hat{x},\hat{p})=\frac{1}{(2\pi)^2}\int d\alpha \hspace{0.05em}  d\beta \hspace{0.05em} dx \hspace{0.05em} dp  \hspace{0.1em} \ A(x,p) e^{i\alpha(\hat{x}-x)+i\beta(\hat{p}-p)},
\end{align}
the expectation value of operator $\hat{A}$ is then given by
\begin{align}
    \langle \hat{A}\rangle=\int \frac{dx \hspace{0.05em} dp}{2 \pi \hbar} \ A(x,p) \star W(x,p)=\int \frac{dx \hspace{0.05em} dp}{2 \pi \hbar} \ A(x,p)  W(x,p)=\int \frac{dx \hspace{0.05em} dp}{2 \pi \hbar} \ W(x,p) \star A(x,p).
\end{align}
This phase-space trace reduces to an ordinary product when only one $\star$ is involved. 

Finally, we consider the Moyal $\star$-product of the $z$-component of angular momentum  $\hat{L_z}$,
\begin{align}
    L_z\star L_z&=(x p_y-y p_x)\star (x p_y-y p_x)\notag
    \\&=\left[(x+\frac{i \hbar}{2}\overrightarrow{\frac{\partial}{\partial p_x}})(p_y-\frac{i \hbar}{2}\overrightarrow{\frac{\partial}{\partial y}})-(y+\frac{i \hbar}{2}\overrightarrow{\frac{\partial}{\partial p_y}})(p_x-\frac{i \hbar}{2}\overrightarrow{\frac{\partial}{\partial x}})\right](x p_y-y p_x)\notag
    \\&=(x p_y-y p_x)^2-\frac{\hbar^2}{2} \notag
    \\&= L_z^2-\frac{\hbar^2}{2}.
\end{align}
One clearly sees that  the quantum correction is at $\mathcal{O}(\hbar^2)$. The expectation value for $L_z^2$ in state $(L,M)=(1,1)$ is given by
\begin{align}
    \langle L_z^2 \rangle&= \int \frac{d^3r \hspace{0.05em} d^3p}{(2 \pi \hbar)^3} \ (L_z\star L_z) \star W_{NLM=111}(r,p) \notag
    \\&= \int \frac{d^3r \hspace{0.05em} d^3p}{(2 \pi \hbar)^3} \ \left[(x p_y-y p_x)^2-\frac{\hbar^2}{2}\right] \left[8(\frac{x^2+y^2}{b^2}-1+b^2(p_x^2+p_y^2)/\hbar^2+2(p_y x-p_x y)/\hbar)e^{-\frac{r^2}{b^2}-b^2p^2/\hbar^2}\right] \notag
    \\&=\hbar^2,
\end{align}
which  corresponds to the correct expectation value $ \langle1,1| \hat{L_z^2} |1,1\rangle =\hbar^2$.

\end{widetext}
%\bibliography{ref}

\begin{thebibliography}{68}%
\makeatletter
\providecommand \@ifxundefined [1]{%
 \@ifx{#1\undefined}
}%
\providecommand \@ifnum [1]{%
 \ifnum #1\expandafter \@firstoftwo
 \else \expandafter \@secondoftwo
 \fi
}%
\providecommand \@ifx [1]{%
 \ifx #1\expandafter \@firstoftwo
 \else \expandafter \@secondoftwo
 \fi
}%
\providecommand \natexlab [1]{#1}%
\providecommand \enquote  [1]{``#1''}%
\providecommand \bibnamefont  [1]{#1}%
\providecommand \bibfnamefont [1]{#1}%
\providecommand \citenamefont [1]{#1}%
\providecommand \href@noop [0]{\@secondoftwo}%
\providecommand \href [0]{\begingroup \@sanitize@url \@href}%
\providecommand \@href[1]{\@@startlink{#1}\@@href}%
\providecommand \@@href[1]{\endgroup#1\@@endlink}%
\providecommand \@sanitize@url [0]{\catcode `\\12\catcode `\$12\catcode
  `\&12\catcode `\#12\catcode `\^12\catcode `\_12\catcode `\%12\relax}%
\providecommand \@@startlink[1]{}%
\providecommand \@@endlink[0]{}%
\providecommand \url  [0]{\begingroup\@sanitize@url \@url }%
\providecommand \@url [1]{\endgroup\@href {#1}{\urlprefix }}%
\providecommand \urlprefix  [0]{URL }%
\providecommand \Eprint [0]{\href }%
\providecommand \doibase [0]{http://dx.doi.org/}%
\providecommand \selectlanguage [0]{\@gobble}%
\providecommand \bibinfo  [0]{\@secondoftwo}%
\providecommand \bibfield  [0]{\@secondoftwo}%
\providecommand \translation [1]{[#1]}%
\providecommand \BibitemOpen [0]{}%
\providecommand \bibitemStop [0]{}%
\providecommand \bibitemNoStop [0]{.\EOS\space}%
\providecommand \EOS [0]{\spacefactor3000\relax}%
\providecommand \BibitemShut  [1]{\csname bibitem#1\endcsname}%
\let\auto@bib@innerbib\@empty
%</preamble>
\bibitem [{\citenamefont {Liang}\ and\ \citenamefont
  {Wang}(2005{\natexlab{a}})}]{Liang:2004ph}%
  \BibitemOpen
  \bibfield  {author} {\bibinfo {author} {\bibfnamefont {Z.-T.}\ \bibnamefont
  {Liang}}\ and\ \bibinfo {author} {\bibfnamefont {X.-N.}\ \bibnamefont
  {Wang}},\ }\href {\doibase 10.1103/PhysRevLett.94.102301} {\bibfield
  {journal} {\bibinfo  {journal} {Phys. Rev. Lett.}\ }\textbf {\bibinfo
  {volume} {94}},\ \bibinfo {pages} {102301} (\bibinfo {year}
  {2005}{\natexlab{a}})},\ \bibinfo {note} {[Erratum: Phys.Rev.Lett. 96, 039901
  (2006)]}\BibitemShut {NoStop}%
\bibitem [{\citenamefont {Liang}\ and\ \citenamefont
  {Wang}(2005{\natexlab{b}})}]{Liang:2004xn}%
  \BibitemOpen
  \bibfield  {author} {\bibinfo {author} {\bibfnamefont {Z.-T.}\ \bibnamefont
  {Liang}}\ and\ \bibinfo {author} {\bibfnamefont {X.-N.}\ \bibnamefont
  {Wang}},\ }\href {\doibase 10.1016/j.physletb.2005.09.060} {\bibfield
  {journal} {\bibinfo  {journal} {Phys. Lett. B}\ }\textbf {\bibinfo {volume}
  {629}},\ \bibinfo {pages} {20} (\bibinfo {year}
  {2005}{\natexlab{b}})}\BibitemShut {NoStop}%
\bibitem [{\citenamefont {Voloshin}(2004)}]{Voloshin:2004ha}%
  \BibitemOpen
  \bibfield  {author} {\bibinfo {author} {\bibfnamefont {S.~A.}\ \bibnamefont
  {Voloshin}},\ }\href@noop {} {\  (\bibinfo {year} {2004})},\ \Eprint
  {http://arxiv.org/abs/nucl-th/0410089} {arXiv:nucl-th/0410089} \BibitemShut
  {NoStop}%
\bibitem [{\citenamefont {Betz}\ \emph {et~al.}(2007)\citenamefont {Betz},
  \citenamefont {Gyulassy},\ and\ \citenamefont {Torrieri}}]{Betz:2007kg}%
  \BibitemOpen
  \bibfield  {author} {\bibinfo {author} {\bibfnamefont {B.}~\bibnamefont
  {Betz}}, \bibinfo {author} {\bibfnamefont {M.}~\bibnamefont {Gyulassy}}, \
  and\ \bibinfo {author} {\bibfnamefont {G.}~\bibnamefont {Torrieri}},\ }\href
  {\doibase 10.1103/PhysRevC.76.044901} {\bibfield  {journal} {\bibinfo
  {journal} {Phys. Rev. C}\ }\textbf {\bibinfo {volume} {76}},\ \bibinfo
  {pages} {044901} (\bibinfo {year} {2007})}\BibitemShut {NoStop}%
\bibitem [{\citenamefont {Becattini}\ \emph {et~al.}(2008)\citenamefont
  {Becattini}, \citenamefont {Piccinini},\ and\ \citenamefont
  {Rizzo}}]{Becattini:2007sr}%
  \BibitemOpen
  \bibfield  {author} {\bibinfo {author} {\bibfnamefont {F.}~\bibnamefont
  {Becattini}}, \bibinfo {author} {\bibfnamefont {F.}~\bibnamefont
  {Piccinini}}, \ and\ \bibinfo {author} {\bibfnamefont {J.}~\bibnamefont
  {Rizzo}},\ }\href {\doibase 10.1103/PhysRevC.77.024906} {\bibfield  {journal}
  {\bibinfo  {journal} {Phys. Rev. C}\ }\textbf {\bibinfo {volume} {77}},\
  \bibinfo {pages} {024906} (\bibinfo {year} {2008})}\BibitemShut {NoStop}%
\bibitem [{\citenamefont {Gao}\ \emph {et~al.}(2008)\citenamefont {Gao},
  \citenamefont {Chen}, \citenamefont {Deng}, \citenamefont {Liang},
  \citenamefont {Wang},\ and\ \citenamefont {Wang}}]{Gao:2007bc}%
  \BibitemOpen
  \bibfield  {author} {\bibinfo {author} {\bibfnamefont {J.-H.}\ \bibnamefont
  {Gao}}, \bibinfo {author} {\bibfnamefont {S.-W.}\ \bibnamefont {Chen}},
  \bibinfo {author} {\bibfnamefont {W.-t.}\ \bibnamefont {Deng}}, \bibinfo
  {author} {\bibfnamefont {Z.-T.}\ \bibnamefont {Liang}}, \bibinfo {author}
  {\bibfnamefont {Q.}~\bibnamefont {Wang}}, \ and\ \bibinfo {author}
  {\bibfnamefont {X.-N.}\ \bibnamefont {Wang}},\ }\href {\doibase
  10.1103/PhysRevC.77.044902} {\bibfield  {journal} {\bibinfo  {journal} {Phys.
  Rev. C}\ }\textbf {\bibinfo {volume} {77}},\ \bibinfo {pages} {044902}
  (\bibinfo {year} {2008})}\BibitemShut {NoStop}%
\bibitem [{\citenamefont {Adamczyk}\ \emph {et~al.}(2017)\citenamefont
  {Adamczyk} \emph {et~al.}}]{STAR:2017ckg}%
  \BibitemOpen
  \bibfield  {author} {\bibinfo {author} {\bibfnamefont {L.}~\bibnamefont
  {Adamczyk}} \emph {et~al.} (\bibinfo {collaboration} {STAR}),\ }\href
  {\doibase 10.1038/nature23004} {\bibfield  {journal} {\bibinfo  {journal}
  {Nature}\ }\textbf {\bibinfo {volume} {548}},\ \bibinfo {pages} {62}
  (\bibinfo {year} {2017})}\BibitemShut {NoStop}%
\bibitem [{\citenamefont {Adam}\ \emph {et~al.}(2018)\citenamefont {Adam} \emph
  {et~al.}}]{STAR:2018gyt}%
  \BibitemOpen
  \bibfield  {author} {\bibinfo {author} {\bibfnamefont {J.}~\bibnamefont
  {Adam}} \emph {et~al.} (\bibinfo {collaboration} {STAR}),\ }\href {\doibase
  10.1103/PhysRevC.98.014910} {\bibfield  {journal} {\bibinfo  {journal} {Phys.
  Rev. C}\ }\textbf {\bibinfo {volume} {98}},\ \bibinfo {pages} {014910}
  (\bibinfo {year} {2018})}\BibitemShut {NoStop}%
\bibitem [{\citenamefont {Adam}\ \emph {et~al.}(2019)\citenamefont {Adam} \emph
  {et~al.}}]{STAR:2019erd}%
  \BibitemOpen
  \bibfield  {author} {\bibinfo {author} {\bibfnamefont {J.}~\bibnamefont
  {Adam}} \emph {et~al.} (\bibinfo {collaboration} {STAR}),\ }\href {\doibase
  10.1103/PhysRevLett.123.132301} {\bibfield  {journal} {\bibinfo  {journal}
  {Phys. Rev. Lett.}\ }\textbf {\bibinfo {volume} {123}},\ \bibinfo {pages}
  {132301} (\bibinfo {year} {2019})}\BibitemShut {NoStop}%
\bibitem [{\citenamefont {Acharya}\ \emph
  {et~al.}(2020{\natexlab{a}})\citenamefont {Acharya} \emph
  {et~al.}}]{ALICE:2019onw}%
  \BibitemOpen
  \bibfield  {author} {\bibinfo {author} {\bibfnamefont {S.}~\bibnamefont
  {Acharya}} \emph {et~al.} (\bibinfo {collaboration} {ALICE}),\ }\href
  {\doibase 10.1103/PhysRevC.101.044611} {\bibfield  {journal} {\bibinfo
  {journal} {Phys. Rev. C}\ }\textbf {\bibinfo {volume} {101}},\ \bibinfo
  {pages} {044611} (\bibinfo {year} {2020}{\natexlab{a}})},\ \bibinfo {note}
  {[Erratum: Phys.Rev.C 105, 029902 (2022)]}\BibitemShut {NoStop}%
\bibitem [{\citenamefont {Karpenko}\ and\ \citenamefont
  {Becattini}(2017)}]{Karpenko:2016jyx}%
  \BibitemOpen
  \bibfield  {author} {\bibinfo {author} {\bibfnamefont {I.}~\bibnamefont
  {Karpenko}}\ and\ \bibinfo {author} {\bibfnamefont {F.}~\bibnamefont
  {Becattini}},\ }\href {\doibase 10.1140/epjc/s10052-017-4765-1} {\bibfield
  {journal} {\bibinfo  {journal} {Eur. Phys. J. C}\ }\textbf {\bibinfo {volume}
  {77}},\ \bibinfo {pages} {213} (\bibinfo {year} {2017})}\BibitemShut
  {NoStop}%
\bibitem [{\citenamefont {Li}\ \emph {et~al.}(2017)\citenamefont {Li},
  \citenamefont {Pang}, \citenamefont {Wang},\ and\ \citenamefont
  {Xia}}]{Li:2017slc}%
  \BibitemOpen
  \bibfield  {author} {\bibinfo {author} {\bibfnamefont {H.}~\bibnamefont
  {Li}}, \bibinfo {author} {\bibfnamefont {L.-G.}\ \bibnamefont {Pang}},
  \bibinfo {author} {\bibfnamefont {Q.}~\bibnamefont {Wang}}, \ and\ \bibinfo
  {author} {\bibfnamefont {X.-L.}\ \bibnamefont {Xia}},\ }\href {\doibase
  10.1103/PhysRevC.96.054908} {\bibfield  {journal} {\bibinfo  {journal} {Phys.
  Rev. C}\ }\textbf {\bibinfo {volume} {96}},\ \bibinfo {pages} {054908}
  (\bibinfo {year} {2017})}\BibitemShut {NoStop}%
\bibitem [{\citenamefont {Sun}\ and\ \citenamefont {Ko}(2017)}]{Sun:2017xhx}%
  \BibitemOpen
  \bibfield  {author} {\bibinfo {author} {\bibfnamefont {Y.}~\bibnamefont
  {Sun}}\ and\ \bibinfo {author} {\bibfnamefont {C.~M.}\ \bibnamefont {Ko}},\
  }\href {\doibase 10.1103/PhysRevC.96.024906} {\bibfield  {journal} {\bibinfo
  {journal} {Phys. Rev. C}\ }\textbf {\bibinfo {volume} {96}},\ \bibinfo
  {pages} {024906} (\bibinfo {year} {2017})}\BibitemShut {NoStop}%
\bibitem [{\citenamefont {Wei}\ \emph {et~al.}(2019)\citenamefont {Wei},
  \citenamefont {Deng},\ and\ \citenamefont {Huang}}]{Wei:2018zfb}%
  \BibitemOpen
  \bibfield  {author} {\bibinfo {author} {\bibfnamefont {D.-X.}\ \bibnamefont
  {Wei}}, \bibinfo {author} {\bibfnamefont {W.-T.}\ \bibnamefont {Deng}}, \
  and\ \bibinfo {author} {\bibfnamefont {X.-G.}\ \bibnamefont {Huang}},\ }\href
  {\doibase 10.1103/PhysRevC.99.014905} {\bibfield  {journal} {\bibinfo
  {journal} {Phys. Rev. C}\ }\textbf {\bibinfo {volume} {99}},\ \bibinfo
  {pages} {014905} (\bibinfo {year} {2019})}\BibitemShut {NoStop}%
\bibitem [{\citenamefont {Deng}\ \emph {et~al.}(2022)\citenamefont {Deng},
  \citenamefont {Huang},\ and\ \citenamefont {Ma}}]{Deng:2021miw}%
  \BibitemOpen
  \bibfield  {author} {\bibinfo {author} {\bibfnamefont {X.-G.}\ \bibnamefont
  {Deng}}, \bibinfo {author} {\bibfnamefont {X.-G.}\ \bibnamefont {Huang}}, \
  and\ \bibinfo {author} {\bibfnamefont {Y.-G.}\ \bibnamefont {Ma}},\ }\href
  {\doibase 10.1016/j.physletb.2022.137560} {\bibfield  {journal} {\bibinfo
  {journal} {Phys. Lett. B}\ }\textbf {\bibinfo {volume} {835}},\ \bibinfo
  {pages} {137560} (\bibinfo {year} {2022})}\BibitemShut {NoStop}%
\bibitem [{\citenamefont {Sun}\ \emph {et~al.}(2025)\citenamefont {Sun},
  \citenamefont {Liu}, \citenamefont {Zheng}, \citenamefont {Chen},
  \citenamefont {Ko},\ and\ \citenamefont {Ma}}]{Sun:2025oib}%
  \BibitemOpen
  \bibfield  {author} {\bibinfo {author} {\bibfnamefont {K.-J.}\ \bibnamefont
  {Sun}}, \bibinfo {author} {\bibfnamefont {D.-N.}\ \bibnamefont {Liu}},
  \bibinfo {author} {\bibfnamefont {Y.-P.}\ \bibnamefont {Zheng}}, \bibinfo
  {author} {\bibfnamefont {J.-H.}\ \bibnamefont {Chen}}, \bibinfo {author}
  {\bibfnamefont {C.~M.}\ \bibnamefont {Ko}}, \ and\ \bibinfo {author}
  {\bibfnamefont {Y.-G.}\ \bibnamefont {Ma}},\ }\href {\doibase
  10.1103/PhysRevLett.134.022301} {\bibfield  {journal} {\bibinfo  {journal}
  {Phys. Rev. Lett.}\ }\textbf {\bibinfo {volume} {134}},\ \bibinfo {pages}
  {022301} (\bibinfo {year} {2025})}\BibitemShut {NoStop}%
\bibitem [{\citenamefont {Deng}\ and\ \citenamefont {Ma}(2025)}]{Deng:2025xfo}%
  \BibitemOpen
  \bibfield  {author} {\bibinfo {author} {\bibfnamefont {X.~G.}\ \bibnamefont
  {Deng}}\ and\ \bibinfo {author} {\bibfnamefont {Y.~G.}\ \bibnamefont {Ma}},\
  }\href@noop {} {\  (\bibinfo {year} {2025})},\ \Eprint
  {http://arxiv.org/abs/2508.19105} {arXiv:2508.19105 [nucl-th]} \BibitemShut
  {NoStop}%
\bibitem [{\citenamefont {Adam}\ \emph {et~al.}(2021)\citenamefont {Adam} \emph
  {et~al.}}]{STAR:2020xbm}%
  \BibitemOpen
  \bibfield  {author} {\bibinfo {author} {\bibfnamefont {J.}~\bibnamefont
  {Adam}} \emph {et~al.} (\bibinfo {collaboration} {STAR}),\ }\href {\doibase
  10.1103/PhysRevLett.126.162301} {\bibfield  {journal} {\bibinfo  {journal}
  {Phys. Rev. Lett.}\ }\textbf {\bibinfo {volume} {126}},\ \bibinfo {pages}
  {162301} (\bibinfo {year} {2021})},\ \bibinfo {note} {[Erratum:
  Phys.Rev.Lett. 131, 089901 (2023)]}\BibitemShut {NoStop}%
\bibitem [{\citenamefont {Abelev}\ \emph {et~al.}(2008)\citenamefont {Abelev}
  \emph {et~al.}}]{STAR:2008lcm}%
  \BibitemOpen
  \bibfield  {author} {\bibinfo {author} {\bibfnamefont {B.~I.}\ \bibnamefont
  {Abelev}} \emph {et~al.} (\bibinfo {collaboration} {STAR}),\ }\href {\doibase
  10.1103/PhysRevC.77.061902} {\bibfield  {journal} {\bibinfo  {journal} {Phys.
  Rev. C}\ }\textbf {\bibinfo {volume} {77}},\ \bibinfo {pages} {061902}
  (\bibinfo {year} {2008})}\BibitemShut {NoStop}%
\bibitem [{\citenamefont {Acharya}\ \emph
  {et~al.}(2020{\natexlab{b}})\citenamefont {Acharya} \emph
  {et~al.}}]{ALICE:2019aid}%
  \BibitemOpen
  \bibfield  {author} {\bibinfo {author} {\bibfnamefont {S.}~\bibnamefont
  {Acharya}} \emph {et~al.} (\bibinfo {collaboration} {ALICE}),\ }\href
  {\doibase 10.1103/PhysRevLett.125.012301} {\bibfield  {journal} {\bibinfo
  {journal} {Phys. Rev. Lett.}\ }\textbf {\bibinfo {volume} {125}},\ \bibinfo
  {pages} {012301} (\bibinfo {year} {2020}{\natexlab{b}})}\BibitemShut
  {NoStop}%
\bibitem [{\citenamefont {Abdallah}\ \emph {et~al.}(2023)\citenamefont
  {Abdallah} \emph {et~al.}}]{STAR:2022fan}%
  \BibitemOpen
  \bibfield  {author} {\bibinfo {author} {\bibfnamefont {M.~S.}\ \bibnamefont
  {Abdallah}} \emph {et~al.} (\bibinfo {collaboration} {STAR}),\ }\href
  {\doibase 10.1038/s41586-022-05557-5} {\bibfield  {journal} {\bibinfo
  {journal} {Nature}\ }\textbf {\bibinfo {volume} {614}},\ \bibinfo {pages}
  {244} (\bibinfo {year} {2023})}\BibitemShut {NoStop}%
\bibitem [{\citenamefont {Sheng}\ \emph {et~al.}(2023)\citenamefont {Sheng},
  \citenamefont {Oliva}, \citenamefont {Liang}, \citenamefont {Wang},\ and\
  \citenamefont {Wang}}]{Sheng:2022wsy}%
  \BibitemOpen
  \bibfield  {author} {\bibinfo {author} {\bibfnamefont {X.-L.}\ \bibnamefont
  {Sheng}}, \bibinfo {author} {\bibfnamefont {L.}~\bibnamefont {Oliva}},
  \bibinfo {author} {\bibfnamefont {Z.-T.}\ \bibnamefont {Liang}}, \bibinfo
  {author} {\bibfnamefont {Q.}~\bibnamefont {Wang}}, \ and\ \bibinfo {author}
  {\bibfnamefont {X.-N.}\ \bibnamefont {Wang}},\ }\href {\doibase
  10.1103/PhysRevLett.131.042304} {\bibfield  {journal} {\bibinfo  {journal}
  {Phys. Rev. Lett.}\ }\textbf {\bibinfo {volume} {131}},\ \bibinfo {pages}
  {042304} (\bibinfo {year} {2023})}\BibitemShut {NoStop}%
\bibitem [{\citenamefont {Acharya}\ \emph {et~al.}(2021)\citenamefont {Acharya}
  \emph {et~al.}}]{ALICE:2020iev}%
  \BibitemOpen
  \bibfield  {author} {\bibinfo {author} {\bibfnamefont {S.}~\bibnamefont
  {Acharya}} \emph {et~al.} (\bibinfo {collaboration} {ALICE}),\ }\href
  {\doibase 10.1016/j.physletb.2021.136146} {\bibfield  {journal} {\bibinfo
  {journal} {Phys. Lett. B}\ }\textbf {\bibinfo {volume} {815}},\ \bibinfo
  {pages} {136146} (\bibinfo {year} {2021})}\BibitemShut {NoStop}%
\bibitem [{\citenamefont {Acharya}\ \emph {et~al.}(2023)\citenamefont {Acharya}
  \emph {et~al.}}]{ALICE:2022dyy}%
  \BibitemOpen
  \bibfield  {author} {\bibinfo {author} {\bibfnamefont {S.}~\bibnamefont
  {Acharya}} \emph {et~al.} (\bibinfo {collaboration} {ALICE}),\ }\href
  {\doibase 10.1103/PhysRevLett.131.042303} {\bibfield  {journal} {\bibinfo
  {journal} {Phys. Rev. Lett.}\ }\textbf {\bibinfo {volume} {131}},\ \bibinfo
  {pages} {042303} (\bibinfo {year} {2023})}\BibitemShut {NoStop}%
\bibitem [{\citenamefont {Acharya}\ \emph {et~al.}()\citenamefont {Acharya}
  \emph {et~al.}}]{ALICE:2025cdf}%
  \BibitemOpen
  \bibfield  {author} {\bibinfo {author} {\bibfnamefont {S.}~\bibnamefont
  {Acharya}} \emph {et~al.} (\bibinfo {collaboration} {ALICE}),\ }\href@noop {}
  {\ }\Eprint {http://arxiv.org/abs/2504.00714} {arXiv:2504.00714 [nucl-ex]}
  \BibitemShut {NoStop}%
\bibitem [{\citenamefont {Becattini}\ and\ \citenamefont
  {Piccinini}(2008)}]{Becattini:2007nd}%
  \BibitemOpen
  \bibfield  {author} {\bibinfo {author} {\bibfnamefont {F.}~\bibnamefont
  {Becattini}}\ and\ \bibinfo {author} {\bibfnamefont {F.}~\bibnamefont
  {Piccinini}},\ }\href {\doibase 10.1016/j.aop.2008.01.001} {\bibfield
  {journal} {\bibinfo  {journal} {Annals Phys.}\ }\textbf {\bibinfo {volume}
  {323}},\ \bibinfo {pages} {2452} (\bibinfo {year} {2008})}\BibitemShut
  {NoStop}%
\bibitem [{\citenamefont {Huang}(2021)}]{Huang:2020xyr}%
  \BibitemOpen
  \bibfield  {author} {\bibinfo {author} {\bibfnamefont {X.-G.}\ \bibnamefont
  {Huang}},\ }\href {\doibase 10.1016/j.nuclphysa.2020.121752} {\bibfield
  {journal} {\bibinfo  {journal} {Nucl. Phys. A}\ }\textbf {\bibinfo {volume}
  {1005}},\ \bibinfo {pages} {121752} (\bibinfo {year} {2021})}\BibitemShut
  {NoStop}%
\bibitem [{\citenamefont {Becattini}\ and\ \citenamefont
  {Lisa}(2020)}]{Becattini:2020ngo}%
  \BibitemOpen
  \bibfield  {author} {\bibinfo {author} {\bibfnamefont {F.}~\bibnamefont
  {Becattini}}\ and\ \bibinfo {author} {\bibfnamefont {M.~A.}\ \bibnamefont
  {Lisa}},\ }\href {\doibase 10.1146/annurev-nucl-021920-095245} {\bibfield
  {journal} {\bibinfo  {journal} {Ann. Rev. Nucl. Part. Sci.}\ }\textbf
  {\bibinfo {volume} {70}},\ \bibinfo {pages} {395} (\bibinfo {year}
  {2020})}\BibitemShut {NoStop}%
\bibitem [{\citenamefont {Huang}\ \emph {et~al.}(2021)\citenamefont {Huang},
  \citenamefont {Liao}, \citenamefont {Wang},\ and\ \citenamefont
  {Xia}}]{Huang:2020dtn}%
  \BibitemOpen
  \bibfield  {author} {\bibinfo {author} {\bibfnamefont {X.-G.}\ \bibnamefont
  {Huang}}, \bibinfo {author} {\bibfnamefont {J.}~\bibnamefont {Liao}},
  \bibinfo {author} {\bibfnamefont {Q.}~\bibnamefont {Wang}}, \ and\ \bibinfo
  {author} {\bibfnamefont {X.-L.}\ \bibnamefont {Xia}},\ }\href {\doibase
  10.1007/978-3-030-71427-7_9} {\bibfield  {journal} {\bibinfo  {journal}
  {Lect. Notes Phys.}\ }\textbf {\bibinfo {volume} {987}},\ \bibinfo {pages}
  {281} (\bibinfo {year} {2021})}\BibitemShut {NoStop}%
\bibitem [{\citenamefont {Gao}\ \emph {et~al.}(2020)\citenamefont {Gao},
  \citenamefont {Ma}, \citenamefont {Pu},\ and\ \citenamefont
  {Wang}}]{Gao:2020vbh}%
  \BibitemOpen
  \bibfield  {author} {\bibinfo {author} {\bibfnamefont {J.-H.}\ \bibnamefont
  {Gao}}, \bibinfo {author} {\bibfnamefont {G.-L.}\ \bibnamefont {Ma}},
  \bibinfo {author} {\bibfnamefont {S.}~\bibnamefont {Pu}}, \ and\ \bibinfo
  {author} {\bibfnamefont {Q.}~\bibnamefont {Wang}},\ }\href {\doibase
  10.1007/s41365-020-00801-x} {\bibfield  {journal} {\bibinfo  {journal} {Nucl.
  Sci. Tech.}\ }\textbf {\bibinfo {volume} {31}},\ \bibinfo {pages} {90}
  (\bibinfo {year} {2020})}\BibitemShut {NoStop}%
\bibitem [{\citenamefont {Becattini}\ \emph {et~al.}(2021)\citenamefont
  {Becattini}, \citenamefont {Liao},\ and\ \citenamefont
  {Lisa}}]{Becattini:2021wqt}%
  \BibitemOpen
  \bibinfo {editor} {\bibfnamefont {F.}~\bibnamefont {Becattini}}, \bibinfo
  {editor} {\bibfnamefont {J.}~\bibnamefont {Liao}}, \ and\ \bibinfo {editor}
  {\bibfnamefont {M.~A.}\ \bibnamefont {Lisa}},\ eds.,\ \href {\doibase
  10.1007/978-3-030-71427-7} {\emph {\bibinfo {title} {{Strongly Interacting
  Matter under Rotation}}}}\ (\bibinfo  {publisher} {Springer},\ \bibinfo
  {year} {2021})\BibitemShut {NoStop}%
\bibitem [{\citenamefont {Hidaka}\ \emph {et~al.}(2022)\citenamefont {Hidaka},
  \citenamefont {Pu}, \citenamefont {Wang},\ and\ \citenamefont
  {Yang}}]{Hidaka:2022dmn}%
  \BibitemOpen
  \bibfield  {author} {\bibinfo {author} {\bibfnamefont {Y.}~\bibnamefont
  {Hidaka}}, \bibinfo {author} {\bibfnamefont {S.}~\bibnamefont {Pu}}, \bibinfo
  {author} {\bibfnamefont {Q.}~\bibnamefont {Wang}}, \ and\ \bibinfo {author}
  {\bibfnamefont {D.-L.}\ \bibnamefont {Yang}},\ }\href {\doibase
  10.1016/j.ppnp.2022.103989} {\bibfield  {journal} {\bibinfo  {journal} {Prog.
  Part. Nucl. Phys.}\ }\textbf {\bibinfo {volume} {127}},\ \bibinfo {pages}
  {103989} (\bibinfo {year} {2022})}\BibitemShut {NoStop}%
\bibitem [{\citenamefont {Chen}\ \emph {et~al.}(2023)\citenamefont {Chen},
  \citenamefont {Liang}, \citenamefont {Ma},\ and\ \citenamefont
  {Wang}}]{Chen:2023hnb}%
  \BibitemOpen
  \bibfield  {author} {\bibinfo {author} {\bibfnamefont {J.}~\bibnamefont
  {Chen}}, \bibinfo {author} {\bibfnamefont {Z.-T.}\ \bibnamefont {Liang}},
  \bibinfo {author} {\bibfnamefont {Y.-G.}\ \bibnamefont {Ma}}, \ and\ \bibinfo
  {author} {\bibfnamefont {Q.}~\bibnamefont {Wang}},\ }\href {\doibase
  10.1016/j.scib.2023.04.001} {\bibfield  {journal} {\bibinfo  {journal} {Sci.
  Bull.}\ }\textbf {\bibinfo {volume} {68}},\ \bibinfo {pages} {874} (\bibinfo
  {year} {2023})}\BibitemShut {NoStop}%
\bibitem [{\citenamefont {Becattini}\ \emph {et~al.}(2024)\citenamefont
  {Becattini}, \citenamefont {Buzzegoli}, \citenamefont {Niida}, \citenamefont
  {Pu}, \citenamefont {Tang},\ and\ \citenamefont {Wang}}]{Becattini:2024uha}%
  \BibitemOpen
  \bibfield  {author} {\bibinfo {author} {\bibfnamefont {F.}~\bibnamefont
  {Becattini}}, \bibinfo {author} {\bibfnamefont {M.}~\bibnamefont
  {Buzzegoli}}, \bibinfo {author} {\bibfnamefont {T.}~\bibnamefont {Niida}},
  \bibinfo {author} {\bibfnamefont {S.}~\bibnamefont {Pu}}, \bibinfo {author}
  {\bibfnamefont {A.-H.}\ \bibnamefont {Tang}}, \ and\ \bibinfo {author}
  {\bibfnamefont {Q.}~\bibnamefont {Wang}},\ }\href {\doibase
  10.1142/9789811294679_0005} {\bibfield  {journal} {\bibinfo  {journal} {Int.
  J. Mod. Phys. E}\ }\textbf {\bibinfo {volume} {33}},\ \bibinfo {pages}
  {2430006} (\bibinfo {year} {2024})}\BibitemShut {NoStop}%
\bibitem [{\citenamefont {Niida}\ and\ \citenamefont
  {Voloshin}(2024)}]{Niida:2024ntm}%
  \BibitemOpen
  \bibfield  {author} {\bibinfo {author} {\bibfnamefont {T.}~\bibnamefont
  {Niida}}\ and\ \bibinfo {author} {\bibfnamefont {S.~A.}\ \bibnamefont
  {Voloshin}},\ }\href {\doibase 10.1142/S0218301324300108} {\bibfield
  {journal} {\bibinfo  {journal} {Int. J. Mod. Phys. E}\ }\textbf {\bibinfo
  {volume} {33}},\ \bibinfo {pages} {2430010} (\bibinfo {year}
  {2024})}\BibitemShut {NoStop}%
\bibitem [{\citenamefont {Chen}\ \emph {et~al.}(2024)\citenamefont {Chen} \emph
  {et~al.}}]{Chen:2024aom}%
  \BibitemOpen
  \bibfield  {author} {\bibinfo {author} {\bibfnamefont {J.}~\bibnamefont
  {Chen}} \emph {et~al.},\ }\href {\doibase 10.1007/s41365-024-01591-2}
  {\bibfield  {journal} {\bibinfo  {journal} {Nucl. Sci. Tech.}\ }\textbf
  {\bibinfo {volume} {35}},\ \bibinfo {pages} {214} (\bibinfo {year}
  {2024})}\BibitemShut {NoStop}%
\bibitem [{\citenamefont {Chen}\ \emph
  {et~al.}(2025{\natexlab{a}})\citenamefont {Chen}, \citenamefont {Liang},
  \citenamefont {Ma}, \citenamefont {Sheng},\ and\ \citenamefont
  {Wang}}]{Chen:2024afy}%
  \BibitemOpen
  \bibfield  {author} {\bibinfo {author} {\bibfnamefont {J.-H.}\ \bibnamefont
  {Chen}}, \bibinfo {author} {\bibfnamefont {Z.-T.}\ \bibnamefont {Liang}},
  \bibinfo {author} {\bibfnamefont {Y.-G.}\ \bibnamefont {Ma}}, \bibinfo
  {author} {\bibfnamefont {X.-L.}\ \bibnamefont {Sheng}}, \ and\ \bibinfo
  {author} {\bibfnamefont {Q.}~\bibnamefont {Wang}},\ }\href {\doibase
  10.1007/s11433-024-2495-1} {\bibfield  {journal} {\bibinfo  {journal} {Sci.
  China Phys. Mech. Astron.}\ }\textbf {\bibinfo {volume} {68}},\ \bibinfo
  {pages} {211001} (\bibinfo {year} {2025}{\natexlab{a}})}\BibitemShut
  {NoStop}%
\bibitem [{\citenamefont {Huang}(2025)}]{Huang:2024ffg}%
  \BibitemOpen
  \bibfield  {author} {\bibinfo {author} {\bibfnamefont {X.-G.}\ \bibnamefont
  {Huang}},\ }\href {\doibase 10.1007/s41365-025-01784-3} {\bibfield  {journal}
  {\bibinfo  {journal} {Nucl. Sci. Tech.}\ }\textbf {\bibinfo {volume} {36}},\
  \bibinfo {pages} {208} (\bibinfo {year} {2025})}\BibitemShut {NoStop}%
\bibitem [{\citenamefont {Liu}\ and\ \citenamefont {Xu}(2024)}]{Liu:2023nkm}%
  \BibitemOpen
  \bibfield  {author} {\bibinfo {author} {\bibfnamefont {R.-J.}\ \bibnamefont
  {Liu}}\ and\ \bibinfo {author} {\bibfnamefont {J.}~\bibnamefont {Xu}},\
  }\href {\doibase 10.1103/PhysRevC.109.014615} {\bibfield  {journal} {\bibinfo
   {journal} {Phys. Rev. C}\ }\textbf {\bibinfo {volume} {109}},\ \bibinfo
  {pages} {014615} (\bibinfo {year} {2024})}\BibitemShut {NoStop}%
\bibitem [{\citenamefont {Liang}\ \emph {et~al.}(2025)\citenamefont {Liang},
  \citenamefont {Lv}, \citenamefont {Kupsc}, \citenamefont {Gou},\ and\
  \citenamefont {Li}}]{Liang:2025owx}%
  \BibitemOpen
  \bibfield  {author} {\bibinfo {author} {\bibfnamefont {Y.-T.}\ \bibnamefont
  {Liang}}, \bibinfo {author} {\bibfnamefont {X.-R.}\ \bibnamefont {Lv}},
  \bibinfo {author} {\bibfnamefont {A.}~\bibnamefont {Kupsc}}, \bibinfo
  {author} {\bibfnamefont {B.}~\bibnamefont {Gou}}, \ and\ \bibinfo {author}
  {\bibfnamefont {H.-B.}\ \bibnamefont {Li}},\ }\href {\doibase
  10.1103/c642-1lzb} {\bibfield  {journal} {\bibinfo  {journal} {Phys. Rev. D}\
  }\textbf {\bibinfo {volume} {112}},\ \bibinfo {pages} {L031502} (\bibinfo
  {year} {2025})}\BibitemShut {NoStop}%
\bibitem [{\citenamefont {Liu}\ \emph {et~al.}()\citenamefont {Liu},
  \citenamefont {Zheng}, \citenamefont {Zhou}, \citenamefont {Chen},
  \citenamefont {Ko}, \citenamefont {Ma}, \citenamefont {Sun},\ and\
  \citenamefont {Zhang}}]{Liu:2025kpp}%
  \BibitemOpen
  \bibfield  {author} {\bibinfo {author} {\bibfnamefont {D.-N.}\ \bibnamefont
  {Liu}}, \bibinfo {author} {\bibfnamefont {Y.-P.}\ \bibnamefont {Zheng}},
  \bibinfo {author} {\bibfnamefont {W.-H.}\ \bibnamefont {Zhou}}, \bibinfo
  {author} {\bibfnamefont {J.-H.}\ \bibnamefont {Chen}}, \bibinfo {author}
  {\bibfnamefont {C.~M.}\ \bibnamefont {Ko}}, \bibinfo {author} {\bibfnamefont
  {Y.-G.}\ \bibnamefont {Ma}}, \bibinfo {author} {\bibfnamefont {K.-J.}\
  \bibnamefont {Sun}}, \ and\ \bibinfo {author} {\bibfnamefont
  {S.}~\bibnamefont {Zhang}},\ }\href@noop {} {\ }\Eprint
  {http://arxiv.org/abs/2508.12193} {arXiv:2508.12193 [nucl-th]} \BibitemShut
  {NoStop}%
\bibitem [{\citenamefont {Tilley}\ \emph {et~al.}(1992)\citenamefont {Tilley},
  \citenamefont {Weller},\ and\ \citenamefont {Hale}}]{Tilley:1992zz}%
  \BibitemOpen
  \bibfield  {author} {\bibinfo {author} {\bibfnamefont {D.~R.}\ \bibnamefont
  {Tilley}}, \bibinfo {author} {\bibfnamefont {H.~R.}\ \bibnamefont {Weller}},
  \ and\ \bibinfo {author} {\bibfnamefont {G.~M.}\ \bibnamefont {Hale}},\
  }\href {\doibase 10.1016/0375-9474(92)90635-W} {\bibfield  {journal}
  {\bibinfo  {journal} {Nucl. Phys. A}\ }\textbf {\bibinfo {volume} {541}},\
  \bibinfo {pages} {1} (\bibinfo {year} {1992})}\BibitemShut {NoStop}%
\bibitem [{\citenamefont {Vovchenko}\ \emph {et~al.}(2020)\citenamefont
  {Vovchenko}, \citenamefont {D\"onigus}, \citenamefont {Kardan}, \citenamefont
  {Lorenz},\ and\ \citenamefont {Stoecker}}]{Vovchenko:2020dmv}%
  \BibitemOpen
  \bibfield  {author} {\bibinfo {author} {\bibfnamefont {V.}~\bibnamefont
  {Vovchenko}}, \bibinfo {author} {\bibfnamefont {B.}~\bibnamefont
  {D\"onigus}}, \bibinfo {author} {\bibfnamefont {B.}~\bibnamefont {Kardan}},
  \bibinfo {author} {\bibfnamefont {M.}~\bibnamefont {Lorenz}}, \ and\ \bibinfo
  {author} {\bibfnamefont {H.}~\bibnamefont {Stoecker}},\ }\href {\doibase
  10.1016/j.physletb.2020.135746} {\bibfield  {journal} {\bibinfo  {journal}
  {Phys. Lett.}\ }\textbf {\bibinfo {volume} {B}},\ \bibinfo {pages} {135746}
  (\bibinfo {year} {2020})}\BibitemShut {NoStop}%
\bibitem [{\citenamefont {Xi}\ \emph {et~al.}(2020)\citenamefont {Xi},
  \citenamefont {Zhang}, \citenamefont {Zhang},\ and\ \citenamefont
  {Ma}}]{Xi:2019vev}%
  \BibitemOpen
  \bibfield  {author} {\bibinfo {author} {\bibfnamefont {B.-S.}\ \bibnamefont
  {Xi}}, \bibinfo {author} {\bibfnamefont {Z.-Q.}\ \bibnamefont {Zhang}},
  \bibinfo {author} {\bibfnamefont {S.}~\bibnamefont {Zhang}}, \ and\ \bibinfo
  {author} {\bibfnamefont {Y.-G.}\ \bibnamefont {Ma}},\ }\href {\doibase
  10.1103/PhysRevC.102.064901} {\bibfield  {journal} {\bibinfo  {journal}
  {Phys. Rev. C}\ }\textbf {\bibinfo {volume} {102}},\ \bibinfo {pages}
  {064901} (\bibinfo {year} {2020})}\BibitemShut {NoStop}%
\bibitem [{\citenamefont {Pilkuhn}(1967)}]{pilkuhn1967interaction}%
  \BibitemOpen
  \bibfield  {author} {\bibinfo {author} {\bibfnamefont {H.}~\bibnamefont
  {Pilkuhn}},\ }\href@noop {} {\emph {\bibinfo {title} {The Interaction of
  Hadrons}}}\ (\bibinfo  {publisher} {North-Holland Publishing Company},\
  \bibinfo {address} {Amsterdam},\ \bibinfo {year} {1967})\BibitemShut
  {NoStop}%
\bibitem [{\citenamefont {Lee}\ \emph {et~al.}(1957)\citenamefont {Lee},
  \citenamefont {Steinberger}, \citenamefont {Feinberg}, \citenamefont
  {Kabir},\ and\ \citenamefont {Yang}}]{Lee:1957he}%
  \BibitemOpen
  \bibfield  {author} {\bibinfo {author} {\bibfnamefont {T.~D.}\ \bibnamefont
  {Lee}}, \bibinfo {author} {\bibfnamefont {J.}~\bibnamefont {Steinberger}},
  \bibinfo {author} {\bibfnamefont {G.}~\bibnamefont {Feinberg}}, \bibinfo
  {author} {\bibfnamefont {P.~K.}\ \bibnamefont {Kabir}}, \ and\ \bibinfo
  {author} {\bibfnamefont {C.-N.}\ \bibnamefont {Yang}},\ }\href {\doibase
  10.1103/PhysRev.106.1367} {\bibfield  {journal} {\bibinfo  {journal} {Phys.
  Rev.}\ }\textbf {\bibinfo {volume} {106}},\ \bibinfo {pages} {1367} (\bibinfo
  {year} {1957})}\BibitemShut {NoStop}%
\bibitem [{\citenamefont {Yang}\ \emph {et~al.}(2018)\citenamefont {Yang},
  \citenamefont {Fang}, \citenamefont {Wang},\ and\ \citenamefont
  {Wang}}]{Yang:2017sdk}%
  \BibitemOpen
  \bibfield  {author} {\bibinfo {author} {\bibfnamefont {Y.-G.}\ \bibnamefont
  {Yang}}, \bibinfo {author} {\bibfnamefont {R.-H.}\ \bibnamefont {Fang}},
  \bibinfo {author} {\bibfnamefont {Q.}~\bibnamefont {Wang}}, \ and\ \bibinfo
  {author} {\bibfnamefont {X.-N.}\ \bibnamefont {Wang}},\ }\href {\doibase
  10.1103/PhysRevC.97.034917} {\bibfield  {journal} {\bibinfo  {journal} {Phys.
  Rev. C}\ }\textbf {\bibinfo {volume} {97}},\ \bibinfo {pages} {034917}
  (\bibinfo {year} {2018})}\BibitemShut {NoStop}%
\bibitem [{\citenamefont {Sheng}\ \emph {et~al.}(2020)\citenamefont {Sheng},
  \citenamefont {Wang},\ and\ \citenamefont {Wang}}]{Sheng:2020ghv}%
  \BibitemOpen
  \bibfield  {author} {\bibinfo {author} {\bibfnamefont {X.-L.}\ \bibnamefont
  {Sheng}}, \bibinfo {author} {\bibfnamefont {Q.}~\bibnamefont {Wang}}, \ and\
  \bibinfo {author} {\bibfnamefont {X.-N.}\ \bibnamefont {Wang}},\ }\href
  {\doibase 10.1103/PhysRevD.102.056013} {\bibfield  {journal} {\bibinfo
  {journal} {Phys. Rev. D}\ }\textbf {\bibinfo {volume} {102}},\ \bibinfo
  {pages} {056013} (\bibinfo {year} {2020})}\BibitemShut {NoStop}%
\bibitem [{\citenamefont {Baltz}\ and\ \citenamefont
  {Dover}(1996)}]{Baltz:1995tv}%
  \BibitemOpen
  \bibfield  {author} {\bibinfo {author} {\bibfnamefont {A.~J.}\ \bibnamefont
  {Baltz}}\ and\ \bibinfo {author} {\bibfnamefont {C.}~\bibnamefont {Dover}},\
  }\href {\doibase 10.1103/PhysRevC.53.362} {\bibfield  {journal} {\bibinfo
  {journal} {Phys. Rev. C}\ }\textbf {\bibinfo {volume} {53}},\ \bibinfo
  {pages} {362} (\bibinfo {year} {1996})}\BibitemShut {NoStop}%
\bibitem [{\citenamefont {Becattini}\ \emph {et~al.}(2013)\citenamefont
  {Becattini}, \citenamefont {Chandra}, \citenamefont {Del~Zanna},\ and\
  \citenamefont {Grossi}}]{Becattini:2013fla}%
  \BibitemOpen
  \bibfield  {author} {\bibinfo {author} {\bibfnamefont {F.}~\bibnamefont
  {Becattini}}, \bibinfo {author} {\bibfnamefont {V.}~\bibnamefont {Chandra}},
  \bibinfo {author} {\bibfnamefont {L.}~\bibnamefont {Del~Zanna}}, \ and\
  \bibinfo {author} {\bibfnamefont {E.}~\bibnamefont {Grossi}},\ }\href
  {\doibase 10.1016/j.aop.2013.07.004} {\bibfield  {journal} {\bibinfo
  {journal} {Annals Phys.}\ }\textbf {\bibinfo {volume} {338}},\ \bibinfo
  {pages} {32} (\bibinfo {year} {2013})}\BibitemShut {NoStop}%
\bibitem [{\citenamefont {Becattini}\ \emph {et~al.}(2017)\citenamefont
  {Becattini}, \citenamefont {Karpenko}, \citenamefont {Lisa}, \citenamefont
  {Upsal},\ and\ \citenamefont {Voloshin}}]{Becattini:2016gvu}%
  \BibitemOpen
  \bibfield  {author} {\bibinfo {author} {\bibfnamefont {F.}~\bibnamefont
  {Becattini}}, \bibinfo {author} {\bibfnamefont {I.}~\bibnamefont {Karpenko}},
  \bibinfo {author} {\bibfnamefont {M.}~\bibnamefont {Lisa}}, \bibinfo {author}
  {\bibfnamefont {I.}~\bibnamefont {Upsal}}, \ and\ \bibinfo {author}
  {\bibfnamefont {S.}~\bibnamefont {Voloshin}},\ }\href {\doibase
  10.1103/PhysRevC.95.054902} {\bibfield  {journal} {\bibinfo  {journal} {Phys.
  Rev. C}\ }\textbf {\bibinfo {volume} {95}},\ \bibinfo {pages} {054902}
  (\bibinfo {year} {2017})}\BibitemShut {NoStop}%
\bibitem [{\citenamefont {Groenewold}(1946)}]{Groenewold:1946kp}%
  \BibitemOpen
  \bibfield  {author} {\bibinfo {author} {\bibfnamefont {H.~J.}\ \bibnamefont
  {Groenewold}},\ }\href {\doibase 10.1016/S0031-8914(46)80059-4} {\bibfield
  {journal} {\bibinfo  {journal} {Physica}\ }\textbf {\bibinfo {volume} {12}},\
  \bibinfo {pages} {405} (\bibinfo {year} {1946})}\BibitemShut {NoStop}%
\bibitem [{\citenamefont {Moyal}(1949)}]{Moyal:1949sk}%
  \BibitemOpen
  \bibfield  {author} {\bibinfo {author} {\bibfnamefont {J.~E.}\ \bibnamefont
  {Moyal}},\ }\href {\doibase 10.1017/S0305004100000487} {\bibfield  {journal}
  {\bibinfo  {journal} {Proc. Cambridge Phil. Soc.}\ }\textbf {\bibinfo
  {volume} {45}},\ \bibinfo {pages} {99} (\bibinfo {year} {1949})}\BibitemShut
  {NoStop}%
\bibitem [{\citenamefont {Curtright}\ and\ \citenamefont
  {Zachos}(2012)}]{Curtright:2011vw}%
  \BibitemOpen
  \bibfield  {author} {\bibinfo {author} {\bibfnamefont {T.~L.}\ \bibnamefont
  {Curtright}}\ and\ \bibinfo {author} {\bibfnamefont {C.~K.}\ \bibnamefont
  {Zachos}},\ }\href {\doibase 10.1142/S2251158X12000069} {\bibfield  {journal}
  {\bibinfo  {journal} {Asia Pac. Phys. Newslett.}\ }\textbf {\bibinfo {volume}
  {1}},\ \bibinfo {pages} {37} (\bibinfo {year} {2012})}\BibitemShut {NoStop}%
\bibitem [{\citenamefont {Xu}(2025)}]{Xu:2025uwd}%
  \BibitemOpen
  \bibfield  {author} {\bibinfo {author} {\bibfnamefont {J.}~\bibnamefont
  {Xu}},\ }\href {\doibase 10.1103/PhysRevC.111.L021602} {\bibfield  {journal}
  {\bibinfo  {journal} {Phys. Rev. C}\ }\textbf {\bibinfo {volume} {111}},\
  \bibinfo {pages} {L021602} (\bibinfo {year} {2025})}\BibitemShut {NoStop}%
\bibitem [{\citenamefont {Liu}\ \emph {et~al.}(2025)\citenamefont {Liu},
  \citenamefont {Xu},\ and\ \citenamefont {Ma}}]{Liu:2025vho}%
  \BibitemOpen
  \bibfield  {author} {\bibinfo {author} {\bibfnamefont {R.-J.}\ \bibnamefont
  {Liu}}, \bibinfo {author} {\bibfnamefont {J.}~\bibnamefont {Xu}}, \ and\
  \bibinfo {author} {\bibfnamefont {Y.-G.}\ \bibnamefont {Ma}},\ }\href
  {\doibase 10.1016/j.physletb.2025.139703} {\bibfield  {journal} {\bibinfo
  {journal} {Phys. Lett. B}\ }\textbf {\bibinfo {volume} {868}},\ \bibinfo
  {pages} {139703} (\bibinfo {year} {2025})}\BibitemShut {NoStop}%
\bibitem [{\citenamefont {Liu}\ and\ \citenamefont {Yin}(2021)}]{Liu:2020dxg}%
  \BibitemOpen
  \bibfield  {author} {\bibinfo {author} {\bibfnamefont {S.~Y.~F.}\
  \bibnamefont {Liu}}\ and\ \bibinfo {author} {\bibfnamefont {Y.}~\bibnamefont
  {Yin}},\ }\href {\doibase 10.1103/PhysRevD.104.054043} {\bibfield  {journal}
  {\bibinfo  {journal} {Phys. Rev. D}\ }\textbf {\bibinfo {volume} {104}},\
  \bibinfo {pages} {054043} (\bibinfo {year} {2021})}\BibitemShut {NoStop}%
\bibitem [{\citenamefont {Tachibana}\ \emph {et~al.}()\citenamefont
  {Tachibana}, \citenamefont {Hagino}, \citenamefont {Yoshida},\ and\
  \citenamefont {Zhao}}]{Tachibana:2025wey}%
  \BibitemOpen
  \bibfield  {author} {\bibinfo {author} {\bibfnamefont {T.}~\bibnamefont
  {Tachibana}}, \bibinfo {author} {\bibfnamefont {K.}~\bibnamefont {Hagino}},
  \bibinfo {author} {\bibfnamefont {K.}~\bibnamefont {Yoshida}}, \ and\
  \bibinfo {author} {\bibfnamefont {Q.}~\bibnamefont {Zhao}},\ }\href@noop {}
  {\ }\Eprint {http://arxiv.org/abs/2507.13597} {arXiv:2507.13597 [nucl-th]}
  \BibitemShut {NoStop}%
\bibitem [{\citenamefont {Tombrello}(1965)}]{Tombrello:1965zz}%
  \BibitemOpen
  \bibfield  {author} {\bibinfo {author} {\bibfnamefont {T.~A.}\ \bibnamefont
  {Tombrello}},\ }\href {\doibase 10.1103/PhysRev.138.B40} {\bibfield
  {journal} {\bibinfo  {journal} {Phys. Rev.}\ }\textbf {\bibinfo {volume}
  {138}},\ \bibinfo {pages} {B40} (\bibinfo {year} {1965})}\BibitemShut
  {NoStop}%
\bibitem [{\citenamefont {Alley}\ and\ \citenamefont
  {Knutson}(1993)}]{Alley:1993zz}%
  \BibitemOpen
  \bibfield  {author} {\bibinfo {author} {\bibfnamefont {M.~T.}\ \bibnamefont
  {Alley}}\ and\ \bibinfo {author} {\bibfnamefont {L.~D.}\ \bibnamefont
  {Knutson}},\ }\href {\doibase 10.1103/PhysRevC.48.1901} {\bibfield  {journal}
  {\bibinfo  {journal} {Phys. Rev. C}\ }\textbf {\bibinfo {volume} {48}},\
  \bibinfo {pages} {1901} (\bibinfo {year} {1993})}\BibitemShut {NoStop}%
\bibitem [{\citenamefont {Wang}\ \emph {et~al.}()\citenamefont {Wang},
  \citenamefont {Zhang}, \citenamefont {Ma}, \citenamefont {Chen},
  \citenamefont {Ko},\ and\ \citenamefont {Sun}}]{Wang:2025wim}%
  \BibitemOpen
  \bibfield  {author} {\bibinfo {author} {\bibfnamefont {R.}~\bibnamefont
  {Wang}}, \bibinfo {author} {\bibfnamefont {Z.}~\bibnamefont {Zhang}},
  \bibinfo {author} {\bibfnamefont {Y.-G.}\ \bibnamefont {Ma}}, \bibinfo
  {author} {\bibfnamefont {L.-W.}\ \bibnamefont {Chen}}, \bibinfo {author}
  {\bibfnamefont {C.~M.}\ \bibnamefont {Ko}}, \ and\ \bibinfo {author}
  {\bibfnamefont {K.-J.}\ \bibnamefont {Sun}},\ }\href@noop {} {\ }\Eprint
  {http://arxiv.org/abs/2507.16613} {arXiv:2507.16613 [nucl-th]} \BibitemShut
  {NoStop}%
\bibitem [{\citenamefont {Sun}\ \emph {et~al.}(2024)\citenamefont {Sun},
  \citenamefont {Wang}, \citenamefont {Ko}, \citenamefont {Ma},\ and\
  \citenamefont {Shen}}]{Sun:2022xjr}%
  \BibitemOpen
  \bibfield  {author} {\bibinfo {author} {\bibfnamefont {K.-J.}\ \bibnamefont
  {Sun}}, \bibinfo {author} {\bibfnamefont {R.}~\bibnamefont {Wang}}, \bibinfo
  {author} {\bibfnamefont {C.~M.}\ \bibnamefont {Ko}}, \bibinfo {author}
  {\bibfnamefont {Y.-G.}\ \bibnamefont {Ma}}, \ and\ \bibinfo {author}
  {\bibfnamefont {C.}~\bibnamefont {Shen}},\ }\href {\doibase
  10.1038/s41467-024-45474-x} {\bibfield  {journal} {\bibinfo  {journal}
  {Nature Commun.}\ }\textbf {\bibinfo {volume} {15}},\ \bibinfo {pages} {1074}
  (\bibinfo {year} {2024})}\BibitemShut {NoStop}%
\bibitem [{\citenamefont {Lv}\ \emph {et~al.}(2024)\citenamefont {Lv},
  \citenamefont {Yu}, \citenamefont {Liang}, \citenamefont {Wang},\ and\
  \citenamefont {Wang}}]{Lv:2024uev}%
  \BibitemOpen
  \bibfield  {author} {\bibinfo {author} {\bibfnamefont {J.-p.}\ \bibnamefont
  {Lv}}, \bibinfo {author} {\bibfnamefont {Z.-h.}\ \bibnamefont {Yu}}, \bibinfo
  {author} {\bibfnamefont {Z.-t.}\ \bibnamefont {Liang}}, \bibinfo {author}
  {\bibfnamefont {Q.}~\bibnamefont {Wang}}, \ and\ \bibinfo {author}
  {\bibfnamefont {X.-N.}\ \bibnamefont {Wang}},\ }\href {\doibase
  10.1103/PhysRevD.109.114003} {\bibfield  {journal} {\bibinfo  {journal}
  {Phys. Rev. D}\ }\textbf {\bibinfo {volume} {109}},\ \bibinfo {pages}
  {114003} (\bibinfo {year} {2024})}\BibitemShut {NoStop}%
\bibitem [{\citenamefont {Chen}\ \emph
  {et~al.}(2025{\natexlab{b}})\citenamefont {Chen}, \citenamefont {Fu},
  \citenamefont {Huang},\ and\ \citenamefont {Ma}}]{Chen:2024hki}%
  \BibitemOpen
  \bibfield  {author} {\bibinfo {author} {\bibfnamefont {H.-L.}\ \bibnamefont
  {Chen}}, \bibinfo {author} {\bibfnamefont {W.-j.}\ \bibnamefont {Fu}},
  \bibinfo {author} {\bibfnamefont {X.-G.}\ \bibnamefont {Huang}}, \ and\
  \bibinfo {author} {\bibfnamefont {G.-L.}\ \bibnamefont {Ma}},\ }\href
  {\doibase 10.1103/g1bh-85h4} {\bibfield  {journal} {\bibinfo  {journal}
  {Phys. Rev. Lett.}\ }\textbf {\bibinfo {volume} {135}},\ \bibinfo {pages}
  {032302} (\bibinfo {year} {2025}{\natexlab{b}})}\BibitemShut {NoStop}%
\bibitem [{\citenamefont {Sheng}\ \emph {et~al.}()\citenamefont {Sheng},
  \citenamefont {Wu}, \citenamefont {Rischke},\ and\ \citenamefont
  {Wang}}]{Sheng:2025puj}%
  \BibitemOpen
  \bibfield  {author} {\bibinfo {author} {\bibfnamefont {X.-L.}\ \bibnamefont
  {Sheng}}, \bibinfo {author} {\bibfnamefont {X.-Y.}\ \bibnamefont {Wu}},
  \bibinfo {author} {\bibfnamefont {D.~H.}\ \bibnamefont {Rischke}}, \ and\
  \bibinfo {author} {\bibfnamefont {X.-N.}\ \bibnamefont {Wang}},\ }\href@noop
  {} {\ }\Eprint {http://arxiv.org/abs/2508.03496} {arXiv:2508.03496 [hep-ph]}
  \BibitemShut {NoStop}%
\bibitem [{\citenamefont {Tang}(2025)}]{Tang:2025oav}%
  \BibitemOpen
  \bibfield  {author} {\bibinfo {author} {\bibfnamefont {A.}~\bibnamefont
  {Tang}},\ }\href {\doibase 10.1016/j.physletb.2025.139820} {\bibfield
  {journal} {\bibinfo  {journal} {Phys. Lett. B}\ }\textbf {\bibinfo {volume}
  {868}},\ \bibinfo {pages} {139820} (\bibinfo {year} {2025})}\BibitemShut
  {NoStop}%
\bibitem [{\citenamefont {STAR}()}]{STAR:2025njp}%
  \BibitemOpen
  \bibfield  {author} {\bibinfo {author} {\bibnamefont {STAR}},\ }\href@noop {}
  {\ }\Eprint {http://arxiv.org/abs/2506.05499} {arXiv:2506.05499 [hep-ex]}
  \BibitemShut {NoStop}%
\bibitem [{\citenamefont {Sun}\ \emph {et~al.}(2018)\citenamefont {Sun},
  \citenamefont {Chen}, \citenamefont {Ko}, \citenamefont {Pu},\ and\
  \citenamefont {Xu}}]{Sun:2018jhg}%
  \BibitemOpen
  \bibfield  {author} {\bibinfo {author} {\bibfnamefont {K.-J.}\ \bibnamefont
  {Sun}}, \bibinfo {author} {\bibfnamefont {L.-W.}\ \bibnamefont {Chen}},
  \bibinfo {author} {\bibfnamefont {C.~M.}\ \bibnamefont {Ko}}, \bibinfo
  {author} {\bibfnamefont {J.}~\bibnamefont {Pu}}, \ and\ \bibinfo {author}
  {\bibfnamefont {Z.}~\bibnamefont {Xu}},\ }\href {\doibase
  10.1016/j.physletb.2018.04.035} {\bibfield  {journal} {\bibinfo  {journal}
  {Phys. Lett. B}\ }\textbf {\bibinfo {volume} {781}},\ \bibinfo {pages} {499}
  (\bibinfo {year} {2018})}\BibitemShut {NoStop}%
\end{thebibliography}

\end{document}